\documentclass{ws-ijmpa}
\usepackage[super,compress]{cite}
\usepackage{graphicx,hyperref}
\numberwithin{equation}{section}
\graphicspath{{figs/}}
\newcommand{\Tr}{\mathop{\mathrm{Tr}}\nolimits}
\newcommand{\Li}[1]{\mathop{\mathrm{Li}}\nolimits_{#1}}
\newcommand{\F}[4]{\,_{#1}F_{#2}\left(\left.\begin{array}{c}#3\end{array}\right|#4\right)}

\begin{document}
\markboth{Andrey Grozin}{QCD cusp anomalous dimension: current status}

%
%

\title{QCD cusp anomalous dimension: current status}

\author{Andrey Grozin}

\address{Budker Institute of Nuclear Physics,
Lavrentiev st.~11, Novosibirsk 630090, Russia\\
A.G.Grozin@inp.nsk.su}

\maketitle


\begin{abstract}
Calculation results for the HQET field anomalous dimension
and the QCD cusp anomalous dimension,
as well as their properties, are reviewed.
The HQET field anomalous dimension $\gamma_h$ is known up to 4 loops.
The cusp anomalous dimension $\Gamma(\varphi)$ is known up to 3 loops,
and its small-angle and large-angle asymptotics --- up to 4 loops.
Some (but not all) color structures at 4 loops
are known with the full $\varphi$ dependence.
Some simple contributions are known at higher loops.
For the $\varphi\to\infty$ asymptotics of $\Gamma(\varphi)$ (the light-like cusp anomalous dimension)
and the $\varphi^2$ term of the small-$\varphi$ expansion (the Bremsstrahlung function),
the $\mathcal{N}=4$ SYM results are equal to the highest-weight parts of the QCD results.
There is an interesting conjecture about the structure of $\Gamma(\varphi)$
which holds up to 3 loops;
at 4 loops it holds for some color structures and breaks down for other ones.
In cases when it holds it related highly non-trivial functions of $\varphi$,
and it cannot be accidental;
however, the reasons of this conjecture and its failures are not understood.
The cusp anomalous dimension at the Euclidean angle $\phi\to\pi$
is related to the static quark-antiquark potential due to conformal symmetry;
in QCD this relation is broken by an anomalous term proportional to the $\beta$ function.

Some new results are also presented.
Using the recent 4-loop result for $\gamma_h$,
here we obtain analytical expressions for some terms
in the 4-loop on-shell renormalization constant of the massive quark field $Z_Q^{\text{os}}$
which were previously known only numerically.
We also present 2 new contribution to $\gamma_h$, $\Gamma(\varphi)$ at 5 loops
and to the quark-antiquark potential at 4 loops.

\end{abstract}


\tableofcontents

\section{Introduction}
\label{S:Intro}

A Wilson line in a gauge theory is the phase factor
for a classical pointlike charged particle (in some representation $R$ of the gauge group)
moving along a world line $C$:
\begin{equation}
W_0 = \biggl\langle P \exp \biggl[i g_0 \int_C dx_\mu A_0^{a\mu}(x) t_R^a\biggr] \biggr\rangle
= Z_W(\alpha_s(\mu),a(\mu)) W(\mu)\,,
\label{Intro:W}
\end{equation}
where $g_0$ and $A_0$ are the bare coupling and gauge field,
$t_R^a$ are the generators of the representation $R$,
$W(\mu)$ is the renormalized Wilson line (we use the $\overline{\text{MS}}$ scheme, $d = 4-2\varepsilon$),
and $a(\mu)$ is the renormalized gauge parameter.
We use the covariant gauge:
\begin{equation*}
\raisebox{-2.2mm}{\begin{picture}(17,8.5)
\put(8.5,4.25){\makebox(0,0){\includegraphics{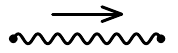}}}
\put(8.5,6){\makebox(0,0)[b]{$k$}}
\put(1,4){\makebox(0,0)[b]{$a$}}
\put(16,4){\makebox(0,0)[b]{$b$}}
\put(1,2){\makebox(0,0)[t]{$\mu$}}
\put(16,2){\makebox(0,0)[t]{$\nu$}}
\end{picture}}
= i \delta^{ab} D_0^{\mu\nu}(k)\,,\quad
D_0^{\mu\nu}(k) = \frac{1}{-k^2} \biggl[g^{\mu\nu} + (1-a_0) \frac{k^\mu k^\nu}{-k^2}\biggr]\,.
\end{equation*}
The renormalized gauge parameter $a(\mu)$ is related to the bare one $a_0$
by the gluon field renormalization constant: $a_0 = Z_A(\alpha_s(\mu),a(\mu)) a(\mu)$.
Wilson lines are widely used in gauge theories, see e.g.\ the textbook\cite{Cherednikov:2020mtu}.
Renormalization of Wilson lines is considered in \cite{Polyakov:1980ca,Dotsenko:1979wb,Brandt:1981kf},
see\cite{Dorn:1986dt} for review and more references.
The $\overline{\text{MS}}$ renormalization constant $Z_W$
accumulates ultraviolet (UV) divergences of the bare Wilson line,
and is determined only by singular points of the line --- its ends and cusps
(we consider only lines without self-intersections);
smooth segments don't contribute.

Infrared (IR) properties of scattering amplitudes in gauge theories are closely related to Wilson
lines\cite{Korchemsky:1985xj,Korchemsky:1985xu,Korchemsky:1985ts,Korchemsky:1986fj,Becher:2009kw},
see\cite{Agarwal:2021ais} for review and references.
IR divergences of a scattering amplitude can be found in the eikonal approximation.
It turns the amplitude to a product of straight semi-infinite Wilson lines along its external momenta.
However, it introduces UV divergences which were absent in the original amplitude.
These UV divergences are equal to the IR divergences of the amplitude with the opposite sign,
because eikonal diagrams contain no dimensionful parameters.

\section{Wilson lines and HQET}
\label{S:HQET}

Wilson lines are non-local objects.
It is easier to understand their properties
in the language of heavy quark effective theory (HQET).
HQET is a usual local quantum field theory,
and renormalization properties of Wilson lines can be obtained
from renormalization of local operators in HQET.
Renormalization theory of local operators is discussed
in many quantum field theory textbooks.

Suppose we have QCD with $n_f$ flavors plus a single heavy colored particle
(in some representation $R$ of the gauge group).
The momentum of this particle can be decomposed as $P = Mv + p$,
where $M$ is its on-shell mass,
$v$ is a reference velocity ($v^2 = 1$),
and $p$ is called the residual momentum of this heavy particle.
If the characteristic residual momentum $p$,
characteristic momenta of light  particles $p_i$
and light particle masses $m_i$ are all small
($p \ll M$, $p_i \ll M$, $m_i \ll M$),
then the system can be described by the HQET Lagrangian
(e.g., see\cite{Neubert:1993mb,Manohar:2000dt,Grozin:2004yc}).
This heavy particle can be, for example, a heavy quark
(its flavor is \emph{not} counted in $n_f$).
At the leading order in $1/M$ the heavy-particle spin does not interact with gluons
and can be freely rotated (heavy quark symmetry).
Moreover, it can be switched off (superflavor symmetry\cite{Georgi:1990ak}).

The HQET Lagrangian is
\begin{equation}
L = h_{v0}^* i D \cdot v h_{v0} + L_{\text{QCD}}\,,\quad
D^\mu h_{v0} = (\partial^\mu - i g_0 A_0^{\mu a} t_R^a) h_{v0}
\label{HQET:L}
\end{equation}
(we can include several HQET fields with several velocities if we need: $\sum_i h_{v_i 0}^* i D \cdot v_i h_{v_i 0}$).
The heavy (static) particle is described by the scalar field $h_v$ in the color representation $R$:
\begin{equation}
h_{v0} = Z_h^{1/2}(\alpha_s(\mu),a(\mu)) h_v(\mu)\,,
\label{HQET:hv}
\end{equation}
where $h_v(\mu)$ is the $\overline{\text{MS}}$ renormalized field.
The momentum-space propagator of the field $h_{v0}$ depends only on the residual energy $\omega = p \cdot v$:
\begin{equation}
\raisebox{-0mm}{\begin{picture}(17,4)
\put(8.5,1){\makebox(0,0){\includegraphics{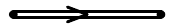}}}
\put(8.5,2){\makebox(0,0)[b]{$p$}}
\end{picture}}
= i S_{h0}(p \cdot v)\,,\quad
S_{h0}(\omega) = \frac{1}{\omega}
\label{HQET:S0}
\end{equation}
(the unit color matrix $\mathbf{1}_R$ is assumed).
The $h_v^* h_v A$ vertex is
\begin{equation}
\raisebox{-0mm}{\begin{picture}(14,10)
\put(7,4){\makebox(0,0){\includegraphics{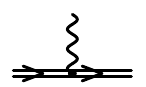}}}
\put(6,7){\makebox(0,0)[r]{$\mu$}}
\put(8,7){\makebox(0,0)[l]{$a$}}
\end{picture}}
= i g_0 v^\mu t_R^a\,.
\label{HQET:vert}
\end{equation}

If we denote the sum of one-particle-irreducible (1PI) self-energy diagrams $-i \Sigma_h(\omega)$
then the full propagator (i.e.\ the sum of all propagator diagrams) is
\begin{equation}
\raisebox{-0mm}{\begin{picture}(17,4)
\put(8.5,1){\makebox(0,0){\includegraphics{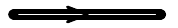}}}
\put(8.5,2){\makebox(0,0)[b]{$\omega$}}
\end{picture}}
= i S_h(\omega)\,,\quad
S_h(\omega) = \frac{1}{\omega - \Sigma_h(\omega)}\,.
\label{HQET:Sh}
\end{equation}
The renormalization factor $Z_h$ is given by
\begin{equation}
\log\frac{S_h(\omega)}{S_{h0}(\omega)} = \log Z_h + \mathcal{O}(\varepsilon^0)\,,
\label{HQET:Zh}
\end{equation}
where $S_h(\omega)$ is expressed via the renormalized quantities $\alpha_s(\mu)$, $a(\mu)$.
The HQET field $h_v(\mu)$ anomalous dimension is
\begin{equation}
\gamma_h(\alpha_s(\mu)) = \frac{d\log Z_h(\alpha_s(\mu),a(\mu))}{d\log\mu}\,.
\label{HQET:gammah}
\end{equation}

We can also use the coordinate space:
\begin{equation}
\raisebox{-0mm}{\begin{picture}(17,4)
\put(8.5,1){\makebox(0,0){\includegraphics{hqet.pdf}}}
\put(1,2){\makebox(0,0)[b]{$0$}}
\put(16,2){\makebox(0,0)[b]{$x$}}
\end{picture}}
= \theta(x \cdot v) \delta(x_\bot)\,.
\label{HQET:S0x}
\end{equation}
In the $v$ rest frame the free propagator is $\delta(\vec{x}^{\,}) S_{h0}(x^0)$,
$i S_{h0}(t) = \theta(t)$.
The full propagator
\begin{equation}
\raisebox{-0mm}{\begin{picture}(17,4)
\put(8.5,1){\makebox(0,0){\includegraphics{hqetf.pdf}}}
\put(1,2){\makebox(0,0)[b]{$x$}}
\put(16,2){\makebox(0,0)[b]{$y$}}
\end{picture}}
=
\raisebox{-0mm}{\begin{picture}(17,4)
\put(8.5,1){\makebox(0,0){\includegraphics{hqet.pdf}}}
\put(1,2){\makebox(0,0)[b]{$x$}}
\put(16,2){\makebox(0,0)[b]{$y$}}
\end{picture}}
\times
\biggl\langle P \exp \biggl[i g_0 \int_x^y dx_\mu A_0^{a\mu}(x) t_R^a\biggr] \biggr\rangle\,,
\label{HQET:W}
\end{equation}
where the integral is taken along the straight line from $x$ to $y$.
In the $v$ rest frame $S_h(t) = S_{h0}(t) W(t)$, where
\begin{equation}
W(t) = \biggl\langle P \exp \biggl[i g_0 \int_0^t dt\,v_\mu A_0^{a\mu}(vt) t^a\biggr] \biggr\rangle
\label{HQET:Wilson}
\end{equation}
is the straight Wilson line along $v$ of the length $t$.
Assuming $t>0$ we have
\begin{equation}
\log W(t) = \log Z_h + \mathcal{O}(\varepsilon^0)\,.
\label{HQET:Zhcoord}
\end{equation}
So, the factor $Z_h$ describes renormalization of a finite-length Wilson line.
In other words, $Z_h^{1/2}$ describes renormalization of an end of a Wilson line.

Transition of a heavy particle with a velocity $v$ into a heavy particle with a velocity $v'$
(e.g., decay of a heavy quark into another heavy quark plus colorless particles)
in HQET framework is described by the current
\begin{equation}
J_0 = h_{v'0}^* h_{v0} = Z_J(\alpha_s(\mu),\varphi) J(\mu)
\label{Intro:J}
\end{equation}
($J(\mu)$ is the $\overline{\text{MS}}$ renormalized current,
$\cosh\varphi = v \cdot v'$).
Its anomalous dimension
\begin{equation}
\Gamma(\alpha_s(\mu),\varphi) = \frac{d\log Z_J(\alpha_s(\mu),\varphi)}{d\log\mu}
\label{Intro:Gamma}
\end{equation}
is called the cusp anomalous dimension.
The current $J$ is colorless, and hence $Z_J$ and $\Gamma$ are gauge invariant.
Dependence of the Isgur--Wise function on $\mu$ is determined\cite{Falk:1990yz} by $\Gamma(\alpha_s,\varphi)$.
The $1/\varepsilon$ IR divergence of massive QCD form factors is given by it.

Let $V(\omega,\omega',\varphi) = 1 + \Lambda(\omega,\omega',\varphi)$
be the sum of 1PI vertex diagrams of $J_0$
(1 is the tree-level diagram;
$\omega$ and $\omega'$ are the incoming residual energy and the outgoing one).
From each diagram for $\Sigma_h$ we can get a set of diagrams for $\Lambda$
by inserting the $J_0$ vertex into each internal HQET line in turn.
The correlator of $J_0$, $h_v^*$, $h_{v'}$ is $V(\omega,\omega',\varphi) \cdot i S_h(\omega) \cdot i S_h(\omega')$;
on the other hand, it is $Z_h Z_J$ times the finite correlator of the 3 renormalized operators.
Recalling the fact that $S_h(\omega)$ is $Z_h$ times the finite renormalized propagator
we have
\begin{equation}
\log V(\omega,\omega',\varphi) = \log Z_J(\varphi) - \log Z_h + \mathcal{O}(\varepsilon^0)\,.
\label{Intro:Z}
\end{equation}
It is sufficient to calculate a single-scale vertex function $V(\omega,\omega,\varphi)$
in order to obtain $Z_J$.
At $\varphi=0$ we have HQET Ward identities
\begin{equation}
\Lambda(\omega,\omega',0) = - \frac{\Sigma_h(\omega) - \Sigma_h(\omega')}{\omega - \omega'}
\quad\text{or}\quad
V(\omega,\omega',0) = \frac{S_h^{-1}(\omega) - S_h^{-1}(\omega')}{\omega - \omega'}\,.
\label{Intro:Ward}
\end{equation}
Therefore $\log V(\omega,\omega',0) = - \log Z_h + \mathcal{O}(\varepsilon^0)$,
$Z_J(\alpha_s,0) = 1$, and
\begin{equation}
\Gamma(\alpha_s,0) = 0\,.
\label{Intro:Gamma0}
\end{equation}

In coordinate space, the Green function $\langle h_{v'0}(x') J_0(0) h_{v0}^*(x)\rangle$
($x^0 < 0 < x^{\prime0}$) is equal to the obvious $\delta$ functions times the Wilson line $W(t,t',\varphi)$:
the straight segment along $v$ of length $t$, the angle $\varphi$,
the straight segment along $v'$ of length $t'$.
The corresponding renormalized Green function is finite, and so
\begin{equation}
\log W(t,t',\varphi) = \log Z_J(\varphi) + \log Z_h + \mathcal{O}(\varepsilon^0)\,.
\label{HQET:Zcoord}
\end{equation}
At $\varphi = 0$ the Wilson line is straight: $W(t,t',0) = W(t+t')$,
and we again obtain $Z_J(0) = 1$.

At $\varphi \to 0$ the cusp anomalous dimension is a regular Taylor series in $\varphi^2$:
\begin{equation}
\Gamma(\alpha_s,\varphi) = \sum_{n=1}^\infty B_n(\alpha_s) \varphi^{2n}\,,
\label{HQET:small}
\end{equation}
$B_1(\alpha_s)$ is usually called the Bremsstrahlung function.
If our Wilson line interacts with QED with $n_f=0$ (i.e.\ the free electromagnetic field),
and we have a classical pointlike charge which is at rest in some reference frame at both $t \to \pm\infty$,
then the energy of the emitted radiation is
\begin{equation}
\Delta E = 2 \pi B_1(\alpha) \int_{-\infty}^{+\infty} dt\,(-a^2(t))\,,
\label{HQET:DE}
\end{equation}
where $a^\mu$ is the acceleration.
In this theory $B_1(\alpha) = \alpha/(3\pi)$ exactly,
and~(\ref{HQET:DE}) is just the classical dipole radiation formula.
The theory is conformally invariant, $\alpha$ does not run.
The formula~(\ref{HQET:DE}) is also valid\cite{Correa:2012at} in $\mathcal{N}=4$ super Yang--Mills (SYM),
which is also conformally invariant.
It cannot be generalized to other theories, because in them $B_1(\alpha_s(\mu))$ depends on $\mu$.

At $\varphi \to \infty$\cite{Korchemsky:1987wg}
\begin{equation}
\Gamma(\alpha_s,\varphi) = K(\alpha_s) \varphi + \mathcal{O}(\varphi^0)\,,
\label{Intro:large}
\end{equation}
where $K$ is called the light-like cusp anomalous dimension.
It is related\cite{Korchemskaya:1992je} to renormalization properties of Wilson lines
with light-like segments.
The coefficients of $1/(1-x)_+$ in DGLAP kernels (as well as those of $1/(x-y)_+$ in ERBL kernels)
are determined\cite{Korchemsky:1988si,Korchemsky:1992xv} by the light-like cusp anomalous dimension.
IR $1/\varepsilon^2$ divergences of form factor of massless particles
are also determined by $K(\alpha_s)$.

If a world line of a finite lengths consists of straight segments with angles $\varphi_i$ between them,
its renormalization factor is $Z_h \prod Z_J(\varphi_i)$
(2 ends contribute $Z_h^{1/2}$ each, each cusp contributes $Z_J(\varphi_i)$).
The same is true if the segments are not straight but just some smooth curves.
We can approximate a smooth curve of length $t$ by a broken line with $N \to \infty$ segments
of length $\Delta t = t/N$ each.
Cusp angles are $\Delta \varphi_i \sim 1/N$,
so, the contributions to $\log Z_J$ are $\sim 1/N^2$ each;
there are $N$ such contributions.
Such Wilson lines can be described by the HQET field $h$
which lives on the world line.
The HQET Lagrangian was actually used as a technical device
for investigating Wilson lines\cite{Gervais:1979fv,Arefeva:1980zd}
(see\cite{Dorn:1986dt} for review).

We can also understand this fact from another point of view.
UV divergences come from small distances, where any smooth line is straight.
The only possible UV divergence is the residual mass $\Sigma_h(0)$.
This is a linear UV divergence.
In dimensional regularization it is discarded:
the residual mass $\Sigma_h(0)$ has dimensionality of mass,
and we cannot construct any non-zero result for it by dimensions counting
(in regularizations based on momentum cutoff $\Lambda_{\text{UV}}$
the UV divergent residual mass is $\propto \Lambda_{\text{UV}}$).

\section{Exponentiation}
\label{S:Exp}

A Wilson line (of any shape) interacting with the free electromagnetic field
(QED with $n_f=0$) is given by the simple exponentiation formula.
Let
\begin{equation}
w(t) =
\raisebox{-2mm}{\begin{picture}(18,7.75)
\put(9,4.875){\makebox(0,0){\includegraphics{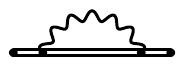}}}
\put(1,2){\makebox(0,0)[t]{$0$}}
\put(17,2){\makebox(0,0)[t]{$t$}}
\put(4,2){\makebox(0,0)[t]{$t_1$}}
\put(14,2){\makebox(0,0)[t]{$t_1$}}
\end{picture}}
\label{Exp:w}
\end{equation}
be the 1-loop contribution to $W(t)$.
Here we use diagrams for Wilson lines, not for HQET propagators,
so that the factors $\theta(x \cdot v) \delta(x_\bot)$ are not included.
This diagram is an integral in $t_1$, $t_2$ such that $0 < t_1 < t_2 < t$.
Let's calculate $w^2$.
It is an integral in $t_1$, $t_2$, $t_1'$, $t_2'$
such that $0 < t_1 < t_2 < t$, $0 < t_1' < t_2' <t$.
This integration region can be subdivided into 6 subregions corresponding to 6 diagrams:
\begin{align}
&\raisebox{-2mm}{\begin{picture}(18,7.75)
\put(9,4.875){\makebox(0,0){\includegraphics{w1.pdf}}}
\put(1,2){\makebox(0,0)[t]{$0$}}
\put(17,2){\makebox(0,0)[t]{$t$}}
\put(4,2){\makebox(0,0)[t]{$t_1$}}
\put(14,2){\makebox(0,0)[t]{$t_2$}}
\end{picture}}
\times
\raisebox{-3.75mm}{\begin{picture}(18,7.75)
\put(9,2.875){\makebox(0,0){\includegraphics{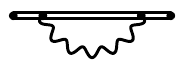}}}
\put(1,5.75){\makebox(0,0)[b]{$0_{\vphantom{\scriptstyle1}}$}}
\put(17,5.75){\makebox(0,0)[b]{$t_{\vphantom{\scriptstyle1}}$}}
\put(4,5.75){\makebox(0,0)[b]{$t_1'$}}
\put(14,5.75){\makebox(0,0)[b]{$t_2'$}}
\end{picture}}
\label{Exp:ww}\\
&{}
= \raisebox{-3.75mm}{\includegraphics{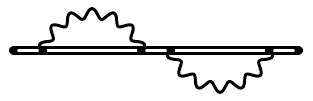}}
+ \raisebox{-3.75mm}{\includegraphics{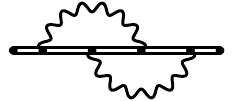}}
+ \raisebox{-3.75mm}{\includegraphics{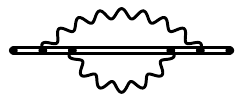}}
\nonumber\\
&{}
+ \raisebox{-3.75mm}{\includegraphics{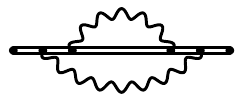}}
+ \raisebox{-3.75mm}{\includegraphics{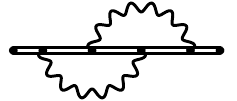}}
+ \raisebox{-3.75mm}{\includegraphics{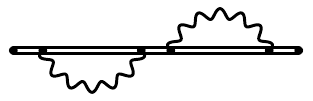}}\,.
\nonumber
\end{align}
This is twice the 2-loop contribution to $W(t)$.
Continuing this drawing exercise we see that the 3-loop contribution is $w^3/3!$, etc.
The exact expression for the full Wilson line is\cite{Yennie:1961ad}
\begin{equation}
\raisebox{-2mm}{\begin{picture}(17,4)
\put(8.5,3){\makebox(0,0){\includegraphics{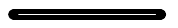}}}
\put(1,2){\makebox(0,0)[t]{$0$}}
\put(16,2){\makebox(0,0)[t]{$t$}}
\end{picture}}
= \exp
\raisebox{-2mm}{\begin{picture}(18,7.75)
\put(9,4.875){\makebox(0,0){\includegraphics{w1.pdf}}}
\put(1,2){\makebox(0,0)[t]{$0$}}
\put(17,2){\makebox(0,0)[t]{$t$}}
\end{picture}}\,,
\quad
W(t) = e^{w(t)}\,.
\label{Exp:QED0}
\end{equation}

So, we have $w = \log Z_h + \mathcal{O}(\varepsilon^0)$,
$\log Z_h$ is \emph{exactly} 1-loop:
\begin{equation}
\gamma_h(\alpha) = 2 (a-3) \frac{\alpha}{4\pi}\,.
\label{Exp:gammah}
\end{equation}
No higher-loop corrections.
The same is true for $\Gamma(\alpha,\varphi)$
because exponentiation works for Wilson lines of any shapes,
including those with cusps
(this exact $\Gamma$ is given below in~(\ref{Hist:Gamma})).

In QED with $n_f>0$ the situation is more complicated:
we don't know exact results, just perturbative series.
However, these series have a simple structure:
\begin{align}
&\raisebox{-0.375mm}{\includegraphics{hqetf0.pdf}}
= \exp \biggl[\raisebox{-0.375mm}{\includegraphics{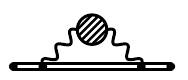}}
+ \raisebox{-0.375mm}{\includegraphics{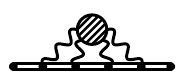}}
+ \cdots\biggr]\,,
\nonumber\\
&W = \exp\bigl[w_2 + w_4 + \cdots\bigr]\,,
\label{Exp:QED1}
\end{align}
where the shaded blobs are the sums of connected diagrams with 2, 4,\ldots{} external photon lines
(odd numbers of lines are forbidden by $C$ conservation):
\begin{align}
&w_2 = \raisebox{-0.375mm}{\includegraphics{w2.pdf}}
\nonumber\\
&{} = \raisebox{-0.375mm}{\includegraphics{w1.pdf}}
+ \raisebox{-0.375mm}{\includegraphics{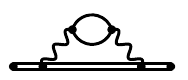}}
+ \raisebox{-0.375mm}{\includegraphics{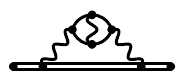}}
+ \raisebox{-0.375mm}{\includegraphics{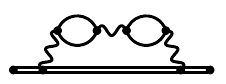}}
+ \cdots\,,
\label{Exp:w2}\\
&w_4 = \raisebox{-0.375mm}{\includegraphics{w4.pdf}}
= \raisebox{-0.375mm}{\includegraphics{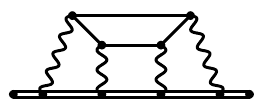}}
+ \cdots\,,
\label{Exp:w4}
\end{align}
and so on (the shaded blobs here are not necessarily 1-particle-irreducible (1PI)).
Diagrams in~(\ref{Exp:w2}), (\ref{Exp:w4}) are called \emph{c-webs} (connected webs).

We have, for example,
\begin{equation*}
\raisebox{-0.375mm}{\includegraphics{w4a.pdf}}
\times \raisebox{-4.125mm}{\includegraphics{w1m.pdf}}
= \cdots
+ \raisebox{-4.125mm}{\includegraphics{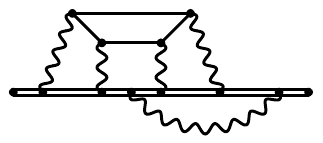}}
+ \cdots\,,
\end{equation*}
and these diagrams (containing 2 c-webs each) are already accounted for
in the expansion of the exponent~(\ref{Exp:QED1}).

The sums of connected diagrams with $2n$ external (off-shell) photon legs at each order
are gauge invariant due to the QED Ward identity,
except the free photon propagator.
Hence all c-webs are gauge invariant except the first one in~(\ref{Exp:w2}).
This 1-loop c-web is linear in $e_0^2 a_0$;
but $e_0^2 a_0 = e^2(\mu) a(\mu)$ because $Z_\alpha = Z_A^{-1}$ in QED.
Therefore the only gauge-dependent term in $\gamma_h$ is the 1-loop term
given by~(\ref{Exp:gammah});
all higher-loop corrections are gauge invariant.

In non-abelian theories\cite{Gatheral:1983cz,Frenkel:1984pz}
\begin{equation}
W = \exp\Bigl[\sum \bar{C}_i w_i\Bigr]\,,
\label{Exp:QCD}
\end{equation}
where $w_i$ are \emph{webs},
and $\bar{C}_i$ are color-connected parts of their color factors $C_i$.
Let's draw all gluon lines in a diagram for $W$
on a single side of its HQET line.
If we remove this HQET line,
we obtain a diagram with gluon external lines
which can be connected (c-web) or not.
When this diagram is connected \emph{if we count line crossings as connections},
the diagram is called a web.
For example,
\begin{equation}
\raisebox{-0.375mm}{\includegraphics{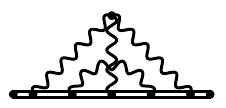}}\quad
\text{is a web, but}\quad
\raisebox{-0.375mm}{\includegraphics{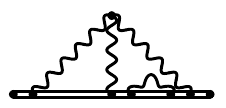}}\quad
\text{is not}.
\label{Exp:web}
\end{equation}

If color factors of each web were the same as for the diagram
where all its c-webs are separated,
the contributions of webs would be accounted for in the exponent
of the sum of c-webs.
This is true in QED.
In non-abelian theories color factors are more complicated.
For example, let's consider the first diagram in~(\ref{Exp:web}).
It is a web but not a c-web: it contains 2 c-webs.
Its color factor $C$ is given by the color diagram
which looks the same as this diagram in~(\ref{Exp:web}).
It is not equal to $C_1 C_2$, the product
of the color factors of these 2 c-webs taken separately.
Let's pull these 2 c-webs apart,
interchanging the vertices on the HQET line belonging to different c-webs
according to the obvious commutator identity
\begin{equation}
\raisebox{-0.375mm}{\includegraphics{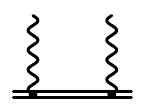}}
- \raisebox{-0.375mm}{\includegraphics{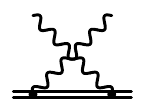}}
= \raisebox{-0.375mm}{\includegraphics{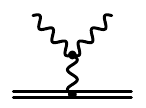}}\,.
\label{Exp:com}
\end{equation}
This color factor $C$ becomes
\begin{equation*}
\raisebox{-0.375mm}{\includegraphics{wt1.pdf}}
= \raisebox{-0.375mm}{\includegraphics{wt2.pdf}}
+ \raisebox{-0.375mm}{\includegraphics{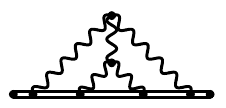}}\,.
\end{equation*}
The first term consists of 2 separate c-webs, it is $C_1 C_2$;
this contribution has already been accounted for
in the expansion of the exponent.
The second term is a connected color diagram,
the connected part $\bar{C}$ of the color factor $C$.
This contribution is not yet accounted for.
Therefore we add this web with the color factor $\bar{C}$
to the exponent in~(\ref{Exp:QCD}).

Let's define
\begin{align}
&\Tr t_R^a t_R^b = T_R \delta^{ab}\,,\quad
t_R^a t_R^a = C_R \mathbf{1}_R\,,\quad
N_R = \Tr \mathbf{1}_R\,,
\nonumber\\
&d_{RR'} = \frac{d_R^{abcd} d_{R'}^{abcd}}{N_R}\,,\quad
d_R^{abcd} = \Tr t_R^{(a} t_R^{b\vphantom{(}} t_R^{c\vphantom{)}} t_R^{d)}\,,
\label{Exp:color}
\end{align}
where brackets mean symmetrization
(note that $\Tr t_R^a t_R^a = C_R N_R = T_R N_A$, and hence $C_R = T_R N_A/N_R$;
in particular, $C_A = T_A$).
For $SU(N_c)$ gauge group with the standard normalization $T_F = \frac{1}{2}$ they are
\begin{align}
&C_F = \frac{N_c^2-1}{2N_c}\,,\quad
C_A = N_c\,,\quad
d_{FF} = \frac{(N_c^2-1) (N_c^4-6N_c^2+18)}{96 N_c^3}\,,
\nonumber\\
&d_{AF} = \frac{N_c}{N_c^2-1} d_{FA} = \frac{N_c (N_c^2+6)}{48}\,,\quad
d_{AA} = \frac{N_c^2 (N_c^2+36)}{24}\,.
\label{Exp:SUN}
\end{align}
In QED $T_F = 1$, $C_R = Z^2$, $C_F = 1$, $C_A = 0$, $d_{RF} = Z^4$, $d_{RA} = 0$,
where $Z$ is the charge of the infinitely heavy particle (in units of $e$).

The possible color structures of $\gamma_h$ and $\Gamma$ without $n_f$ are,
due to the non-abelian exponentiation,
$C_R$ at 1 loop;
$C_R C_A$ at 2 loops;
$C_R C_A^2$ at 3 loops;
$C_R C_A^3$ and $d_{RA}$ at 4 loops.
Color structures containing $C_F$ are forbidden.
As soon as there is at least 1 quark loop in the diagram
(i.e.\ at least 1 power of $n_f$ in the color structure),
all possible color factors are allowed:
gluons attached to quark loops are not restricted by exponentiation.
Up to 3 loops all color structures are proportional to $C_R$.
The $R$ dependence of $\gamma_h$ and $\Gamma$ is given by the factor $C_R$
--- Casimir scaling.
At 4 loops the quartic Casimirs $d_{RA}$, $d_{RF}$ appear.
They are not proportional to $C_R$, and Casimir scaling breaks down.

\section{History and the current status of calculations}
\label{S:Hist}

The one-loop cusp anomalous dimension follows from classical electrodynamics.
When an infinitely heavy charged  particle instantly changes its velocity,
it either remains itself (probability $|F|^2$) or emits one or more photons.
Up to the order $\alpha$
\begin{equation*}
\biggl|\raisebox{-3.7mm}{\includegraphics[scale=0.4]{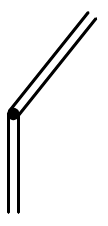}}
+ \raisebox{-3.7mm}{\includegraphics[scale=0.4]{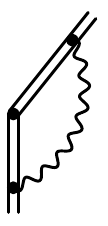}}\biggr|^2
+ \int \biggl|\raisebox{-3.7mm}{\includegraphics[scale=0.4]{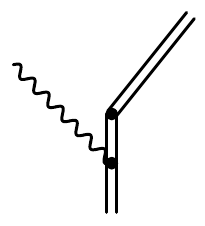}}
+ \raisebox{-3.7mm}{\includegraphics[scale=0.4]{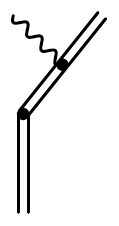}}\biggr|^2
= 1\,.
\end{equation*}
In classical electrodynamics the spectrum of the emitted radiation is\cite{LL,Jackson}
\begin{equation*}
dE = \frac{e^2}{2 \pi^2} (\varphi \coth\varphi - 1)\,d\omega\,,
\end{equation*}
and hence the probability of photon emission is $dE/\omega$.
In dimensional regularization, by dimensions counting, this probability becomes $C(\varepsilon) dE/\omega^{1+2\varepsilon}$
with $C(0)=1$.
Hence the form factor is
\begin{equation*}
F = 1 - \frac{C(\varepsilon)}{2} \int_\lambda^\infty
\frac{e^2}{2 \pi^2} (\varphi \coth\varphi - 1)
\frac{d\omega}{\omega^{1+2\varepsilon}}
= 1 - 2 \frac{\alpha}{4\pi\varepsilon} (\varphi \coth\varphi - 1 + \mathcal{O}(\varepsilon))
\end{equation*}
(where $\lambda$ is an IR cutoff),
and the cusp anomalous dimension is
\begin{equation}
\Gamma = 4 \frac{\alpha}{4\pi} (\varphi \coth\varphi - 1)\,.
\label{Hist:Gamma}
\end{equation}
The one-loop cusp anomalous dimension should be included in The Guinness Book of Records
as the anomalous dimension known for a longest time (probably, $> 100$ years).
In QCD it includes the obvious extra factor $C_R$.
Of course, there are very many ways to obtain~(\ref{Hist:Gamma}),
using momentum or coordinate space.

After a wrong calculation\cite{Aoyama:1981ev},
the 2-loop $\gamma_h$ (with $n_f=0$) has been calculated in\cite{Knauss:1984rx}.
The full result (with $n_f$) has been obtained in\cite{Broadhurst:1991fy}
as a by-product of a 2-loop calculation of the on-shell renormalization constant $Z_Q^{\text{os}}$
of the field of a massive quark,
essentially from the requirement that the renormalized QCD/HQET matching constant~(\ref{Gamma:dec}) is finite
(though this was not explicitly stated in the article),
and reproduced\cite{Ji:1991pr,Broadhurst:1991fz} by direct HQET calculations.
At 3 loops it has been calculated in\cite{Melnikov:2000zc},
again as a by-product of the $Z_Q^{\text{os}}$ calculation, now at 3 loops;
the result has been confirmed\cite{Chetyrkin:2003vi} by a direct HQET calculation.

The first attempt\cite{Knauss:1984rx} to calculate the cusp anomalous dimension at 2 loops
(at $n_f=0$) was unsuccessful:
the authors were unable to eliminate complicated double and triple integrals.
A usable result (at $n_f=0$) has been obtained in\cite{Korchemsky:1985xu,Korchemsky:1985xj,Korchemsky:1987wg}.
It contains 3 single integrals
(2 of them were calculated\cite{Grozin:2004yc} in terms of $\Li2$ and $\Li3$, the formula~(7.15)).
The (rather simple) $n_f$ term was added in\cite{Korchemsky:1988si}.
The calculation\cite{Korchemsky:1985xu,Korchemsky:1985xj,Korchemsky:1987wg}
was repeated several times\cite{Kilian:1992tj,Kidonakis:2009ev}.
The most nice form of the result (no integrals, just $\Li2$ and $\Li3$) is\cite{Kidonakis:2009ev}
\begin{align}
&\Gamma = 4 C_R \frac{\alpha_s}{4\pi} \biggl\{\varphi\coth\varphi-1
+ \frac{\alpha_s}{4\pi} \biggl[C_A \biggl[\frac{2}{3} \pi^2 - \frac{49}{9} + 2 \varphi^2
\nonumber\\
&\quad{} + \coth\varphi \biggl(2 \Li2(e^{-2\varphi}) - 4 \varphi \log(1-e^{-2\varphi})
- \frac{\pi^2}{3} - \frac{2}{3} \pi^2 \varphi + \frac{67}{9} \varphi - 2 \varphi^2 - \frac{2}{3} \varphi^3\biggr)
\nonumber\\
&\quad{} + \coth^2\varphi \biggl(2 \Li3(e^{-2\varphi}) + 2 \varphi \Li2(e^{-2\varphi})
- 2 \zeta_3 + \frac{\pi^2}{3} \varphi + \frac{2}{3} \varphi^3\biggr)
\biggr]
\nonumber\\
&{} - \frac{20}{9} T_F n_f (\varphi\coth\varphi-1)\biggr]
+ \mathcal{O}(\alpha_s^2)\biggr\}
\nonumber\\
&{} = 4 C_R \frac{\alpha_s}{4\pi} \biggl\{\varphi\coth\varphi-1
\nonumber\\
&{} + \frac{\alpha_s}{4\pi} \biggl[C_A \biggl[2 \biggl(1 + \frac{2}{3} \varphi^2\biggr)
- \frac{1}{3} (\varphi\coth\varphi-1) \biggl(2 \pi^2 - \frac{67}{3} + 2 \varphi^2\biggr)
\nonumber\\
&\quad{} + \coth\varphi (\varphi\coth\varphi+1) \bigl(\Li2(1-e^{2\varphi}) - \Li2(1-e^{-2\varphi})\bigr)
\nonumber\\
&\quad{} - 2 \coth^2\varphi \bigl(\Li3(1-e^{2\varphi}) + \Li3(1-e^{-2\varphi})\bigr)\biggr]
\nonumber\\
&{} - \frac{20}{9} T_F n_f (\varphi\coth\varphi-1)\biggr]
+ \mathcal{O}(\alpha_s^2)\biggr\}
\label{Hist:Gamma2}
\end{align}
(the last form is explicitly even in $\varphi$).

At 3 loops the cusp anomalous dimension has been obtained in\cite{Grozin:2014hna,Grozin:2015kna}
(the results are given in Sect.~\ref{S:Conj} below).
Results for supersymmetric QCD extensions have been also obtained\cite{Grozin:2015kna}.

The light-like cusp anomalous dimension~(\ref{Intro:large}) at 3 loops
has been obtained in the course of calculating the 3-loop DGLAP evolution kernels\cite{Moch:2004pa}
(recently confirmed in\cite{Blumlein:2021enk})
and confirmed in massless form-factor calculations\cite{Moch:2005id,Moch:2005tm}.

\begin{table}[ht]
\tbl{Four-loop contributions to $\gamma_h$, $\Gamma$ and its limiting cases.}
{\begin{tabular}{|l|l|l|l|l|l|l|}
\hline
color & example & $\gamma_h$ & $\varphi\ll1$ & $\Gamma(\varphi)$ & 1L & $\varphi\gg1$ \\
\hline
$C_R (T_F n_f)^3$ & \raisebox{-2mm}{\includegraphics[scale=0.4]{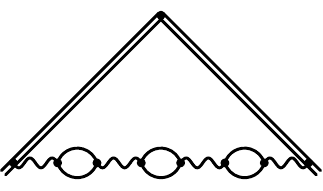}} &
\cite{Broadhurst:1994se} & \cite{Grozin:2004yc} & \cite{Grozin:2004yc} & \checkmark & \cite{Gracey:1994nn,Beneke:1995pq} \\
\hline
$C_R C_F (T_F n_f)^2$ & \raisebox{-2mm}{\includegraphics[scale=0.4]{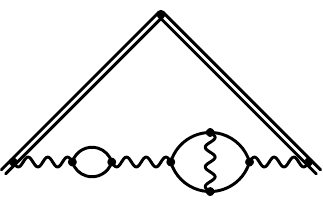}} &
\cite{Grozin:2015kna,Grozin:2016ydd} & \cite{Grozin:2015kna,Grozin:2016ydd} & \cite{Grozin:2015kna,Grozin:2016ydd} & \checkmark &
\cite{Grozin:2015kna,Grozin:2016ydd,Ruijl:2016pkm} \\
$C_R C_A (T_F n_f)^2$ & \raisebox{-2mm}{\includegraphics[scale=0.4]{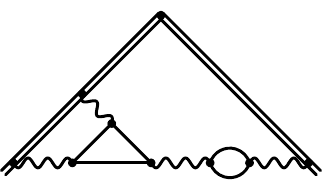}} &
\cite{Marquard:2018rwx,Bruser:2019auj} & \cite{Bruser:2019auj} & \cite{Bruser:2019auj} & & \cite{Ruijl:2016pkm,Henn:2016men,Davies:2016jie} \\
\hline
$C_R C_F^2 T_F n_f$ & \raisebox{-2mm}{\includegraphics[scale=0.4]{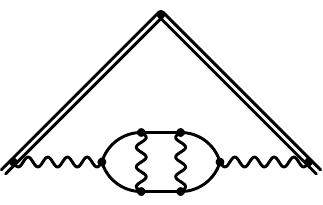}} &
\cite{Grozin:2018vdn} & \cite{Grozin:2018vdn} & \cite{Grozin:2018vdn} & \checkmark & \cite{Grozin:2018vdn} \\
$C_R C_F C_A T_F n_f$ & \raisebox{-2mm}{\includegraphics[scale=0.4]{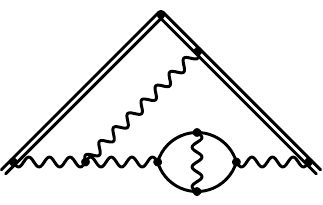}} &
\cite{Bruser:2019auj} & \cite{Bruser:2019auj} & \cite{Bruser:2019auj} & & \nocite{Henn:2019swt}\cite{Bruser:2019auj,vonManteuffel:2020vjv} \\
$C_R C_A^2 T_F n_f$ & \raisebox{-2mm}{\includegraphics[scale=0.4]{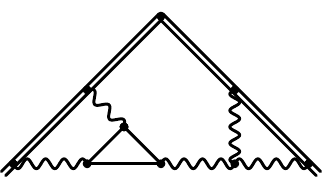}} &
\cite{Bruser:2019auj} & \cite{Bruser:2019auj} & & & \cite{Henn:2019swt,vonManteuffel:2020vjv} \\
$d_{RF} n_f$ & \raisebox{-2mm}{\includegraphics[scale=0.4]{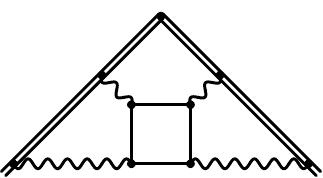}} &
\cite{Grozin:2017css} & \cite{Grozin:2017css,Bruser:2019auj} & \cite{Bruser:2020bsh} & & \cite{Lee:2019zop,Henn:2019rmi} \\
$n_f^1$, $N_c \to \infty$ & &
\cite{Bruser:2019auj} & \cite{Bruser:2019auj} & & & \cite{Henn:2016men,Henn:2016wlm,Moch:2017uml} \\
\hline
$C_R C_A^3$ & \raisebox{-2mm}{\includegraphics[scale=0.4]{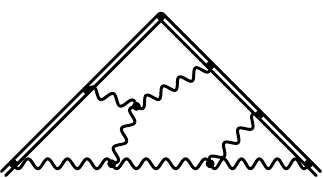}} &
\cite{Grozin:2022wse} & \cite{Grozin:2022wse} & & & \cite{Henn:2019swt,vonManteuffel:2020vjv} \\
$d_{RA}$ & \raisebox{-2mm}{\includegraphics[scale=0.4]{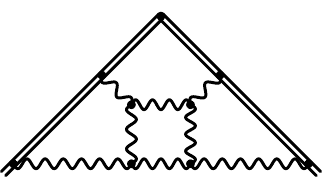}} &
\cite{Grozin:2022wse} & \cite{Grozin:2022wse} & & & \cite{Henn:2019swt,vonManteuffel:2020vjv} \\
$n_f^0$, $N_c \to \infty$ & &
\cite{Grozin:2022wse} & \cite{Grozin:2022wse} & & & \cite{Henn:2016wlm,Moch:2017uml} \\
\hline
\end{tabular}}
\label{T:4loop}
\end{table}

The status of 4-loop calculations is summarized in table~\ref{T:4loop}.
The column 1L shows the color structures which have
the simple 1-loop $\varphi$ dependence $\varphi \coth\varphi - 1$.
For several color structures the exact angle dependence is not known.
A simple interpolation formula (in terms of the variable $\beta=\tanh(\varphi/2)$)
for the 4-loop cusp anomalous dimension has been proposed\cite{Kidonakis:2023lgc}.
It is based on the known asymptotics $\beta\to0$ and $\beta\to1$.
Such approximate formulas give a good precision at 2 and 3 loops\cite{Kidonakis:2009ev,Kidonakis:2016voy},
so, it seems reasonable to hope that it also works well at 4 loops.

Some simple classes of contributions are also known at higher loops (Sects.~\ref{S:abel}--\ref{S:5l}).

\section{HQET field anomalous dimension}
\label{S:Gamma}

\begin{figure}[ht]
\begin{center}
\begin{picture}(119,87)
\put(16,84){\makebox(0,0){\includegraphics{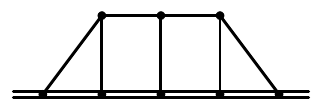}}}
\put(53,84){\makebox(0,0){\includegraphics{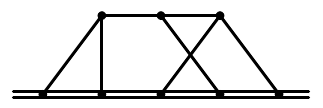}}}
\put(90,84){\makebox(0,0){\includegraphics{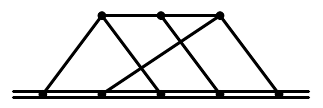}}}
\put(16,69){\makebox(0,0){\includegraphics{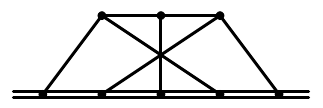}}}
\put(53,69){\makebox(0,0){\includegraphics{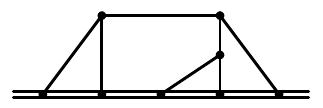}}}
\put(90,69){\makebox(0,0){\includegraphics{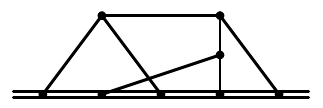}}}
\put(16,52){\makebox(0,0){\includegraphics{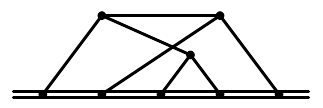}}}
\put(53,53){\makebox(0,0){\includegraphics{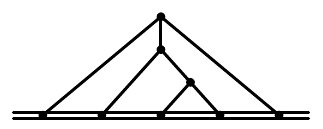}}}
\put(90,53){\makebox(0,0){\includegraphics{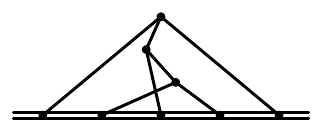}}}
\put(13,37){\makebox(0,0){\includegraphics{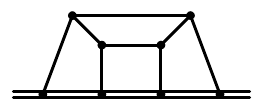}}}
\put(44,37){\makebox(0,0){\includegraphics{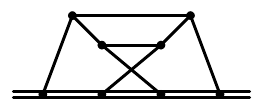}}}
\put(75,37){\makebox(0,0){\includegraphics{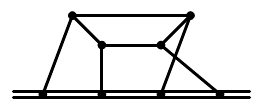}}}
\put(106,37){\makebox(0,0){\includegraphics{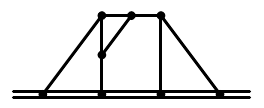}}}
\put(13,20){\makebox(0,0){\includegraphics{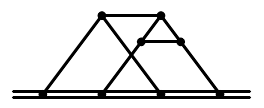}}}
\put(44,21){\makebox(0,0){\includegraphics{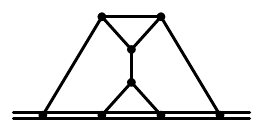}}}
\put(75,21){\makebox(0,0){\includegraphics{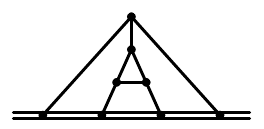}}}
\put(10,5){\makebox(0,0){\includegraphics{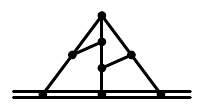}}}
\put(35,5){\makebox(0,0){\includegraphics{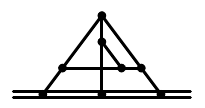}}}
\put(60,5){\makebox(0,0){\includegraphics{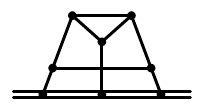}}}
\end{picture}
\end{center}
\caption{Families of 4-loop HQET self-energy Feynman integrals.
Double lines are HQET ones, solid lines are massless.}
\label{F:T}
\end{figure}

\begin{figure}[p]
\begin{picture}(122.0,198.59615)
\put(13.0,5.0){\makebox(0,0){\includegraphics{t15}}}
\put(43.0,5.0){\makebox(0,0){\includegraphics{t16}}}
\put(12.0,18.45){\makebox(0,0){\includegraphics{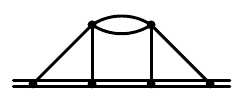}}}
\put(39.0,19.0){\makebox(0,0){\includegraphics{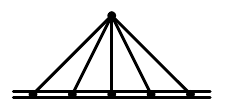}}}
\put(67.0,19.0){\makebox(0,0){\includegraphics{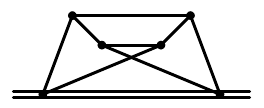}}}
\put(97.0,19.0){\makebox(0,0){\includegraphics{t14}}}
\put(11.0,33.0){\makebox(0,0){\includegraphics{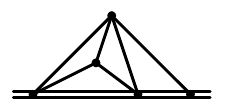}}}
\put(38.0,32.45){\makebox(0,0){\includegraphics{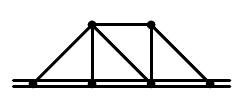}}}
\put(66.0,32.45){\makebox(0,0){\includegraphics{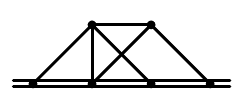}}}
\put(94.0,32.45){\makebox(0,0){\includegraphics{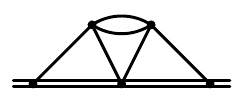}}}
\put(12.0,47.5){\makebox(0,0){\includegraphics{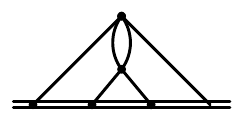}}}
\put(42.0,47.5){\makebox(0,0){\includegraphics{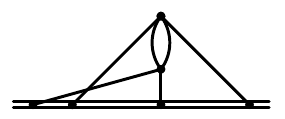}}}
\put(71.0,47.0){\makebox(0,0){\includegraphics{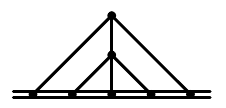}}}
\put(97.0,47.0){\makebox(0,0){\includegraphics{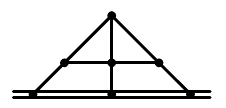}}}
\put(12.0,62.125){\makebox(0,0){\includegraphics{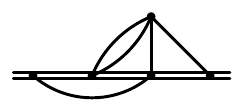}}}
\put(43.0,63.625){\makebox(0,0){\includegraphics{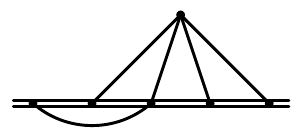}}}
\put(77.0,63.625){\makebox(0,0){\includegraphics{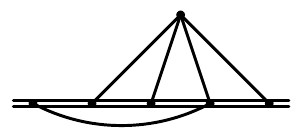}}}
\put(108.0,64.75){\makebox(0,0){\includegraphics{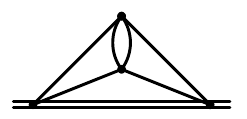}}}
\put(12.0,83.125){\makebox(0,0){\includegraphics{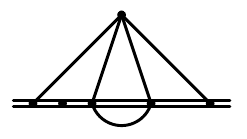}}}
\put(40.0,83.125){\makebox(0,0){\includegraphics{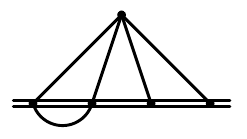}}}
\put(68.0,83.125){\makebox(0,0){\includegraphics{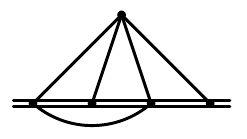}}}
\put(99.0,80.875){\makebox(0,0){\includegraphics{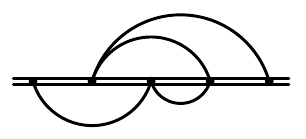}}}
\put(12.0,100.375){\makebox(0,0){\includegraphics{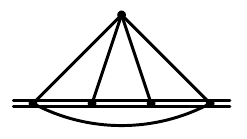}}}
\put(40.0,101.5){\makebox(0,0){\includegraphics{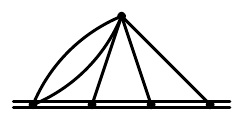}}}
\put(68.0,101.5){\makebox(0,0){\includegraphics{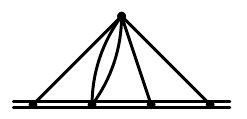}}}
\put(96.0,100.375){\makebox(0,0){\includegraphics{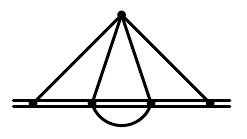}}}
\put(12.0,116.5){\makebox(0,0){\includegraphics{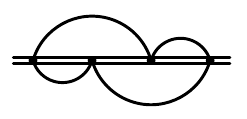}}}
\put(40.0,116.5){\makebox(0,0){\includegraphics{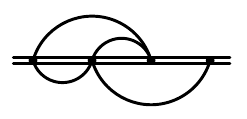}}}
\put(68.0,116.5){\makebox(0,0){\includegraphics{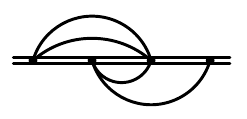}}}
\put(99.0,121.0){\makebox(0,0){\includegraphics{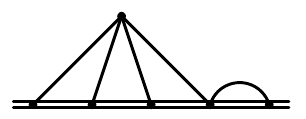}}}
\put(15.0,138.25){\makebox(0,0){\includegraphics{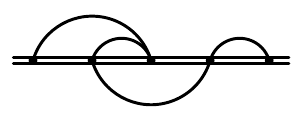}}}
\put(46.0,137.125){\makebox(0,0){\includegraphics{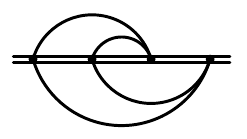}}}
\put(74.0,139.25){\makebox(0,0){\includegraphics{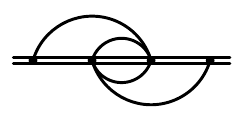}}}
\put(103.0,138.25){\makebox(0,0){\includegraphics{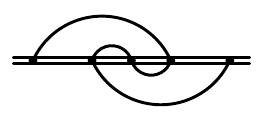}}}
\put(9.0,154.0){\makebox(0,0){\includegraphics{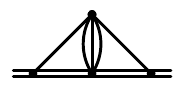}}}
\put(31.0,154.0){\makebox(0,0){\includegraphics{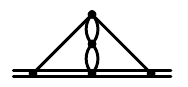}}}
\put(53.0,154.0){\makebox(0,0){\includegraphics{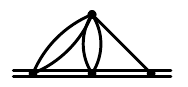}}}
\put(75.0,152.875){\makebox(0,0){\includegraphics{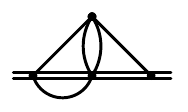}}}
\put(97.0,152.875){\makebox(0,0){\includegraphics{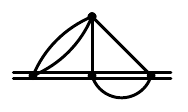}}}
\put(12.0,166.0){\makebox(0,0){\includegraphics{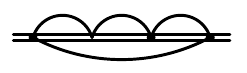}}}
\put(38.0,166.0){\makebox(0,0){\includegraphics{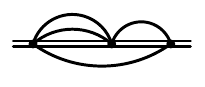}}}
\put(61.0,165.625){\makebox(0,0){\includegraphics{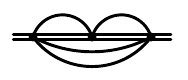}}}
\put(86.0,169.0){\makebox(0,0){\includegraphics{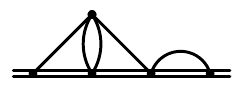}}}
\put(111.0,167.875){\makebox(0,0){\includegraphics{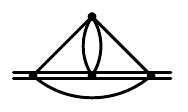}}}
\put(12.0,183.25){\makebox(0,0){\includegraphics{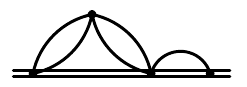}}}
\put(40.0,180.25){\makebox(0,0){\includegraphics{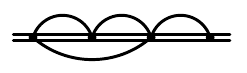}}}
\put(65.0,183.25){\makebox(0,0){\includegraphics{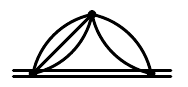}}}
\put(87.0,183.298075){\makebox(0,0){\includegraphics{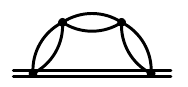}}}
\put(109.0,182.125){\makebox(0,0){\includegraphics{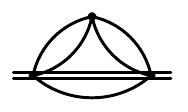}}}
\put(15.0,196.47115){\makebox(0,0){\includegraphics{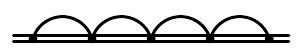}}}
\put(46.0,195.34615){\makebox(0,0){\includegraphics{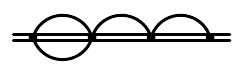}}}
\put(72.0,195.34615){\makebox(0,0){\includegraphics{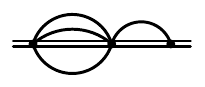}}}
\put(95.0,195.34615){\makebox(0,0){\includegraphics{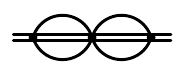}}}
\put(115.0,195.34615){\makebox(0,0){\includegraphics{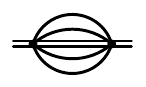}}}
\end{picture}
\caption{Master integrals for 4-loop HQET self-energy diagrams.}
\label{F:M1}
\end{figure}

It is convenient to calculate the HQET self-energy $\Sigma_h(\omega)$ at $\omega<0$, below the mass shell,
where the result is analytical.
The power of $(-2\omega)$ in each term of the perturbative expansion is fixed by dimensions counting,
so, we can set $\omega = -\frac{1}{2}$ during the calculation.
Many diagrams have linear dependent HQET denominators
which can be killed by partial fractioning.
At 3 loops all 4 families of integrals reduce\cite{Grozin:2000jv} to 8 master integrals:
5 trivial, 2 are expressed\cite{Grozin:2000jv,Grozin:2003ak,Beneke:1994sw}
via the hypergeometric functions ${}_3F_2$ of unit argument,
and for the last one several terms of $\varepsilon$ expansion are known\cite{Czarnecki:2001rh}
(see\cite{Grozin:2008tp} for review).
At 4 loops we are left with 19 families (Fig.~\ref{F:T}).
The families 10--12 were considered in\cite{Grozin:2017css},
and 1 (and one sub-family, i.e.\ a family with a contracted line) in\cite{Bruser:2019auj}.
There are 54 master integrals (Fig.~\ref{F:M1}):
13 recursively-1-loop (expressible via $\Gamma$ functions);
10 can be calculated using the formulas from\cite{Grozin:2000jv,Grozin:2003ak,Beneke:1994sw}
(in one case the hypergeometric function happens to be expressible via $\Gamma$ functions);
for 2 integrals several terms of $\varepsilon$ expansions are known from\cite{Czarnecki:2001rh}.
$\varepsilon$ expansions of all master integrals up to weight 12
have been obtained\cite{Lee:2022art} using the DRA method\cite{Lee:2009dh}.

The complete result for the HQET field anomalous dimension up to 4 loops is\cite{Grozin:2022wse}
\begin{align}
&\gamma_h = 2 C_R (a-3) \frac{\alpha_s}{4\pi}
+ C_R \biggl(\frac{\alpha_s}{4\pi}\biggr)^{\!2}
\biggl[C_A \biggl(\frac{a^2}{2} + 4 a - \frac{179}{6}\biggr) + \frac{32}{3} T_F n_f\biggr]
\nonumber\displaybreak\\
&{} + C_R \biggl(\frac{\alpha_s}{4\pi}\biggr)^{\!3}
\biggl\{C_A^2 \biggl[\frac{5}{8} a^3 + \frac{3}{4} \biggl(\zeta_3 + \frac{13}{4}\biggr) a^2
+ \biggl(6 \zeta_3 - \frac{4}{45} \pi^4 + \frac{271}{16}\biggr) a
\nonumber\\
&\qquad{}
- \frac{123}{4} \zeta_3 - \frac{4}{15} \pi^4 - \frac{23815}{216}\biggr]
\nonumber\\
&\quad{}
+ C_A T_F n_f \biggl(- \frac{17}{2} a + 96 \zeta_3 + \frac{782}{27}\biggr)
- 6 C_F T_F n_f (16 \zeta_3 - 17)
+ \frac{160}{27} (T_F n_f)^2\biggr\}
\nonumber\\
&{} + \biggl(\frac{\alpha_s}{4\pi}\biggr)^{\!4}
\biggl\{C_R C_A^3 \biggl[\biggl(\frac{5}{3} \zeta_5 + \frac{\zeta_3}{6} + \frac{19}{2}\biggr) \frac{a^4}{16}
+ \biggl(21 \zeta_3 - \frac{\pi^4}{30} + \frac{149}{3}\biggr) \frac{a^3}{16}
\nonumber\\
&\qquad{}
- \biggl(\frac{169}{4} \zeta_5 - \frac{13}{3} \pi^2 \zeta_3 - \frac{653}{4} \zeta_3
+ \frac{121}{90} \pi^4 - \pi^2 - \frac{6707}{48}\biggr) \frac{a^2}{12}
\nonumber\\
&\qquad{}
- \biggl(\frac{272}{9} \zeta_5 - \frac{11}{4} \zeta_3^2 - \frac{164}{27} \pi^2 \zeta_3 - \frac{3839}{48} \zeta_3
- \frac{472}{8505} \pi^6 + \frac{9109}{4320} \pi^4 + \frac{2}{9} \pi^2
\nonumber\\
&\qquad{}
- \frac{1690475}{15552}\biggr) a
+ \frac{3859}{12} \zeta_5 + \frac{451}{4} \zeta_3^2 - \frac{709}{36} \pi^2 \zeta_3 - \frac{212237}{288} \zeta_3
\nonumber\\
&\qquad{}
+ \frac{850}{1701} \pi^6 - \frac{10501}{2160} \pi^4 + \frac{781}{36} \pi^2 - \frac{471001}{648}\biggr]
\nonumber\displaybreak\\
&\quad{}
+ d_{RA} \biggl[(5 \zeta_5 - 7 \zeta_3) \frac{a^4}{4} - 3 \zeta_3 a^3 - \frac{3}{2} (25 \zeta_5 - \zeta_3) a^2
\nonumber\\
&\qquad{}
+ \biggl(540 \zeta_5 + 24 \zeta_3^2 - \frac{128}{3} \pi^2 \zeta_3 - 11 \zeta_3
- \frac{884}{2835} \pi^6 + \frac{16}{3} \pi^2\biggr) a
\nonumber\\
&\qquad{}
- \frac{4815}{4} \zeta_5 - 384 \zeta_3^2 + \frac{320}{3} \pi^2 \zeta_3 + \frac{569}{4} \zeta_3
+ \frac{224}{405} \pi^6 + \frac{128}{15} \pi^4 - \frac{16}{3} \pi^2\biggr]
\nonumber\\
&\quad{}
+ C_R C_A^2 T_F n_f \biggl[\biggl(- \frac{7}{3} \zeta_3 + \frac{\pi^4}{180} - \frac{109}{36}\biggr) a^2
\nonumber\\
&\qquad{}
+ \biggl(\frac{4}{9} \zeta_5 - \frac{16}{27} \pi^2 \zeta_3 - 82 \zeta_3 + \frac{\pi^4}{15} - \frac{37957}{1944}\biggr) a
\nonumber\\
&\qquad{}
- \frac{1534}{3} \zeta_5 - 96 \zeta_3^2 + \frac{104}{9} \pi^2 \zeta_3 + \frac{5506}{3} \zeta_3
- \frac{3097}{540} \pi^4 - \frac{16}{3} \pi^2 + \frac{30617}{81}\biggr]
\nonumber\\
&\quad{}
+ C_R C_F C_A T_F n_f \biggl[\biggl(88 \zeta_3 + \frac{4}{15} \pi^4 - \frac{767}{6}\biggr) a
- 480 \zeta_5 - 928 \zeta_3 + \frac{88}{15} \pi^4 + \frac{21703}{27}\biggr]
\nonumber\\
&\quad{}
+ 16 C_R C_F^2 T_F n_f \biggl(60 \zeta_5 - 37 \zeta_3 - \frac{35}{3}\biggr)
\nonumber\\
&\quad{}
+ 64 d_{RF} n_f \biggl(- 5 \zeta_5 + \frac{8}{3} \pi^2 \zeta_3 + 4 \zeta_3 - \frac{8}{3} \pi^2\biggr)
\nonumber\\
&\quad{}
+ 2 C_R C_A (T_F n_f)^2 \biggl[\frac{4}{3} \biggl(4 \zeta_3 - \frac{269}{81}\biggr) a
- 192 \zeta_3 + \frac{16}{15} \pi^4 - \frac{1027}{81}\biggr]
\nonumber\\
&\quad{}
+ 32 C_R C_F (T_F n_f)^2 \biggl(12 \zeta_3 - \frac{\pi^4}{15} - \frac{103}{27}\biggr)
\nonumber\\
&\quad{}
- \frac{256}{27} C_R (T_F n_f)^3 (3 \zeta_3 - 1)\biggr\}
+ \cdots
\label{Gamma:gamma}
\end{align}
The terms up to $\alpha_s^3$ agree with\cite{Melnikov:2000zc,Chetyrkin:2003vi}.
Some color structures of the $\alpha_s^4$ contribution were known earlier, see table~\ref{T:4loop}.
In QED the only gauge-dependent term is the 1-loop one;
if $n_f=0$, all contributions but the 1-loop one vanish (Sect.~\ref{S:Exp}).

Curiously, the difference of $\gamma_h$ (with $R = F$) and $\gamma_q$
is gauge invariant up to 2 loops,
linear in $a$ at $\alpha_s^3$ and quadratic in $a$ at $\alpha_s^4$:
\begin{align}
&\gamma_h - \gamma_q = - 6 C_F \frac{\alpha_s}{4\pi}
+ C_F \biggl(\frac{\alpha_s}{4\pi}\biggr)^{\!2} \biggl(3 C_F - \frac{127}{3} C_A + \frac{44}{3} T_F n_f\biggr)
\nonumber\\
&{} + C_F \biggl(\frac{\alpha_s}{4\pi}\biggr)^{\!3}
\biggl\{
C_A^2 \biggl[\biggl(\frac{9}{2} \zeta_3 - \frac{4}{45} \pi^4 + \frac{1}{2}\biggr) a
  - \frac{27}{2} \zeta_3 - \frac{4}{15} \pi^4 - \frac{6410}{27}\biggr]
\nonumber\\
&\quad{}
- C_F C_A \biggl(24 \zeta_3 - \frac{143}{2}\biggr)
- 3 C_F^2
\nonumber\\
&\quad{}
+ 8 C_A T_F n_f \biggl(12 \zeta_3 + \frac{313}{27}\biggr)
- 96 C_F T_F n_f (\zeta_3 - 1)
+ \frac{40}{27} (T_F n_f)^2\biggr\}
\nonumber\displaybreak\\
&{} + \biggl(\frac{\alpha_s}{4\pi}\biggr)^{\!4} \biggl\{
C_F C_A^3 \biggl[\biggl(\frac{7}{4} \zeta_5 + \frac{13}{9} \pi^2 \zeta_3 + \frac{43}{6} \zeta_3
  - \frac{191}{540} \pi^4 + \frac{\pi^2}{3} + \frac{5}{3}\biggr) \frac{a^2}{4}
\nonumber\\
&\qquad{} + \biggl(\frac{637}{36} \zeta_5 + \frac{11}{4} \zeta_3^2 + \frac{164}{27} \pi^2 \zeta_3 + \frac{139}{6} \zeta_3
  + \frac{472}{8505} \pi^6 - \frac{4409}{2160} \pi^4 - \frac{2}{9} \pi^2 + \frac{433}{72}\biggr) a
\nonumber\\
&\qquad{} + \frac{27871}{48} \zeta_5 + \frac{451}{4} \zeta_3^2 - \frac{709}{36} \pi^2 \zeta_3 - \frac{39089}{72} \zeta_3
  + \frac{850}{1701} \pi^6 - \frac{6389}{1080} \pi^4 + \frac{781}{36} \pi^2
\nonumber\\
&\qquad{}
  - \frac{1233089}{648}\biggr]
\nonumber\\
&\quad{}
- C_F^2 C_A^2 \biggl[(10 \zeta_5 - 7 \zeta_3 + 6) a
  + 1570 \zeta_5 - 421 \zeta_3 - \frac{22}{15} \pi^4 - \frac{23885}{18}\biggr]
\nonumber\\
&\quad{}
+ C_F^3 C_A \biggl(2880 \zeta_5 - 1696 \zeta_3 - \frac{5131}{6}\biggr)
- C_F^4 \biggl(1280 \zeta_5 - 800 \zeta_3 - \frac{1027}{4}\biggr)
\nonumber\\
&\quad{}
+ 2 d_{RA} \biggl[- 2 (5 \zeta_5 + \zeta_3) a^2
\nonumber\\
&\qquad{}
  + \biggl(325 \zeta_5 + 12 \zeta_3^2 - \frac{64}{3} \pi^2 \zeta_3 - 56 \zeta_3 - \frac{442}{2835} \pi^6 + \frac{8}{3} \pi^2 - 1\biggr) a
\nonumber\\
&\qquad{}
- 495 \zeta_5 - 192 \zeta_3^2 + \frac{160}{3} \pi^2 \zeta_3 - 68 \zeta_3
  + \frac{112}{405} \pi^6 + \frac{64}{15} \pi^4 - \frac{8}{3} \pi^2 + 67\biggr]
\nonumber\\
&\quad{}
+ C_F C_A^2 T_F n_f \biggl[\biggl(\frac{4}{9} \zeta_5 - \frac{16}{27} \pi^2 \zeta_3 - 8 \zeta_3
  + \frac{11}{45} \pi^4 - \frac{4}{9}\biggr) a
\nonumber\\
&\qquad{}
- \frac{1054}{3} \zeta_5 - 96 \zeta_3^2 + \frac{104}{9} \pi^2 \zeta_3 + \frac{5515}{3} \zeta_3
  - \frac{289}{54} \pi^4 - \frac{16}{3} \pi^2 + \frac{85730}{81}\biggr]
\nonumber\\
&\quad{}
- 2 C_F^2 C_A T_F n_f \biggl(400 \zeta_5 + 464 \zeta_3 - \frac{8}{3} \pi^4 - \frac{10931}{27}\biggr)
\nonumber\\
&\quad{}
+ 8 C_F^3 T_F n_f \biggl(120 \zeta_5 - 58 \zeta_3 - \frac{89}{3}\biggr)
\nonumber\\
&\quad{}
- 64 d_{FF} n_f \biggl(5 \zeta_5 - \frac{8}{3} \pi^2 \zeta_3 - 4 \zeta_3 + \frac{8}{3} \pi^2 + 4\biggr)
\nonumber\\
&\quad{}
- 32 C_F C_A (T_F n_f)^2 \biggl(14 \zeta_3 - \frac{\pi^4}{15} + \frac{229}{81}\biggr)
+ 32 C_F^2 (T_F n_f)^2 \biggl(14 \zeta_3 - \frac{\pi^4}{15} - \frac{160}{27}\biggr)
\nonumber\\
&\quad{}
- \frac{16}{9} C_F (T_F n_f)^3 \biggl(16 \zeta_3 - \frac{13}{9}\biggr)
\biggr\} + \cdots
\label{Gamma:hq}
\end{align}
(see\cite{Luthe:2017ttg,Chetyrkin:2017bjc} for $\gamma_q$).
In QED the only gauge-dependent term in $\gamma_h$ is the 1-loop one,
the same is true for $\gamma_q$ (see\cite{Grozin:2010wa}),
and $\gamma_h-\gamma_q$ is gauge invariant to all orders.
The on-shell renormalization constant of a heavy quark field $Z_Q^{\text{os}}$
has the same pattern of $a$ dependence.

As a check of calculations,
the $hhg$ vertex at $p_g = 0$ has also been calculated in\cite{Grozin:2022wse}.
It has a single structure $\Gamma(\omega) v^\mu t_R^a$.
Infrared divergences are absent,
therefore $\log \Gamma(\omega) = \log Z_\Gamma(\alpha_s,a) + \mathcal{O}(\varepsilon^0)$,
where $Z_\Gamma$ contains ultraviolet divergences.
We have $Z_\alpha(\alpha_s) = Z_A^{-1}(\alpha_s,a) (Z_\Gamma(\alpha_s,a) Z_h(\alpha_s,a))^{-2}$,
and hence
\begin{align}
\beta(\alpha_s) &{}= \frac{1}{2} \frac{d\log Z_\alpha(\alpha_s)}{d\log\mu}
= \sum_{L=1}^\infty \beta_{L-1} \biggl(\frac{\alpha_s}{4\pi}\biggr)^{\!L}
\nonumber\\
&{} = - \gamma_\Gamma(\alpha_s,a) - \gamma_h(\alpha_s,a) - \frac{1}{2} \gamma_A(\alpha_s,a)
\label{Gamma:beta}
\end{align}
The well known result for the 4-loop $\beta$ function\cite{vanRitbergen:1997va,Czakon:2004bu}
is reproduced using the 4-loop $\gamma_A$\cite{vanRitbergen:1997va,Czakon:2004bu}.
This is a strong check of the calculation\cite{Grozin:2022wse} of $\gamma_h$.

If we consider QCD with $n_f$ light flavors and 1 heavy flavor $Q$,
then, in situations accessible for the HQET approach,
the QCD heavy quark field $Q$ can be expressed via the HQET field $h_v$
via the matching relation\cite{Grozin:2010wa,Grozin:2020jvt}
\begin{equation}
Q(\mu) = z(\mu) h_v(\mu) + \mathcal{O}\biggl(\frac{1}{M}\biggr)\,.
\label{Gamma:dec}
\end{equation}
The matching coefficient $z(\mu)$ can be obtained from the on-shell renormalization constant
of the field $Q$ in the $(n_f+1)$-flavor QCD
\begin{equation}
Z_Q^{\text{os}} = 1 + \sum_{L=1}^\infty
\biggl(4 \frac{g_0^2 M^{-2\varepsilon}}{(4\pi)^{d/2} \varepsilon} e^{-\gamma\varepsilon}\biggr)^{\!L} Z_L\,,\quad
Z_L = \sum_{n=0}^\infty Z_{L,n}(\xi) \varepsilon^n\,,
\label{Gamma:Z}
\end{equation}
where $g_0 = g_0^{(n_f+1)}$, $\xi = 1 - a_0^{(n_f+1)}$, $M$ is the on-shell mass of $Q$.
The renormalized matching coefficient $z(\mu)$ must be finite.
This requirement was used\cite{Grozin:2020jvt} to obtain analytical expressions
for $Z_{4,n}$ with $n<4$, see tables~I and~II.
However, analytical expressions for the color structures $C_F C_A^3$ and $d_{FA}$ in $Z_{4,3}$
remained unknown (table~II), because the corresponding terms in $\gamma_h$ were not known.
Now we are in a position to complete the table~II of\cite{Grozin:2020jvt}:
\begin{align}
&Z_{4,3} = C_F C_A^3 \biggl[
\frac{1}{16} \biggl(\frac{2387}{3} a_4 + \frac{2387}{72} a_1^4 + \frac{2849}{36} \pi^2 a_1^2 - \frac{15907}{72} \pi^2 a_1
    - \frac{102349}{384} \zeta_5 + \frac{231}{256} \zeta_3^2
\nonumber\\
&\qquad{} + \frac{44671}{576} \pi^2 \zeta_3 + \frac{5389931}{6912} \zeta_3
    + \frac{787}{181440} \pi^6 - \frac{5659447}{414720} \pi^4 - \frac{275029}{10368} \pi^2 - \frac{155786381}{124416}\biggr)
\nonumber\\
&\quad{} + \frac{\xi}{256} \biggl(\frac{1757}{192} \zeta_5 - \frac{11}{32} \zeta_3^2 + \frac{169}{48} \pi^2 \zeta_3 - \frac{3389}{96} \zeta_3
    - \frac{59}{8505} \pi^6 + \frac{34433}{51840} \pi^4 - \frac{353}{144} \pi^2 - \frac{6629}{576}\biggr)
\nonumber\\
&\quad{} - \frac{\xi^2}{8192} \biggl(\frac{203}{12} \zeta_5 + \frac{17}{3} \pi^2 \zeta_3 - \frac{355}{6} \zeta_3
    + \frac{59}{60} \pi^4 - \frac{13}{3} \pi^2 - 19\biggr)\biggr]
\nonumber\\
&{} - d_{FA} \biggl[
\frac{1}{16} \biggl(\frac{45}{16} \zeta_5 + \frac{45}{16} \zeta_3^2 - \frac{\pi^2}{2} \zeta_3 + \frac{63}{32} \zeta_3
    - \frac{19}{10080} \pi^6 - \frac{\pi^4}{15} - \frac{33}{32}\biggr)
\nonumber\\
&\quad{} + \frac{\xi}{16} \biggl(\frac{305}{64} \zeta_5 + \frac{3}{16} \zeta_3^2 - \frac{\pi^2}{3} \zeta_3 - \frac{15}{16} \zeta_3
    - \frac{221}{90720} \pi^6 + \frac{\pi^2}{24} - \frac{1}{64}\biggr)
+ \frac{\xi^2}{512} (5 \zeta_5 + \zeta_3)\biggr] + \cdots
\nonumber\\
&{} = C_F C_A^3 (- 123.3401041 + 0.1197511751 \xi - 0.005818521661 \xi^2)
\nonumber\\
&{} + d_{FA} (0.3701524967 + 0.1134626128 \xi - 0.01247401500 \xi^2) + \cdots
\label{Gamma:Z43}
\end{align}
where $a_n = \Li{n}\left(\frac{1}{2}\right)$ (in particular, $a_1 = \log 2$),
and dots mean other color structures (table~II in\cite{Grozin:2020jvt}).
The corresponding numerical results from the tables~V, VI, VII of\cite{Marquard:2018rwx} are
\begin{align*}
&Z_{4,3} = C_F C_A^3 \bigl[- 123.354 \pm 0.086 + (0.11976 \pm 0.00013) \xi - (0.005817 \pm 0.000025) \xi^2\bigr]\\
&{} + d_{FA} \bigl[0.40 \pm 0.21 + (0.1135 \pm 0.0025) \xi - (0.01250 \pm 0.00061) \xi^2\bigr] + \cdots
\end{align*}
Our analytical results~(\ref{Gamma:Z43}) agree with them within the stated uncertainties.
This is a good check of the new\cite{Grozin:2022wse} $C_R C_A^3$ and $d_{RA}$ terms in~(\ref{Gamma:gamma}).

\section{Cusp anomalous dimension at small angles}
\label{S:Cusp}

When calculating $V(\omega,\omega,\varphi)$, we write
\begin{equation}
v' = v + \delta v\,,\quad
\delta v = v (\cosh\varphi-1) + n \sinh\varphi\,,\quad
v \cdot n = 0\,,\quad
n^2 = -1\,.
\label{Calc:dv}
\end{equation}
We expand the integrands in $\delta v$ and average over $n$ direction
in the $(d-1)$-dimensional subspace orthogonal to $v$
(this method of calculating the small-angle cusp anomalous dimension
was first used at 2 loops\cite{Bagan:1993js}).

The complete result for the first 2 terms of the small-angle expansion~(\ref{HQET:small})
of the cusp anomalous dimension up to 4 loops is\cite{Grozin:2022wse}
\begin{align}
&\Gamma(\alpha_s,\varphi) = 4 \frac{\alpha_s}{4\pi} (\varphi \coth\varphi - 1) \biggl\{ C_R
- \frac{2}{3} C_R \frac{\alpha_s}{4\pi}
\biggl[C_A \biggl(\pi^2 - \frac{47}{3}\biggr) + \frac{10}{3} T_F n_f\biggr]
\nonumber\\
&{} + C_R \biggl(\frac{\alpha_s}{4\pi}\biggr)^{\!2}
\biggl[\frac{C_A^2}{3} \biggl(10 \zeta_3 + 2 \pi^4 - \frac{340}{9} \pi^2 + \frac{473}{2}\biggr)
\nonumber\\
&\quad{}
- \frac{2}{3} C_A T_F n_f \biggl(28 \zeta_3 - \frac{40}{9} \pi^2 + \frac{389}{9}\biggr)
+ C_F T_F n_f \biggl(16 \zeta_3 - \frac{55}{3}\biggr)
- \frac{16}{27} (T_F n_f)^2\biggr]
\nonumber\\
&{} + \biggl(\frac{\alpha_s}{4\pi}\biggr)^{\!3}  \biggl[
- \frac{C_R C_A^3}{3} \biggl(310 \zeta_5 + \frac{64}{3} \pi^2 \zeta_3 - \frac{4756}{9} \zeta_3
  + \frac{20}{9} \pi^6 - \frac{4841}{90} \pi^4 + \frac{35906}{81} \pi^2
\nonumber\\
&\quad{}
  - \frac{89011}{54}\biggr)
- \frac{16}{3} d_{RA} \pi^2 \biggl(34 \zeta_3 - \frac{2}{15} \pi^4 - \frac{8}{3} \pi^2 + \frac{1}{3}\biggr)
\nonumber\\
&\quad{}
+ \frac{C_R C_A^2 T_F n_f}{3} \biggl(440 \zeta_5 + \frac{224}{3} \pi^2 \zeta_3 - \frac{14444}{9} \zeta_3
  - \frac{88}{9} \pi^4 + \frac{14768}{81} \pi^2 - \frac{48161}{54}\biggr)
\nonumber\\
&\quad{}
+ C_R C_F C_A T_F n_f \biggl(80 \zeta_5 - \frac{64}{3} \pi^2 \zeta_3 + \frac{2720}{9} \zeta_3
  - \frac{44}{45} \pi^4 + \frac{220}{9} \pi^2 - \frac{25943}{81}\biggr)
\nonumber\\
&\quad{}
- 2 C_R C_F^2 T_F n_f \biggl(80 \zeta_5 - \frac{148}{3} \zeta_3 - \frac{143}{9}\biggr)
+ \frac{16}{3} d_{RF} n_f \pi^2 \biggl(16 \zeta_3 - \frac{5}{3} \pi^2 - \frac{10}{3}\biggr)
\nonumber\\
&\quad{} + \frac{C_R C_A (T_F n_f)^2}{27} \biggl(2240 \zeta_3 - \frac{56}{5} \pi^4 - \frac{608}{9} \pi^2 + \frac{1835}{3}\biggr)
\nonumber\\
&\quad{} - \frac{8}{9} C_R C_F (T_F n_f)^2 \biggl(80 \zeta_3 - \frac{2}{5} \pi^4 - \frac{299}{9}\biggr)
    + \frac{64}{27} C_R (T_F n_f)^3 \biggl(2 \zeta_3 - \frac{1}{3}\biggr)
\biggr] \biggr\}
\nonumber\displaybreak\\
&{} + \varphi^4 \biggl(\frac{\alpha_s}{4\pi}\biggr)^{\!2} \biggl\{ \frac{4}{135} C_R C_A
- \frac{16}{3} C_R C_A \frac{\alpha_s}{4\pi}
\biggl[C_A \biggl(\frac{23}{25} \zeta_3 - \frac{\pi^2}{27} - \frac{1531}{2025}\biggr) + \frac{2}{81} T_F n_f\biggr]
\nonumber\\
&{} + \frac{4}{9} \biggl(\frac{\alpha_s}{4\pi}\biggr)^{\!2} \biggl[
\frac{C_R C_A^3}{15} \biggl(2816 \zeta_5 + \frac{1864}{15} \pi^2 \zeta_3 - \frac{1194292}{225} \zeta_3
    - \frac{907}{225} \pi^4 - \frac{16858}{405} \pi^2
\nonumber\\
&\quad{}
+ \frac{5696611}{2025}\biggr)
- 16 d_{RA} \biggl(14 \zeta_5 - \frac{268}{75} \pi^2 \zeta_3 - \frac{566}{45} \zeta_3
    + \frac{53}{225} \pi^4 + \frac{848}{405} \pi^2 - \frac{8}{27}\biggr)
\nonumber\\
&\quad{} + \frac{2}{15} C_R C_A^2 T_F n_f \biggl(32 \zeta_5 - \frac{64}{15} \pi^2 \zeta_3 + \frac{39368}{75} \zeta_3
    - \frac{514}{225} \pi^4 + \frac{361}{135} \pi^2 - \frac{874967}{2025}\biggr)
\nonumber\\
&\quad{} + \frac{2}{3} C_R C_F C_A T_F n_f \biggl(\frac{16}{5} \zeta_3 - \frac{11}{3}\biggr)
\nonumber\\
&\quad{}
- 16 d_{RF} n_f \biggl(16 \zeta_5 - \frac{16}{75} \pi^2 \zeta_3 - \frac{284}{25} \zeta_3
  - \frac{8}{75} \pi^4 + \frac{23}{25} \pi^2 + \frac{23}{25}\biggr)
\nonumber\\
&\quad{}
+ \frac{304}{1215} C_R C_A (T_F n_f)^2
\biggr] \biggr\} + \mathcal{O}(\varphi^6,\alpha_s^5)\,.
\label{Cusp:QCD}
\end{align}
The terms up to $\alpha_s^3$ agree with the results\cite{Grozin:2014hna,Grozin:2015kna} expanded in $\varphi^2$
(see~(\ref{Conj:small})).
Some color structures of the $\alpha_s^4$ contribution were known earlier, see table~\ref{T:4loop}.
For some of them, the full $\varphi$ dependence is known;
for some, one more term in the $\varphi^2$ expansion is known\cite{Bruser:2019auj}
in addition to the terms presented in~(\ref{Cusp:QCD}).
In the abelian case $d_{RF} n_f \alpha^4$ is the only term
which does not have the simple 1-loop angle dependence\cite{Grozin:2018vdn}
(note that there is a typo in the formula~(4.2) in this paper).

In the first curly bracket in~(\ref{Cusp:QCD}),
the highest weight of constants in the $\alpha_s^L$ term is $2(L-1)$.
In the second curly bracket terms with this highest weight are absent.

In $\mathcal{N}=4$ supersymmetric Yang--Mills theory (SYM) supersymmetric Wilson lines are usually discussed.
They interact not only with gluons but also with scalars.
A cusp on a supersymmetric Wilson line is characterized by the geometric angle $\varphi$
and the internal angle $\vartheta$ (we'll consider the case $\vartheta=0$).
In the case of $SU(N_c)$ gauge group in the large $N_c$ limit
the Bremsstrahlung function (the $\varphi^2$ term) is known exactly in coupling\cite{Correa:2012at}.
The full $\varphi$ dependence is known up to 4 loops\cite{Henn:2013wfa}.
The result has the structure
\begin{equation*}
\Gamma = \sum_{L=1}^\infty \Gamma_L \biggl(\frac{N_c \alpha_s}{2 \pi}\biggr)^{\!L}\,,\quad
\Gamma_L = \sum_{n=1}^L \Gamma_{Ln} \tanh^n\frac{\varphi}{2}\,.
\end{equation*}
Up to 3 loops the large-$N_c$ results are sufficient for reconstructing the complete results
for an arbitrary gauge groups via Casimirs.
However, at 4 loops $d_{RA}/(C_R C_A^3) = 1/24 + \mathcal{O}(1/N_c^2)$
(the $1/N_c^2$ term depends on $R$),
and we know only a certain linear combination of the coefficients of $C_R C_A^3$ and $d_{RA}$.
The Bremsstrahlung function is known for an arbitrary gauge group via Casimirs
up to (in principle) an arbitrarily high order\cite{Fiol:2018yuc}.
Expanding the results of\cite{Henn:2013wfa} up to $\varphi^4$
and replacing the $\varphi^2$ term by the result of\cite{Fiol:2018yuc}
we have
\begin{align}
&\Gamma = \frac{\alpha_s}{2\pi} \varphi \tanh\frac{\varphi}{2} \biggl[ C_R
- \frac{1}{6} C_R C_A \pi \alpha_2 + \frac{1}{24} C_RC_A^2 (\pi \alpha_s)^2
\nonumber\\
&\quad{} - \biggl(\frac{5}{24} C_R C_A^3 - \frac{d_{RA}}{5}\biggr) \frac{(\pi \alpha_s)^3}{18} \biggr]
\nonumber\\
&{} + \frac{N_c \alpha_s \varphi^4}{2 \pi} \biggl[ \frac{1}{6}
- \frac{3}{2} \zeta_3 \frac{N_c \alpha_s}{2 \pi}
+ \biggl(\frac{45}{4} \zeta_5 + \frac{2}{3} \pi^2 \zeta_3 - \frac{2}{45} \pi^4\biggr)
\biggl(\frac{N_c \alpha_s}{2 \pi}\biggr)^{\!2} \biggr]
\nonumber\\
&{} + \mathcal{O}(\varphi^6,\alpha_s^5)\,.
\label{Cusp:super}
\end{align}
The $\mathcal{O}(\varphi)$ terms in $\Gamma_{L1}$ contain only maximum-weight contributions,
and produce the first bracket in~(\ref{Cusp:super});
it is homogeneous in weight.
The $\mathcal{O}(\varphi^3)$ terms in $\Gamma_{L1}$
and the $\mathcal{O}(\varphi^2)$ terms in $\Gamma_{L2}$
contain only lower-weight contributions,
and produce the second square bracket in~(\ref{Cusp:super});
it is not homogeneous and only known in the $N_c\to\infty$ limit.
If we keep only maximum-weight terms, this second bracket vanishes,
just like the second curly bracket in the QCD result~(\ref{Cusp:QCD}).
If we keep only maximum-weight terms in the first curly bracket in~(\ref{Cusp:QCD}),
we obtain exactly the first square bracket in the supersymmetric result~(\ref{Cusp:super}).
So, the principle of maximal transcendentality\cite{Kotikov:2002ab,Kotikov:2004er}
works for the Bremsstrahlung function up to 4 loops.

The HQET field anomalous dimension $\gamma_h$~(\ref{Gamma:gamma})
has the same pattern of weights as the first curly bracket in~(\ref{Cusp:QCD}).
If we retain only the highest weights $2(L-1)$ in $\gamma_h$~(\ref{Gamma:gamma}),
we get
\begin{align*}
&2 C_F (a-3) \frac{\alpha_s}{4\pi}
- \frac{4}{15} \biggl(\frac{a}{3} + 1\biggr) \pi^4 \biggl(\frac{\alpha_s}{4\pi}\biggr)^{\!3}\\
&{} + \biggl(\frac{\alpha_s}{4\pi}\biggr)^{\!4} \biggl\{
C_R C_A^3 \biggl[\biggl(\frac{11}{4} \zeta_3^2 + \frac{472}{8505} \pi^6\biggr) a
+ \frac{451}{4} \zeta_3^2 + \frac{850}{1701} \pi^6\biggr]\\
&\quad{} + 4 d_{RA} \biggl[\biggl(6 \zeta_3^2 - \frac{221}{2835} \pi^6\biggr) a
- 8 \biggl(12 \zeta_3^2 - \frac{7}{405} \pi^6\biggr)\biggr]
- 96 \zeta_3^2 C_R C_A T_F n_f
\biggr\} + \cdots
\end{align*}
(the $\alpha_s^2$ term is absent because the corresponding term in $\gamma_h$ contains no $\pi^2$).
All terms are linear in $a$ here.
It would be interesting to understand if this expression is somehow related to the anomalous dimension
of an end of Wilson line in the $\mathcal{N}=4$ SYM.
The term with $n_f$ does not look encouraging in this respect.

\section{Light-like cusp anomalous dimension}
\label{S:Light}

The full result up to 4 loops has been obtained in\cite{Henn:2019swt}
and confirmed in\cite{vonManteuffel:2020vjv} from form factor calculations:
\begin{align}
&K(\alpha_s) = 4 \frac{\alpha_s}{4\pi} \biggl\{ C_R
- C_R \frac{\alpha_s}{4\pi}
\biggl[\frac{C_A}{3} \biggl(\pi^2 - \frac{67}{3}\biggr) + \frac{20}{9} T_F n_f\biggr]
\nonumber\\
&{} + C_R \biggl(\frac{\alpha_s}{4\pi}\biggr)^{\!2} \biggl[
\frac{C_A^2}{3} \biggl(22 \zeta_3 + \frac{11}{15} \pi^4 - \frac{134}{9} \pi^2 + \frac{245}{2}\biggr)
\nonumber\\
&\quad{} - \frac{2}{3} C_A T_F n_f \biggl(28 \zeta_3 - \frac{20}{9} \pi^2 + \frac{209}{9}\biggr)
+ C_F T_F n_f \biggl(16 \zeta_3 - \frac{55}{3}\biggr)
- \frac{16}{27} (T_F n_f)^2
\biggr]
\nonumber\\
&{} + \biggl(\frac{\alpha_s}{4\pi}\biggr)^{\!3} \biggl[
- C_R C_A^3 \biggl(\frac{902}{9} \zeta_5 + 4 \zeta_3^2 + \frac{44}{9} \pi^2 \zeta_3 - \frac{5236}{27} \zeta_3
+ \frac{626}{2835} \pi^6 - \frac{451}{90} \pi^4
\nonumber\\
&\qquad{} + \frac{11050}{243} \pi^2 - \frac{42139}{162}\biggr)
+ 8 d_{RA} \biggl(\frac{110}{3} \zeta_5 - 12 \zeta_3^2 + \frac{4}{3} \zeta_3 - \frac{31}{945} \pi^6 - \frac{2}{3} \pi^2\biggr)
\nonumber\\
&\quad{} + \frac{C_R C_A^2 T_F n_f}{9} \biggl(1048 \zeta_5 + 112 \pi^2 \zeta_3 - \frac{11552}{3} \zeta_3
- \frac{44}{15} \pi^4 + \frac{5080}{27} \pi^2 - \frac{24137}{18}\biggr)
\nonumber\\
&\quad{} + C_R C_F C_A T_F n_f \biggl(80 \zeta_5 - \frac{32}{3} \pi^2 \zeta_3 + \frac{1856}{9} \zeta_3
- \frac{44}{45} \pi^4 + \frac{110}{9} \pi^2 - \frac{17033}{81}\biggr)
\nonumber\\
&\quad{} - 2 C_R C_F^2 T_F n_f \biggr(80 \zeta_5 - \frac{148}{3} \zeta_3 - \frac{143}{9}\biggr)
- \frac{32}{3} d_{RF} n_f \biggl(10 \zeta_5 + 2 \zeta_3 - \pi^2\biggr)
\nonumber\\
&\quad{} + \frac{C_R C_A (T_F n_f)^2}{27} \biggl(2240 \zeta_3 - \frac{56}{5} \pi^4 - \frac{304}{9} \pi^2 + \frac{923}{3}\biggr)
\nonumber\\
&\quad{} - \frac{8}{9} C_R C_F (T_F n_f)^2 \biggl(80 \zeta_3 - \frac{2}{5} \pi^4 - \frac{299}{9}\biggr)
\nonumber\\
&\quad{} + \frac{64}{27} C_R (T_F n_f)^3 \biggl(2 \zeta_3 - \frac{1}{3}\biggr)
\biggr]
+ \mathcal{O}(\alpha_s^4) \biggr\}
\label{Light:QCD}
\end{align}
The terms up to $\alpha_s^3$ agree with\cite{Moch:2004pa}.
Some color structures of the $\alpha_s^4$ contribution were known earlier,
see table~\ref{T:4loop}.
The 4-loop $C_R C_F C_A T_F n_f$ in\cite{Henn:2019swt} was derived from a conjecture,
see Sect.~\ref{S:Conj} for details;
in\cite{vonManteuffel:2020vjv} it was confirmed by a direct calculation.
In QED with $n_f=0$ only the 1-loop term remains.

In $\mathcal{N}=4$ SYM with $SU(N_c)$ gauge group in the large $N_c$ limit
the light-like anomalous dimension is known\cite{Beisert:2006ez} exactly in $g^2 N_c$;
results up to 4 loops were derived in\cite{Bern:2006ew,Cachazo:2006az,Henn:2013wfa}.
Up to 3 loops these results are sufficient for reconstructing the full result
for an arbitrary gauge group expressed via Casimirs.
At 4 loops there are 2 different Casimirs;
the full analytical result has been obtained in\cite{Henn:2019swt,Huber:2019fxe}:
\begin{align}
&K(\alpha_s) = 4 \frac{\alpha_s}{4\pi} \biggl\{ C_R
- \frac{\pi^2}{3} C_R C_A \frac{\alpha_s}{4\pi}
+ \frac{11}{45} \pi^4 C_R C_A^2 \biggl(\frac{\alpha_s}{4\pi}\biggr)^{\!2}
\nonumber\\
&{} - \biggl(\frac{\alpha_s}{4\pi}\biggr)^{\!3} \biggl[
2 C_R C_A^3 \biggl(2 \zeta_3^2 + \frac{313}{2835} \pi^6\biggr)
+ 8 d_{RA} \biggl(12 \zeta_3^2 + \frac{31}{945} \pi^6\biggr)
\biggr] + \mathcal{O}(\alpha_s^4) \biggr\}
\label{Light:super}
\end{align}
If we keep only maximum-weight terms in the QCD result~(\ref{Light:QCD}),
we obtain exactly the SYM result~(\ref{Light:super}).
So, the principle of maximal transcendentality\cite{Kotikov:2002ab,Kotikov:2004er}
works for the light-like anomalous dimension up to 4 loops.

\section{A conjecture which sometimes works}
\label{S:Conj}

An interesting property of $\Gamma$ up to 3 loops has been noticed in\cite{Grozin:2014hna,Grozin:2015kna}.
Let's introduce a new coupling $A$ instead of $\alpha_s/(4\pi)$:
\begin{align}
&K(\alpha_s) = 4 C_R A\,,\quad
A = \frac{\alpha_s}{4\pi}
\biggl[1 + K_2 \frac{\alpha_s}{4\pi} + K_3 \biggl(\frac{\alpha_s}{4\pi}\biggr)^{\!2} + \cdots\biggr]\,,
\label{Conj:A}\\
&K_2 = C_A K_A + T_F n_f K_f\,,
\nonumber\\
&K_3 = C_A^2 K_{AA} + C_A T_F n_f K_{Af} + C_F T_F n_f K_{Ff} + (T_F n_f)^2 K_{ff}\,,\,\ldots
\nonumber
\end{align}
where $K(\alpha_s)$ is the light-like cusp anomalous dimension (see~(\ref{Light:QCD})),
and re-express $\Gamma(\alpha_s,\varphi)$ via it:
\begin{equation}
\Gamma(\alpha_s,\varphi) = \Omega(A,\varphi)
= C_R \biggl[\Omega_1(\varphi) A + \Omega_2(\varphi) A^2 + \Omega_3(\varphi) A^3 + \cdots\biggr]\,.
\label{Conj:Gamma}
\end{equation}
Then the function $\Omega$ does not depend of $n_f$,
i.e.\ on the number of matter spinor fields in fundamental representation.
Moreover, it remains the same in a generic gauge theory with any number of fermions and scalars
(including supersymmetric QCD extensions).
It contains only the adjoint-representation color structures:
\begin{equation}
\Omega_2(\varphi) = C_A \Omega_A(\varphi)\,,\quad
\Omega_3(\varphi) = C_A^2 \Omega_{AA}(\varphi)\,,\,\ldots
\label{Conj:Omega3}
\end{equation}

If we write $\Gamma(\alpha_s,\varphi)$ as
\begin{align*}
&\Gamma(\alpha_s,\varphi) = C_R \frac{\alpha_s}{4\pi}
\biggl[\Gamma_1(\varphi) + \Gamma_2(\varphi) \frac{\alpha_s}{4\pi}
+ \Gamma_3(\varphi) \biggl(\frac{\alpha_s}{4\pi}\biggr)^{\!2} + \cdots\biggr]\,,\\
&\Gamma_2(\varphi) = C_A \Gamma_A(\varphi) + T_F n_f \Gamma_f(\varphi)\,,\\
&\Gamma_3(\varphi) = C_A^2 \Gamma_{AA}(\varphi) + C_A T_F n_f \Gamma_{Af}(\varphi)
+ C_F T_F n_f \Gamma_{Ff}(\varphi) + (T_F n_f)^2 \Gamma_{ff}(\varphi)\,,\,\ldots\,,
\end{align*}
then
\begin{align}
&\Gamma_1(\varphi) = \Omega_1(\varphi)\,,\quad
\Gamma_A(\varphi) = \Omega_A(\varphi) + K_A \Omega_1(\varphi)\,,
\label{Conj:12}\\
&\Gamma_{AA}(\varphi) = \Omega_{AA}(\varphi) + 2 K_A \Omega_A(\varphi) + K_{AA} \Omega_1(\varphi)\,,
\label{Conj:Glue3}\\
&\Gamma_f(\varphi) = K_f \Omega_1(\varphi)\,,\quad
\Gamma_{Ff}(\varphi) = K_{Ff} \Omega_1(\varphi)\,,\quad
\Gamma_{ff}(\varphi) = K_{ff} \Omega_1(\varphi)\,,
\label{Conj:1L3}\\
&\Gamma_{Af}(\varphi) = 2 K_f \Omega_A(\varphi) + K_{Af} \Omega_1(\varphi)\,,
\label{Conj:3}
\end{align}
For purely gluonic structures $\Gamma_X$ ($X=1$, $A$, $AA$),
using~(\ref{Conj:Glue3}) we can express $\Omega_X$ via $\Gamma_X$ and lower-loop results.
For the abelian structures $X=1$, $f$, $Ff$, $ff$, the terms
$\Gamma_X$ are given by diagrams with a single 2-leg c-webs (see Sect.~\ref{S:abel}),
and hence have the pure 1-loop angle dependence $\Omega_1(\varphi) = 4 (\varphi \coth\varphi - 1)$.
In these cases, the coefficient of $\Omega_1$ is fixed by the $\varphi\to\infty$ limit to be $K_X$;
the relations~(\ref{Conj:1L3}) hold by construction, and contain no interesting information.
The relation~(\ref{Conj:3}) is the only interesting one.
It expresses the contribution of highly non-trivial 3-loop diagrams
\begin{equation*}
\includegraphics{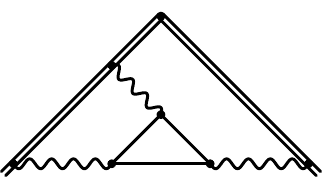}
\end{equation*}
to $\Gamma$ via lower-loop contributions.
The $\varphi$ dependence of $\Gamma_{Af}$ is very non-trivial;
it seems absolutely impossible that the relation~(\ref{Conj:3}) holds accidentally,
it must have some explanation (which is not yet known).

The result of the calculation\cite{Grozin:2014hna,Grozin:2015kna} is
\begin{align}
&\Omega_1 = 4 \tilde{A}_1\,,\quad
\Omega_A = 8 \biggl(\tilde{A}_3 + \tilde{A}_2 + \frac{\pi^2}{6} \tilde{A}_1\biggr)\,,
\label{Conj:Omega}\\
&\Omega_{AA} = 16 \biggl(\tilde{A}_5 + \tilde{A}_4 - \tilde{A}_2 + \tilde{B}_5 + \tilde{B}_3
+ \frac{\pi^2}{3} (\tilde{A}_3 + \tilde{A}_2) - \frac{\pi^4}{180} \tilde{A}_1\biggr)\,,
\nonumber
\end{align}
where $\tilde{A}_i = A_i(x)-A_i(1)$, $\tilde{B}_i = B_i(x)-B_i(1)$, $x = e^{-\varphi}$,
\begin{align*}
&A_1(x) = \frac{\xi}{2} H_{1}(y)\,,\\
&A_2(x) = \frac{1}{2} H_{1,1}(y) + \frac{\pi^2}{3}
- \xi \biggl[H_{0,1}(y) + \frac{1}{2} H_{1,1}(y)\biggr]\,,\\
&A_3(x) = - \xi \biggl[\frac{1}{4} H_{1,1,1}(y) + \frac{\pi^2}{6} H_{1}(y)\biggr]
+ \xi^2 \biggl[\frac{1}{2} H_{1,0,1}(y) + \frac{1}{4} H_{1,1,1}(y)\biggr]\,,\\
&A_4(x) = - \frac{1}{4} H_{1,1,1,1}(y) - \frac{\pi^2}{6} H_{1,1}(y)
+ \xi \biggl[2 H_{1,1,0,1}(y) + \frac{3}{2} H_{0,1,1,1}(y) + \frac{7}{4} H_{1,1,1,1}(y)\\
&\quad{} + \frac{\pi^2}{3} H_{0,1}(y) + \frac{\pi^2}{6} H_{1,1}(y) + 3 \zeta_3 H_{1}(y)\biggr]
- \xi^2 \biggl[2 H_{1,0,0,1}(y) + 2 H_{0,1,0,1}(y)\\
&\quad{} + 2 H_{1,1,0,1}(y) + H_{1,0,1,1}(y) + H_{0,1,1,1}(y) + \frac{3}{2} H_{1,1,1,1}(y)\biggr]\,,\\
&A_5(x) = \xi \biggl[\frac{5}{8} H_{1,1,1,1,1}(y) + \frac{\pi^2}{4} H_{1,1,1}(y) + \frac{\pi^4}{12} H_{1}(y)\biggr]
- \xi^2 \biggl[H_{1,1,1,0,1}(y)\\
&\quad{} + \frac{3}{4} H_{1,0,1,1,1}(y) + H_{0,1,1,1,1}(y) + \frac{11}{8} H_{1,1,1,1,1}(y)
+ \frac{\pi^2}{6} H_{1,0,1}(y) + \frac{\pi^2}{3} H_{0,1,1}(y)\\
&\quad{} + \frac{\pi^2}{4} H_{1,1,1}(y) + \frac{3}{2} \zeta_3 H_{1,1}(y)\biggr]
+ \xi^3 \biggl[H_{1,1,0,0,1}(y) + H_{1,0,1,0,1}(y) + H_{1,1,1,0,1}(y)\\
&\quad{} + \frac{1}{2} H_{1,1,0,1,1}(y)
+ \frac{1}{2} H_{1,0,1,1,1}(y) + \frac{3}{4} H_{1,1,1,1,1}(y)\biggr]\,,\displaybreak\\
&B_3(x) = - H_{1,0,1}(y) + \frac{1}{2} H_{0,1,1}(y) - \frac{1}{4} H_{1,1,1}(y)\\
&\quad{} + \xi \biggl[2 H_{0,0,1}(y) + H_{1,0,1}(y) + H_{0,1,1}(y) + \frac{1}{4} H_{1,1,1}(y)\biggr]\,,\\
&B_5(x) = \frac{x}{1-x^2} \biggl[- 4 H_{-1,0,-1,0,0}(x) + 4 H_{-1,0,1,0,0}(x) - 4 H_{1,0,-1,0,0}(x)\\
&\quad{} + 4 H_{1,0,1,0,0}(x) + 4 H_{-1,0,0,0,0}(x) + 4 H_{1,0,0,0,0}(x)\\
&\quad{} + 2 \zeta_3 H_{-1,0}(x) + 2 \zeta_3 H_{1,0}(x) - \frac{\pi^4}{60} H_{-1}(x) - \frac{\pi^4}{60} H_{1}(x)\biggr]\,,
\end{align*}
$\xi = (1+x^2)/(1-x^2)$, $y = 1-x^2$.
Here $H_{\cdots}(x)$ are harmonic polylogarithms\cite{Remiddi:1999ew}.
Symbolic manipulations and numerical evaluation of these functions
is available in \texttt{Mathematica}\cite{Maitre:2005uu,Maitre:2007kp}
and \texttt{Maple}\cite{Frellesvig:2018lmm}.
Numerical evaluation of multiple polylogarithms (including harmonic ones)
was implemented\cite{Vollinga:2004sn} in \texttt{C++}
and used in \texttt{GiNaC}\cite{Bauer:2000cp} (\url{https://ginac.de/}).
Numerical evaluation of harmonic polylogarithms was implemented in \texttt{Fortran}
up to weight 4\cite{Gehrmann:2001pz,Buehler:2011ev}
and then up to weight 8\cite{Ablinger:2018sat}.

At large $\varphi$ we have
\begin{align}
&A_1 = \varphi + \cdots\,,\quad
A_2 = \frac{\pi^2}{6} + \cdots\,,\quad
A_3 = - \frac{\pi^2}{6} \varphi - \zeta_3 + \cdots\,,\quad
A_4 = \frac{19}{180} \pi^4 + \cdots\,,
\nonumber\\
&A_5 = \frac{11}{180} \pi^4 \varphi + \frac{9}{2} \zeta_5 - \frac{\pi^2}{5} \zeta_3 + \cdots\,,\quad
B_3 = \frac{7}{2} \zeta_3 + \cdots\,,\quad
B_5 = 0 + \cdots\,,
\label{Conj:large1}
\end{align}
where dots mean exponentially suppressed terms.
Subtracting $A_i$, $B_i$ at $\varphi=0$ (see~(\ref{Conj:small1})) to obtain $\tilde{A}_i$, $\tilde{B}_i$,
we get
\begin{align}
&\Omega_1 = 4 (\varphi-1)+\cdots\,,\quad
\Omega_A = 8 (1-\zeta_3)+\cdots\,,
\nonumber\\
&\Omega_{AA} = 8 (9 \zeta_5 - \pi^2 \zeta_3 - 2 \zeta_3 + \pi^2 - 4) + \cdots\,.
\label{Conj:large}
\end{align}
By construction, only the 1-loop term $\Omega_1$ contains the linearly growing contribution $\varphi$,
all higher terms are $\mathcal{O}(\varphi^0)$.
Using the formulas~(\ref{Conj:12}--\ref{Conj:3}),
it is easy to reconstruct the $\mathcal{O}(\varphi^0)$ terms
in the large-$\varphi$ asymptotics~(\ref{Intro:large}) up to 3 loops.

At small $\varphi$ we have
\begin{align}
&A_1 = 1
 + \frac{\varphi^2}{3}
 - \frac{\varphi^4}{45}
 + \frac{2}{945} \varphi^6
 - \frac{\varphi^8}{4725} \varphi^8
 + \frac{2}{93555} \varphi^{10}
 + \mathcal{O}(\varphi^{12})\,,
\nonumber\\
&A_2 = \frac{\pi^2}{3} - 2
 + \frac{\varphi^2}{9}
 - \frac{14}{675} \varphi^4
 + \frac{304}{99225} \varphi^6
 - \frac{22}{55125} \varphi^8
 + \frac{26104}{540280125} \varphi^{10}
 + \mathcal{O}(\varphi^{12})\,,
\nonumber\\
&A_3 = - \frac{\pi^2}{3} + 1
 - \biggl(\pi^2 - \frac{7}{2}\biggr) \frac{\varphi^2}{9}
 + \biggl(\frac{\pi^2}{3} - \frac{2}{5}\biggr) \frac{\varphi^4}{45}
 - \frac{2}{2835} \biggl(\pi^2 + \frac{19}{35}\biggr) \varphi^6
\nonumber\\
&\quad{}
 + \biggl(\pi^2 + \frac{67}{35}\biggr) \frac{\varphi^8}{14175}
 - \frac{2}{93555} \biggl(\frac{\pi^2}{3} + \frac{1949}{1925}\biggr) \varphi^{10}
 + \mathcal{O}(\varphi^{12})\,,
\nonumber\displaybreak\\
&A_4 = 2 \biggl(3 \zeta_3 + \frac{\pi^2}{3} - 3\biggr)
 + \biggl(2 \zeta_3 - \frac{\pi^2}{27} - \frac{91}{54}\biggr) \varphi^2
 - \biggl(2 \zeta_3 - \frac{14}{135} \pi^2 - \frac{1789}{1350}\biggr) \frac{\varphi^4}{15}
\nonumber\\
&\quad{}
 + \biggl(4 \zeta_3 - \frac{304}{945} \pi^2 - \frac{250121}{66150}\biggr) \frac{\varphi^6}{315}
 - \biggl(2 \zeta_3 - \frac{22}{105} \pi^2 - \frac{296647}{99225}\biggr) \frac{\varphi^8}{1575}
\nonumber\\
&\quad{}
 + \biggl(4 \zeta_3 - \frac{26104}{51975} \pi^2 - \frac{352666739}{40020750}\biggr) \frac{\varphi^{10}}{31185}
 + \mathcal{O}(\varphi^{12})\,,
\nonumber\\
&A_5 = - 3 \zeta_3 + \frac{\pi^4}{6} - \frac{2}{3} \pi^2 + 2
 - \biggl(2 \zeta_3 - \frac{\pi^4}{18} + \frac{5}{27} \pi^2 - \frac{65}{54}\biggr) \varphi^2
\nonumber\\
&\quad{}
 - \biggl(\zeta_3 + \frac{\pi^4}{54} - \frac{41}{405} \pi^2 - \frac{1649}{2025}\biggr) \frac{\varphi^4}{5}
 + \biggl(2 \zeta_3 + \frac{\pi^4}{45} - \frac{349}{1575} \pi^2 - \frac{6401}{18375}\biggr) \frac{\varphi^6}{63}
\nonumber\\
&\quad{}
 - \biggl(\zeta_3 + \frac{\pi^4}{126} - \frac{32}{245} \pi^2 + \frac{38959}{138915}\biggr) \frac{\varphi^8}{225}
\nonumber\\
&\quad{}
 + \biggl(2 \zeta_3 + \frac{\pi^4}{81} - \frac{5683}{18711} \pi^2 + \frac{232902262}{180093375}\biggr) \frac{\varphi^{10}}{3465}
 + \mathcal{O}(\varphi^{12})\,,
\nonumber\\
&B_3 = 4
 + \frac{5}{54} \varphi^2
 - \frac{889}{40500} \varphi^4
 + \frac{80299}{20837250} \varphi^6
 - \frac{357533}{625117500} \varphi^8
 + \frac{10632271}{138671898750} \varphi^{10}
\nonumber\\
&\quad{}
 + \mathcal{O}(\varphi^{12})\,,
\nonumber\\
&B_5 = \frac{3}{2} \zeta_3
 - \biggl(\zeta_3 + \frac{1}{6}\biggr) \frac{\varphi^2}{3}
 + \biggl(11 \zeta_3 + \frac{31}{12}\biggr) \frac{\varphi^4}{225}
 - \biggl(202 \zeta_3 + \frac{143}{3}\biggr) \frac{\varphi^6}{33075}
\nonumber\\
&\quad{}
 + \biggl(13 \zeta_3 + \frac{7739}{2916}\biggr) \frac{\varphi^8}{18375}
 - \biggl(\frac{2026}{7} \zeta_3 + \frac{1261}{27}\biggr) \frac{\varphi^{10}}{3675375}
 + \mathcal{O}(\varphi^{12})\,,
\label{Conj:small1}
\end{align}
and hence
\begin{align}
&\Omega_1 = 4 \biggl(
 \frac{\varphi^2}{3}
 - \frac{\varphi^4}{45}
 + \frac{2}{945} \varphi^6
 - \frac{\varphi^8}{4725}
 + \frac{2}{93555} \varphi^{10}
\biggr)
 + \mathcal{O}(\varphi^{12})\,,
\nonumber\\
&\Omega_A = 4 \biggl[
 - \biggl(\frac{\pi^2}{9} - 1\biggr) \varphi^2
 + (\pi^2 - 8) \frac{\varphi^4}{135}
 - \frac{2}{2835} \biggl(\pi^2 - \frac{38}{5}\biggr) \varphi^6
 + \biggl(\pi^2 - \frac{262}{35}\biggr) \frac{\varphi^8}{14175}
\nonumber\\
&\quad{}
 - \frac{2}{280665} \biggl(\pi^2 - \frac{262}{35}\biggr) \varphi^{10}
\biggr]
 + \mathcal{O}(\varphi^{12})\,,
\nonumber\\
&\Omega_{AA} = 4 \biggl[
 - \biggl(4 \zeta_3 - \frac{\pi^4}{5} + \frac{2}{3} \pi^2 + \frac{20}{3}\biggr) \frac{\varphi^2}{3}
 - \biggl(\frac{256}{5} \zeta_3 + \frac{\pi^4}{5} - \frac{28}{9} \pi^2 - \frac{706}{15}\biggr) \frac{\varphi^4}{45}
\nonumber\\
&\quad{}
 + \frac{2}{945} \biggl(\frac{2536}{35} \zeta_3 + \frac{\pi^4}{5} - \frac{62}{9} \pi^2 - \frac{10826}{315}\biggr) \varphi^6
\nonumber\\
&\quad{}
 - \biggl(\frac{368}{49} \zeta_3 + \frac{\pi^4}{63} - \frac{76}{81} \pi^2 - \frac{33398}{35721}\biggr) \frac{\varphi^8}{375}
\nonumber\\
&\quad{}
 + \frac{2}{467775} \biggl(\frac{32248}{55} \zeta_3 + \pi^4 - \frac{3994}{45} \pi^2 + \frac{36182402}{363825}\biggr) \varphi^{10}
\biggr]
 + \mathcal{O}(\varphi^{12})\,.
\label{Conj:small}
\end{align}
It is easy to reconstruct small $\varphi$ expansions of all color structures of $\Gamma$
up to 3 loops up to $\varphi^{10}$ (more terms can be added if desired).

If has been conjectured\cite{Grozin:2014hna,Grozin:2015kna}
that this structure holds at higher orders.
At 4 loops we have
\begin{align*}
&K_4 = C_A^3 K_{AAA} + \frac{d_{RA}}{C_R} K_{dRA}\\
&{} + C_A^2 T_F n_f K_{AAf} + C_F C_A T_F n_f K_{FAf} + C_F^2 T_F n_f K_{FFf} + \frac{d_{RF}}{C_R} n_f K_{dRF}\\
&{} + C_A (T_F n_f)^2 K_{Aff} + C_F (T_F n_f)^2 K_{Fff} + (T_F n_f)^3 K_{fff}\,.
\end{align*}
The quartic-Casimir terms here do not look nice because the ``universal'' coupling $A$ depends on $R$.
We should, probably, suppose
\begin{equation*}
\Omega_4(\varphi) = C_A^3 \Omega_{AAA}(\varphi) + \frac{d_{RA}}{C_R} \Omega_{dRA}(\varphi)
\end{equation*}
(it is also $R$-dependent).
Then for
\begin{align*}
&\Gamma_4 = C_A^3 \Gamma_{AAA} + \frac{d_{dRA}}{C_R} \Gamma_{RA}\\
&{} + C_A^2 T_F n_f \Gamma_{AAf} + C_F C_A T_F n_f \Gamma_{FAf} + C_F^2 T_F n_f \Gamma_{FFf} + \frac{d_{RF}}{C_R} n_f \Gamma_{dRF}\\
&{} + C_A (T_F n_f)^2 \Gamma_{Aff} + C_F (T_F n_f)^2 \Gamma_{Fff} + (T_F n_f)^3 \Gamma_{fff}
\end{align*}
the conjecture results in
\begin{align}
&\Gamma_{AAA} = \Omega_{AAA} + 3 K_A \Omega_{AA} + (2 K_{AA} + K_A^2) \Omega_A + K_{AAA} \Omega_1\,,
\nonumber\\
&\Gamma_{dRA} = \Omega_{dRA} + K_{dRA} \Omega_1\,,
\label{Conj:Glue4}\\
&\Gamma_{FFf} = K_{FFf} \Omega_1\,,\quad
\Gamma_{Fff} = K_{Fff} \Omega_1\,,\quad
\Gamma_{fff} = K_{fff} \Omega_1\,,
\label{Conj:1L4}\\
&\Gamma_{AAf} = 3 K_f \Omega_{AA} + 2 (K_{Af} + K_A K_f) \Omega_A + K_{AAF} \Omega_1\,,
\nonumber\\
&\Gamma_{dRF} = K_{RF} \Omega_1\,,
\label{Conj:Wrong}\\
&\Gamma_{FAf} = 2 K_{Ff} \Omega_A + K_{FAf} \Omega_1\,,
\nonumber\\
&\Gamma_{Aff} = (2 K_{ff} + K_f^2) \Omega_A + K_{Aff} \Omega_1\,,
\label{Conj:Right}
\end{align}
The abelian terms $\Gamma_{FFf}$, $\Gamma_{Fff}$, $\Gamma_{fff}$ are given by diagrams
containing a single 2-leg c-web,
and hence have the 1-loop $\varphi$ dependence $\Omega_1(\varphi)$
(see Sect.~\ref{S:abel}).
So, the relations~(\ref{Conj:1L4}) hold by construction.

The first 2 terms of the small $\varphi$ expansion of $\Gamma_{dRF}$ have been obtained in\cite{Grozin:2017css}.
It has been proved that the relation~(\ref{Conj:Wrong}) for $\Gamma_{dRF}$ does not hold.
Later the 3-rd term of this expansion\cite{Bruser:2019auj},
the large $\varphi$ limit\cite{Lee:2019zop,Henn:2019rmi},
and, finally, the full $\varphi$ dependence\cite{Bruser:2020bsh}
have been obtained.
The $\varphi$ dependence is \emph{extremely} complicated,
and certainly does not satisfy the relation in~(\ref{Conj:Wrong}).
The relation~(\ref{Conj:Wrong}) for $\Gamma_{AAf}$ is also wrong,
as demonstrated in\cite{Bruser:2019auj} by the calculation of 2 terms in the small $\varphi$ expansion.
These 2 structures get contributions from diagrams containing a light-quark box.
Maybe, such diagrams are the reason of breaking the conjecture.

However, the 2 remaining structures, $\Gamma_{FAf}$ and $\Gamma_{Aff}$,
pass all existing tests, and seem to agree\cite{Bruser:2019auj} with the relations~(\ref{Conj:Right}).
For $\Gamma_{Aff}$, 3 terms of the small $\varphi$ expansion\cite{Bruser:2019auj}
and the large $\varphi$ limit\cite{Ruijl:2016pkm,Henn:2016men,Davies:2016jie} are known,
so that there are 3 analytical checks of the corresponding relation in~(\ref{Conj:Right}).
For $\Gamma_{FAf}$, 2 terms of the small $\varphi$ expansion are known\cite{Bruser:2019auj};
the large $\varphi$ limit $K_{FAf}$ was only known numerically at the moment\cite{Moch:2017uml}.
So, there was only 1 analytical and 1 numerical check.
The analytical form of $K_{FAf}$ has been predicted\cite{Bruser:2019auj} on the basis of the conjecture,
and later confirmed\cite{vonManteuffel:2020vjv} by a direct calculation.
It seems that there can be little doubt that the relations~(\ref{Conj:Right})
for the full $\varphi$ dependence of $\Gamma_{FAf}$ and $\Gamma_{Aff}$ are valid.
If we believe in this statement, we can get many terms of small $\varphi$ expansions
of these structures using~(\ref{Conj:small}),
and their large $\varphi$ asymptotics including the $\mathcal{O}(\varphi^0)$ terms
using~(\ref{Conj:large}).

\section{Euclidean angle near $\pi$}
\label{S:pi}

In the paper\cite{Kilian:1993nk} the authors have noticed
that the 2-loop cusp anomalous dimension $\Gamma$ at Euclidean angle $\phi\to\pi$
is related to the 1-loop static quark-antiquark potential $V(r)$:
\begin{equation}
\Gamma(\pi-\delta) = \frac{r\,V(r)}{\delta}\,.
\label{pi:Mannel}
\end{equation}
The proof of this relations to all orders given in this paper is incorrect:
we shall see that it breaks down for the 3-loop $\Gamma$.

In fact, this relation follows from conformal symmetry\cite{Grozin:2014hna,Grozin:2015kna}
(which is broken in QCD by the conformal anomaly).
Let's consider Euclidean space with the metric
\begin{equation*}
ds^2 = dx_0^2 + d\vec{x}^{\,2}\,.
\end{equation*}
In spherical coordinates
\begin{equation*}
x_0 = r \cos\delta\,,\quad
\vec{x} = r \vec{n} \sin\delta\,,\quad
ds^2 = dr^2 + r^2 (d\delta^2 + \sin^2\delta\,d\vec{n}^{\,2})\,.
\end{equation*}
We assume $\delta\ll1$ and introduce the new coordinates $y$ by
\begin{equation*}
r = e^{y_0}\,,\quad
\vec{y} = \delta \vec{n}\,,\quad
ds^2 = e^{2 y_0} \left(dy_0^2 + d\vec{y}^{\,2}\right)\,.
\end{equation*}
This metric is conformally flat.

Let's consider a Wilson line in $x$ space having the shape of a small angle $\delta$
(the $x_0$ axis is directed upwards).
We introduce an UV cutoff $x_0^{\text{UV}}$ close to the angle
and an IR cutoff $x_0^{\text{UV}}$ far from it.
Then
\begin{equation}
\log W = \Gamma \log\frac{x_0^{\text{IR}}}{x_0^{\text{UV}}}\,.
\label{pi:x}
\end{equation}
In $y$ space it looks like a pair of antiparallel lines
at a distance $\vec{y}$ from each other:
\begin{equation*}
\begin{picture}(28,21)
\put(5.5,10.5){\makebox(0,0){\includegraphics{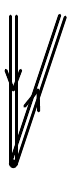}}}
\put(2.5,3){\makebox(0,0)[r]{$x_0^{\text{UV}}$}}
\put(2.5,18){\makebox(0,0)[r]{$x_0^{\text{IR}}$}}
\put(14,10.5){\makebox(0,0){$\Rightarrow$}}
\put(22.5,10.5){\makebox(0,0){\includegraphics{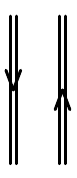}}}
\put(25.5,3){\makebox(0,0)[l]{$y_0^{\text{UV}}$}}
\put(25.5,18){\makebox(0,0)[l]{$y_0^{\text{IR}}$}}
\end{picture}
\end{equation*}
and
\begin{equation}
\log W = V(\vec{y}^{\,}) \left( y_0^{\text{IR}} - y_0^{\text{UV}} \right)\,.
\label{pi:y}
\end{equation}
If our theory is conformally invariant, these Wilson lines coincide, and
\begin{equation}
\Gamma(\pi-\delta) = \frac{y V(y)}{\delta}\,,
\label{pi:coord}
\end{equation}
where $y = |\vec{y}^{\,}|$.
Due to conformal symmetry, $V(y) = \text{const}/y$.
We can also re-write this relation in momentum space:
\begin{equation}
\Gamma(\pi-\delta) = \frac{\vec{q}^{\,2} V(\vec{q}^{\,})}{4\pi \delta}
\label{pi:mom}
\end{equation}
(the momentum-space potential is $V(\vec{q}^{\,}) = \text{const}/\vec{q}^{\,2}$).
In $\mathcal{N}=4$ SYM (which is conformally symmetric)
the 3-loop $\Gamma(\pi-\delta)$\cite{Grozin:2014hna,Grozin:2015kna}
agrees with the 2-loop $V$\cite{Prausa:2013qva}.

Let's introduce the conformal anomaly $\Delta(\alpha_s)$ by
\begin{equation}
4 \pi \Delta(\alpha_s(|\vec{q}^{\,}|)) = \bigl[\delta\,\Gamma(\alpha_s(|\vec{q}^{\,}|),\pi-\delta)\bigr]_{\delta \to 0}
- \frac{\vec{q}^{\,2} V(\alpha_s(|\vec{q}^{\,}|),\vec{q}^{\,})}{4\pi}\,.
\label{pi:Delta}
\end{equation}
In QCD at 3 loops $\Delta$ can be obtained from the general result~(\ref{Conj:Omega})
using the asymptotics $\phi=\pi-\delta$, $\delta\to0$ of $A_i$, $B_i$ (see~(A.13) in\cite{Bruser:2019auj}).
The result is\cite{Grozin:2015kna}
\begin{equation}
\Delta(\alpha_s) = \frac{4}{27} \beta_0 C_R (47 C_A - 28 T_F n_f)
\biggl(\frac{\alpha_s}{4\pi}\biggr)^{\!3} + \mathcal{O}(\alpha_s^4)\,.
\label{pi:3}
\end{equation}
It vanishes when $\beta_0=0$.
In QCD (as well as in QED and many other gauge theories)
conformal symmetry is anomalous, broken by the $\beta$ function.
Therefore, it seems reasonable to assume\cite{Grozin:2015kna} that,
similarly to the Crewther relation\cite{Broadhurst:1993ru,Crewther:1997ux,Braun:2003rp},
the conformal anomaly has the form
\begin{equation}
\Delta(\alpha_s) = \beta(\alpha_s) C(\alpha_s)\,.
\label{pi:beta}
\end{equation}
In addition to the $\alpha_s^2$ term~(\ref{pi:3}) of $C$,
several color structures of the $\alpha_s^3$ term are known\cite{Bruser:2019auj}
(the 3-loop $V(\vec{q}^{\,})$\cite{Smirnov:2009fh,Anzai:2009tm,Lee:2016cgz} is used):
\begin{align}
&C(\alpha_s) = \frac{4}{27} C_R (47 C_A - 28 T_F n_f)
\biggl(\frac{\alpha_s}{4\pi}\biggr)^{\!2}
\nonumber\\
&{} + 4 C_R \biggl[x_{AA} C_A^2
- \biggl(5 \zeta_3 + \frac{\pi^4}{6} - \frac{79}{648}\biggr) C_A T_F n_f
+ \frac{2}{3} \biggl(19 \zeta_3 + \frac{\pi^4}{10} - \frac{1711}{48}\biggr) C_F T_F n_f
\nonumber\\
&\quad{}
+ \frac{8}{9} \biggl(\zeta_3 + \frac{58}{27}\biggr) (T_F n_f)^2
\biggr]
\biggl(\frac{\alpha_s}{4\pi}\biggr)^{\!3}
+ \mathcal{O}(\alpha_s^4)\,,
\label{pi:C}
\end{align}
where $x_{AA}$ is unknown (and, in fact, ill-defined, see Sect.~\ref{S:logd}).

The coefficient of $C_F$ in the $\alpha_s^2$ term of $C$,
as well as that of $C_F^2$ in the $\alpha_s^3$ term of $C$, vanish;
this follows from a more general result (Sect.~\ref{S:CF}).
The coefficient of $(T_F n_f)^2$ in the $\alpha_s^3$ term of $C$
follows from $\Gamma_{fff}$ which is known (Sect.~\ref{S:Lb0});
the factorized form~(\ref{pi:beta}) requires
a definite result for $\Gamma_{Aff}$ in the limit $\delta \to 0$
which agrees with the conjecture~(\ref{Conj:Right}).
This is one more confirmation of this conjecture for $\Gamma_{Aff}$
(if we believe in~(\ref{pi:beta})).
The coefficient of $C_F T_F n_f$ in the $\alpha_s^3$ term of $C$
follows from $\Gamma_{Fff}$ which is known (Sect.~\ref{S:NLb0});
the factorized form~(\ref{pi:beta}) requires
a definite result for $\Gamma_{FAf}$ in the limit $\delta \to 0$
which agrees with the conjecture~(\ref{Conj:Right}).
This is one more confirmation of this conjecture for $\Gamma_{FAf}$
(if we believe in~(\ref{pi:beta})).
The coefficient of $C_A T_F n_f$ in the $\alpha_s^3$ term of $C$
follows from the conjectured~(\ref{Conj:Right}) result for $\Gamma_{Aff}$
in the limit $\delta \to 0$.
If we believe in~(\ref{pi:beta}), there is no $d_{FF} n_f$ term in $\Delta(\alpha_s)$ at 4 loops;
the full $\varphi$ dependence of this structure is known\cite{Bruser:2020bsh},
but it is \emph{extremely} complicated,
and this conjecture has not been explicitly checked yet.

In the recent paper\cite{Kataev:2022iqf} it was proposed to represent $C(\alpha_s)$ in the form
\begin{equation}
C(\alpha_s) = \sum_{n=0}^\infty C_n(\alpha_s) \bigl[\beta(\alpha_s)\bigr]^n\,,
\label{pi:Kataev}
\end{equation}
where $C_n(\alpha_s)$ are series in $\alpha_s$ whose coefficients don't contain $T_F n_f$.
In fact, an arbitrary series
\begin{equation}
C(\alpha_s) = \sum_{n=1}^\infty P_n(T_F n_f) \biggl(\frac{\alpha_s}{4\pi}\biggr)^{\!n+1}
\label{pi:Arb}
\end{equation}
(where $P_n(x)$ is a polynomial of degree $n$) can be reduced to the form~(\ref{pi:Kataev})
by a simple algorithm.
In $P_1$ we express $T_F n_f$ via
\begin{equation*}
\beta(\alpha_s) - \sum_{n=1}^\infty \beta_n \biggl(\frac{\alpha_s}{4\pi}\biggr)^{\!n+1}
\end{equation*}
(note that $\beta_{n\ge1}$ is a polynomial in $T_F n_f$ of degree $n$)
and update $P_{\ge2}(T_F n_f)$ by incorporating this sum.
Then we repeat the same step for $P_2$, and so on.
At the $N$-th step, 3 kinds of terms appear:
\begin{itemize}
\item $\bigl[\beta(\alpha_s)\bigr]^m$ with coefficients not containing $\beta_{\ge1}$ ($m\in[0,N]$);
these terms become a part of the final result.
\item $\bigl[\beta(\alpha_s)\bigr]^m$ times some series having the form~(\ref{pi:Arb})
($m\in[1,N-1]$); for these series, we call the algorithm recursively.
\item terms without $\beta(\alpha_s)$ containing $\beta_{\ge1}$;
they are absorbed into $P_{\ge N+1}(T_F n_f)$.
\end{itemize}
As a result, we get the desired form~(\ref{pi:Kataev}), and this representation is unique.
This is not a physical statement, but a simple algebraic fact.
Let's stress that here we discuss the dependence on $T_F n_f$ only;
$n_f$ can appear in other contexts, such as $d_F^{abcd} n_f$ (light-by-light).

\subsection{$C_R C_A^3 \alpha_s^4 \log(\delta)/\delta$ term}
\label{S:logd}

At 4 loops a $C_R C_A^3 \alpha_s^4 \log(\delta)/\delta$ term appears\cite{Bruser:2018aud} in $\Gamma(\pi-\delta)$.
It is similar to the 3-loop $\log(\mu r)$ term in the potential\cite{Brambilla:1999qa,Brambilla:1999xf}.

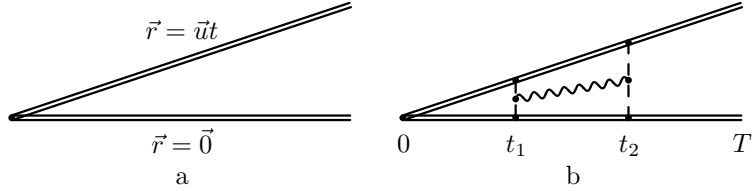
\begin{figure}[ht]
\begin{center}
\begin{picture}(99,24)
\put(23.5,16.5){\makebox(0,0){\includegraphics{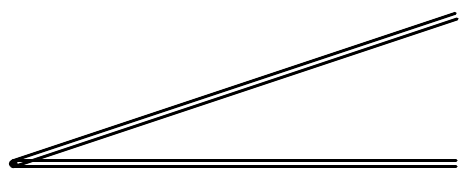}}}
\put(23.5,7.5){\makebox(0,0)[t]{$\vec{r} = \vec{0}\strut$}}
\put(23.5,18){\makebox(0,0)[b]{$\vec{r} = \vec{u} t\strut$}}
\put(23.5,0){\makebox(0,0)[b]{a}}
\put(75.5,16.5){\makebox(0,0){\includegraphics{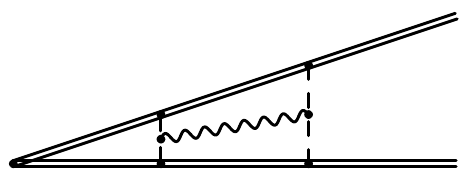}}}
\put(53,7.5){\makebox(0,0)[t]{$0\strut$}}
\put(68,7.5){\makebox(0,0)[t]{$t_1\strut$}}
\put(83,7.5){\makebox(0,0)[t]{$t_2\strut$}}
\put(98,7.5){\makebox(0,0)[t]{$T\strut$}}
\put(75.5,0){\makebox(0,0)[b]{b}}
\end{picture}
\end{center}
\caption{The Wilson line describing production of a heavy quark--antiquark pair
with a small relative velocity $\vec{u}$ (a);
The first transverse-gluon contribution (b).}
\label{F:us}
\end{figure}

Let's consider a cusped Wilson line in Minkowski space (Fig.~\ref{F:us}a).
It is formed by the static quark and antiquark world lines
$\vec{r}=\vec{0}$ and $\vec{r} = \vec{u} t$,
where $\vec{u}$ is a small relative velocity ($u=|\vec{u}^{\,}|\ll1$).
At the end of the calculation we'll analytically continue the result
to Euclidean space ($u = i\delta$).
We neglect all terms suppressed by powers of $u$.
It is convenient to use Coulomb gauge.
The static quark and antiquark interact by exchanging instantaneous Coulomb gluons:
\begin{equation}
V(\vec{q}^{\,}) = - C_F \frac{g_0^2}{\vec{q}^{\,2}}\,,\quad
V(\vec{r}^{\,}) = - C_F \kappa_0 \frac{g_0^2}{4\pi} \frac{1}{r^{1-2\varepsilon}}
\label{us:V0}
\end{equation}
(the power of $r$ is obvious from dimensions counting).
Here and below $\kappa_i = 1 + \mathcal{O}(\varepsilon)$
are some normalization factors (we don't need their exact form).

Transverse gluons interact only with Coulomb ones,
but not with the static quarks.
The first transverse-gluon contribution is shown in Fig.~\ref{F:us}b.
Here $T$ is an infrared cutoff.
We use the method of regions to analyze this contribution.
In the ultrasoft region $t_1 \sim t_2 \sim t_2-t_1$;
Coulomb gluons characteristic momentum is $q \sim 1/(ut_{1,2})$,
and the transverse gluon characteristic momentum is $k \sim 1/t_{1,2} \ll q$.
In the soft region $t_2-t_1 \sim ut_{1,2}$, and $k \sim 1/(t_2-t_1) \sim q$.
To determine the coefficient of the logarithm in the $1/\delta$ term in $\Gamma(\pi-\delta)$,
it turns out to be sufficient to consider the ultrasoft region.
Neglecting $k$ in the 3-gluon vertex,
we obtain in momentum and coordinate spaces
\begin{equation}
\raisebox{-10mm}{\begin{picture}(13,22)
\put(6,11){\makebox(0,0){\includegraphics{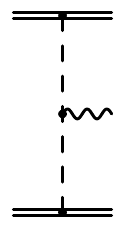}}}
\put(5,14){\makebox(0,0)[r]{$q$}}
\put(5,8){\makebox(0,0)[r]{$q$}}
\put(5,19){\makebox(0,0)[r]{$a_1$}}
\put(5,3){\makebox(0,0)[r]{$a_2$}}
\put(8,10){\makebox(0,0)[t]{0}}
\put(11,12){\makebox(0,0)[b]{$i$}}
\put(11,10){\makebox(0,0)[t]{$a$}}
\end{picture}}
= f^{a a_1 a_2} g_0^3 \frac{2 q^i}{(\vec{q}^{\,2})^2}\,,\quad
\raisebox{-10mm}{\begin{picture}(13,22)
\put(6,11){\makebox(0,0){\includegraphics{us.pdf}}}
\put(5,19){\makebox(0,0)[r]{$\vec{r}$}}
\put(5,3){\makebox(0,0)[r]{0}}
\put(11,12){\makebox(0,0)[b]{$i$}}
\end{picture}}
= i f^{a a_1 a_2} \kappa_0 \frac{g_0^3}{4\pi} \frac{r^i}{r^{1-2\varepsilon}}\,.
\label{us:vertex}
\end{equation}

The ratio of the Wilson line (Fig.~\ref{F:us}b) to the one without the transverse-gluon correction
is $1 + R_{\text{us}} + R_{\text{soft}}$.
The ultrasoft contribution is
\begin{equation}
R_{\text{us}} = \int_0^T dt_2 \int_0^{t_2} dt_1\,K(t_1,t_2)\,,
\label{us:WK}
\end{equation}
where
\begin{equation}
K(t_1,t_2) = \frac{1}{4} C_F C_A^2 \kappa_0^2 \frac{g_0^6}{(4\pi)^2}
\frac{r_1^i}{r_1^{1-2\varepsilon}} \frac{r_2^j}{r_2^{1-2\varepsilon}} D^{ij}(v(t_2-t_1))
\exp\left[- i \int_{t_1}^{t_2} dt\,\Delta V(ut)\right]
\label{us:K}
\end{equation}
($v=(1,\vec{0}^{\,})$ is the 4-velocity of our small dipole).
During the time interval between $t_1$ and $t_2$,
the static quark--antiquark pair is in the adjoint color state instead of the singlet one,
and their leading-order interaction potential $V_o(r)$ is obtained from the expression
for the singlet potential $V(r)$~(\ref{us:V0}) by replacing the color factor $C_F$ with $C_F - C_A/2$.
Therefore, we get the integral of $\Delta V(r) = V_o(r) - V(r)$.
The characteristic sizes of the regions of the transverse gluon emission and absorption
are $\sim ut_{1,2}$;
we neglect them, so that this gluon propagates between the points $v t_1$ and $v t_2$:
\begin{equation}
D^{ij}(vt) = 8 (i/2)^{2\varepsilon} \frac{\Gamma(2-\varepsilon)}{3-2\varepsilon}
\frac{t^{-2+2\varepsilon}}{(4\pi)^{2-\varepsilon}} \delta^{ij}\,.
\label{us:Dt}
\end{equation}
We obtain
\begin{equation}
K(t_1,t_2) = \frac{2}{3} C_F C_A^2 \kappa_1 \frac{g_0^6}{(4\pi)^4} u^{4\varepsilon}
t_1^{2\varepsilon} t_2^{2\varepsilon} (t_2-t_1)^{-2+2\varepsilon}
\exp\left[- \frac{i}{4} C_A \kappa_0 \frac{g_0^2}{4\pi}
\frac{t_2^{2\varepsilon} - t_1^{2\varepsilon}}{\varepsilon u^{1-2\varepsilon}}\right]\,.
\label{us:K2}
\end{equation}

Now we consider just a single Coulomb gluon exchange between $t_1$ and $t_2$:
\begin{equation}
K^{(1)}(t_1,t_2) = - \frac{i}{6} C_F C_A^3 \kappa_2 \frac{g_0^8}{(4\pi)^5}
\frac{t_1^{2\varepsilon} t_2^{2\varepsilon} (t_2^{2\varepsilon}-t_1^{2\varepsilon}) (t_2-t_1)^{-2+2\varepsilon}}{\varepsilon u^{1-6\varepsilon}}\,.
\label{us:K1}
\end{equation}
Calculating the integral~(\ref{us:WK}) by the substitutions $t_1 = x t_2$ we obtain
\begin{equation}
\int_0^1 dx\,x^{2\varepsilon} (1-x^{2\varepsilon}) (1-x)^{-2+2\varepsilon}
= \frac{\Gamma(1+2\varepsilon)}{1-2\varepsilon}
\left[3 \frac{\Gamma(1+4\varepsilon)}{\Gamma(1+6\varepsilon)}
- 2 \frac{\Gamma(1+2\varepsilon)}{\Gamma(1+4\varepsilon)}\right]
= 1 + \mathcal{O}(\varepsilon)\,,
\label{us:Intx}
\end{equation}
and
\begin{equation}
R_{\text{us}}^{(1)} = - \frac{i}{48} C_F C_A^3 \kappa_3 \frac{g_0^8}{(4\pi)^5} \frac{T^{8\varepsilon}}{\varepsilon^2 u^{1-6\varepsilon}}\,.
\label{us:Wres}
\end{equation}

The soft contribution is nearly local in time ($t_2-t_1 \sim u t_{1,2} \ll t_{1,2}$),
and can be described by a soft potential.
For a single coulomb exchange between $t_1$ and $t_2$, it is
\begin{equation}
V_{\text{soft}}^{(1)}(r) = c C_F C_A^3 \frac{g_0^8}{r^{1-8\varepsilon}}
\label{us:Vs1}
\end{equation}
by counting dimensions, so that
\begin{equation}
R_{\text{soft}}^{(1)} = - i \int_0^T dt\,V_{\text{soft}}^{(1)}(ut)
= - i c C_F C_A^3 \frac{g_0^8 T^{8\varepsilon}}{8 \varepsilon u^{1-8\varepsilon}}\,.
\label{us:Rs1}
\end{equation}
The double pole $1/\varepsilon^2$ should cancel
in $R^{(1)} = R_{\text{us}}^{(1)} + R_{\text{soft}}^{(1)}$;
this fixes the $1/\varepsilon$ term in $c$, and we obtain
\begin{equation}
R^{(1)} = - \frac{i}{48} C_F C_A^3 \frac{g_0^8 T^{8\varepsilon}}{(4\pi)^5}
\frac{\kappa_3 u^{6\varepsilon} - \kappa_4 u^{8\varepsilon}}{\varepsilon^2 u}
= \frac{i}{24} C_F C_A^3 \frac{\alpha_s^4(\mu) (\mu T)^{8\varepsilon}}{4\pi}
\frac{\log u + \text{const}}{\varepsilon u}\,.
\label{us:R1}
\end{equation}
This leads to the following contribution to $\Gamma$:
\begin{equation}
\Delta \Gamma = - \frac{i}{3} C_F C_A^3 \frac{\alpha_s^4}{4\pi}
\frac{\log u + \text{const}}{u}\,.
\label{us:M}
\end{equation}
Finally, analytically continuing it to Euclidean space
($\varphi_E = \pi + i\varphi_M$, $\varphi_M = u$)
and replacing $C_F \to C_R$ we obtain\cite{Bruser:2018aud}
\begin{equation}
\Delta \Gamma(\pi-\delta) = - \frac{1}{3} C_F C_A^3 \frac{\alpha_s^4}{4\pi}
\frac{\log\delta + \text{const}}{\delta}\,.
\label{us:E}
\end{equation}

This contribution does not allow us to take the limit $\delta \to 0$ in~(\ref{pi:Delta}).
Higher orders in $\alpha_s$ will contain higher powers of $\log\delta$.
Hopefully, summing terms with the leading powers of $\log\delta$ to all orders
will produce a finite result for $\delta\,\Gamma(\pi-\delta)$,
in which $\log\alpha_s$ will appear in place of $\log\delta$.

\section{Higher-loop abelian results: 2-leg c-webs}
\label{S:abel}

In general, $\log W$ is given by the sum of webs~(\ref{Exp:QCD}).
For some families of abelian color structures only 2-leg c-webs~(\ref{Exp:w2}) contribute.
We can work in QED.
The full photon propagator is
\begin{align}
&\raisebox{-2.2mm}{\begin{picture}(17,8.5)
\put(8.5,4.25){\makebox(0,0){\includegraphics{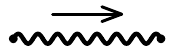}}}
\put(8.5,6){\makebox(0,0)[b]{$k$}}
\end{picture}}
= i D^{\mu\nu}(k)\,,
\nonumber\\
&D^{\mu\nu}(k) = \frac{D_0^{\mu\nu}(k)}{1 - \Pi(k^2)}\,,\quad
D_0^{\mu\nu}(k) = \frac{1}{-k^2} \biggl(g^{\mu\nu} + \frac{k^\mu k^\nu}{-k^2}\biggr)\,,
\label{abel:D}
\end{align}
(only the 0-order propagator is gauge dependent;
here and below we use Landau gauge).
The photon self energy is the sum of 1PI diagrams:
\begin{align}
&\raisebox{-3mm}{\includegraphics{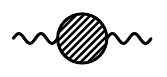}}
= i (k^2 g^{\mu\nu} - k^\mu k^\nu) \Pi(k^2)\,,\quad
\Pi(k^2) = \sum_{L=1}^\infty \Pi_{L-1}(k^2)\,,
\nonumber\\
&\Pi_{L-1}(k^2) = \Pi_{L-1} \bigl[A_0 (-k^2)^{-2\varepsilon}\bigr]^L\,,\quad
A_0 = e^{-\gamma\varepsilon} \frac{e_0^2}{(4\pi)^{d/2}}\,.
\label{abel:Pi}
\end{align}
The $\overline{\text{MS}}$ charge renormalization is
\begin{equation}
A_0 = \mu^{2\varepsilon} \frac{\alpha(\mu)}{4\pi} Z_\alpha(\alpha(\mu))\,.
\label{abel:MS}
\end{equation}

We can select a subset $S$ of abelian color structures of the gluon propagator
which satisfies 2 conditions:
\begin{enumerate}
\item the contributions of the color structures $C_R \times S$ to bare $\log W$
are given by 2-leg c-webs only (see~(\ref{Exp:QED1}));
\item no color factor $C \not\in S$ being multiplied by a color factor in $Z_\alpha$
can produce a color factor $C' \in S$.
\end{enumerate}

The contribution of the color structures $S$ to the gluon propagator is
\begin{align}
&D_S^{\mu\nu}(k) = \sum_{L=0}^\infty d_S^{(L)} D_L^{\mu\nu}(k) A_0^L\,,
\nonumber\\
&D_L^{\mu\nu}(k) = \frac{1}{(-k^2)^{1+L\varepsilon}}
\biggl(g^{\mu\nu} + \frac{k^\mu k^\nu}{-k^2}\biggr)\,,
\label{abel:Dk}
\end{align}
where $d_S^{(L)}$ are products of some color structures of $\Pi_{L'}$.
In coordinate space
\begin{align}
&D_{L-1}^{\mu\nu}(x) = \frac{i}{(4\pi)^{d/2}}
\frac{\Gamma(1-u)}{\Gamma(2+u-\varepsilon)}
\frac{2^{1-2u}}{(-x^2)^{1-u}}
\nonumber\\
&\quad{} \times
\biggl[(1+2(u-\varepsilon)) g^{\mu\nu} - 2 (1-u) \frac{x^\mu x^\nu}{-x^2}\biggr]\,.
\label{abel:Dx}
\end{align}
Here and below
\begin{equation}
u = L\varepsilon\,.
\label{abel:u}
\end{equation}

\subsection{HQET field anomalous dimension}

For a straight Wilson line in Euclidean time ($t = -i\tau$)
\begin{align}
&\raisebox{-1.25mm}{\begin{picture}(18,6)
\put(9,4.125){\makebox(0,0){\includegraphics{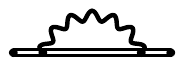}}}
\put(1,1.25){\makebox(0,0)[t]{$0$}}
\put(17,1.25){\makebox(0,0)[t]{$t$}}
\end{picture}}
=
\raisebox{-1.25mm}{\begin{picture}(17,4.25)
\put(8.5,2.25){\makebox(0,0){\includegraphics{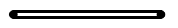}}}
\put(1,1.25){\makebox(0,0)[t]{$0$}}
\put(16,1.25){\makebox(0,0)[t]{$t$}}
\end{picture}}
\times w_S(t)\,,
\label{abel:wt}\\
&w_S(\tau) = \sum_{L=1}^\infty d_S^{(L-1)} w_L
\bigl[A_0 (\tau/2)^{2\varepsilon} e^{\gamma\varepsilon}\bigr]^L\,,\quad
w_L = - e^{\gamma(\varepsilon-2u)} \frac{(3-2\varepsilon) \Gamma(-u)}{(1-2u) \Gamma(2+u-\varepsilon)}\,.
\nonumber
\end{align}
A finite $t$ provides an IR cutoff;
the region $t_2-t_1 \to 0$ gives an UV divergence.
Re-expressing~(\ref{abel:MS}) $w_S(\tau)$ via $\alpha_s(\mu)$
(say, with $\mu = (2/\tau) e^{-\gamma}$)
and using $w_S(\tau) = (\log Z_h)_S + \mathcal{O}(\varepsilon^0)$~(\ref{HQET:Zhcoord}),
we obtain $(\log Z_h)_S$ (it does not depend on $\tau$) and hence $(\gamma_h)_S$.

Alternatively, we can work in momentum space.
The Fourier image of $w_S(t)$~(\ref{abel:wt}) is
\begin{align}
&\raisebox{-1.25mm}{\begin{picture}(22,7)
\put(11,4.125){\makebox(0,0){\includegraphics{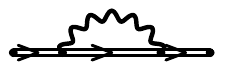}}}
\put(3.5,1.25){\makebox(0,0)[t]{$\omega$}}
\put(18.5,1.25){\makebox(0,0)[t]{$\omega$}}
\end{picture}}
=
\raisebox{-1.25mm}{\begin{picture}(17,4.25)
\put(8.5,2.25){\makebox(0,0){\includegraphics{hqet.pdf}}}
\put(8.5,1.25){\makebox(0,0)[t]{$\omega$}}
\end{picture}}
\times \tilde{w}_S(\omega)\,,
\label{abel:ww}\\
&\tilde{w}_S(\omega) = \sum_{L=1}^\infty d_S^{(L-1)} \tilde{w}_L
\bigl[A_0 (-2\omega)^{-2\varepsilon}\bigr]^L\,,\quad
\tilde{w}_L = - e^{\gamma\varepsilon} \frac{(3-2\varepsilon) \Gamma(-u) \Gamma(1+2u)}{(1-2u) \Gamma(2+u-\varepsilon)}\,.
\nonumber
\end{align}
This result can be derived in momentum space using\cite{Grozin:2004yc}
\begin{align}
&\frac{1}{i\pi^{d/2}} \int \frac{d^d k}{[-2(k\cdot v+\omega)]^{n_1} (-k^2)^{n_2}}
= I(n_1,n_2) (-2\omega)^{d-n_1-2n_2}\,,
\nonumber\\
&I(n_1,n_2) = \frac{\Gamma(n_1 + 2 n_2 - d) \Gamma\bigl(\frac{d}{2} - n_2\bigr)}{\Gamma(n_1) \Gamma(n_2)}\,,
\label{abel:I}
\end{align}
or by Fourier transforming~(\ref{abel:wt}).
Now a non-zero $\omega$ provides an IR cutoff;
the region $k \to \infty$ gives an UV divergence which coincides with that of~(\ref{abel:wt}).
Re-expressing~(\ref{abel:MS}) $\tilde{w}_S(\omega)$ via $\alpha_s(\mu)$
(say, with $\mu = -2\omega$)
and using $\tilde{w}_S(\omega) = (\log Z_h)_S + \mathcal{O}(\varepsilon^0)$,
we obtain $(\log Z_h)_S$ (it does not depend on $\omega$) and hence $(\gamma_h)_S$.

The renormalization constant can be written as
\begin{equation}
(\log Z_h)_S = \sum_{n=1}^\infty \frac{z_{hn}}{\varepsilon^n}\,,\quad
z_{hn} = \mathcal{O}(\alpha_s^n)\,.
\label{abel:zn}
\end{equation}
Only $z_{h1}$ is needed in order to obtain
\begin{equation}
\gamma_h(\alpha_s) = - 2 \frac{d z_{h1}(\alpha_s)}{d\log\alpha_s}\,;
\label{abel:z1}
\end{equation}
higher $z_{hn}$ contain no new information,
and are uniquely reconstructed from $z_{h1}$ using self-consistency conditions.

\subsection{Cusp anomalous dimension}

For a cusped Wilson line we have ($-t < 0 < t'$)
\begin{align}
&(\log W(t,t',\varphi))_S = w_S(t,t',\varphi)
\label{abel:cusp}\\
&{} =
\raisebox{-7mm}{\begin{picture}(30,15)
\put(15,7){\makebox(0,0){\includegraphics{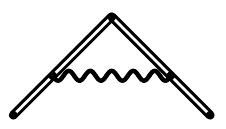}}}
\put(15,12.5){\makebox(0,0)[b]{$0$}}
\put(4.5,2){\makebox(0,0)[r]{$-vt$}}
\put(25.5,2){\makebox(0,0)[l]{$v't'$}}
\end{picture}}
+
\raisebox{-7mm}{\begin{picture}(30,15)
\put(15,7){\makebox(0,0){\includegraphics{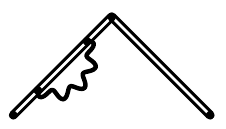}}}
\put(15,12.5){\makebox(0,0)[b]{$0$}}
\put(4.5,2){\makebox(0,0)[r]{$-vt$}}
\put(25.5,2){\makebox(0,0)[l]{$v't'$}}
\end{picture}}
+
\raisebox{-7mm}{\begin{picture}(30,15)
\put(15,7){\makebox(0,0){\includegraphics{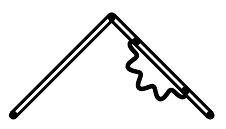}}}
\put(15,12.5){\makebox(0,0)[b]{$0$}}
\put(4.5,2){\makebox(0,0)[r]{$-vt$}}
\put(25.5,2){\makebox(0,0)[l]{$v't'$}}
\end{picture}}
\,,
\nonumber
\end{align}
and
\begin{equation}
w_S(t,t',\varphi) - w_S(t,t',0) =
\raisebox{-7mm}{\begin{picture}(30,15)
\put(15,7){\makebox(0,0){\includegraphics{v1.pdf}}}
\put(15,12.5){\makebox(0,0)[b]{$0$}}
\put(4.5,2){\makebox(0,0)[r]{$-vt$}}
\put(25.5,2){\makebox(0,0)[l]{$v't'$}}
\end{picture}}
-
\raisebox{-4.5mm}{\begin{picture}(26,8.5)
\put(13,3.25){\makebox(0,0){\includegraphics{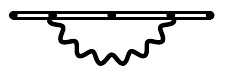}}}
\put(13,6){\makebox(0,0)[b]{$0$}}
\put(3,6){\makebox(0,0)[b]{$-vt$}}
\put(23,6){\makebox(0,0)[b]{$vt'$}}
\end{picture}}
\,.
\label{abel:cusp2}
\end{equation}
Let's denote ($-t<-t_1<0<t_2<t'$)
\begin{align}
&
\raisebox{-7mm}{\begin{picture}(30,15)
\put(15,7){\makebox(0,0){\includegraphics{v1.pdf}}}
\put(15,12.5){\makebox(0,0)[b]{$0$}}
\put(4.5,2){\makebox(0,0)[r]{$-vt$}}
\put(25.5,2){\makebox(0,0)[l]{$v't'$}}
\put(9,7){\makebox(0,0)[r]{$-vt_1$}}
\put(22,7){\makebox(0,0)[l]{$v't_2$}}
\end{picture}}
=
\raisebox{-7mm}{\begin{picture}(30,15)
\put(15,7){\makebox(0,0){\includegraphics{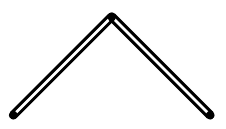}}}
\put(15,12.5){\makebox(0,0)[b]{$0$}}
\put(4.5,2){\makebox(0,0)[r]{$-vt$}}
\put(25.5,2){\makebox(0,0)[l]{$v't'$}}
\end{picture}}
\times V_S(t,t',\varphi)\,,
\label{abel:V}\\
&V_S(\tau,\tau',\varphi) = \sum_{L=1}^\infty d_S^{(L-1)} V_L(\tau'/\tau,\varphi)
\bigl[A_0 (\tau\tau'/4)^\varepsilon e^{\gamma\varepsilon}\bigr]^L\,.
\nonumber
\end{align}
When expressed via the renormalized quantities,
$V_S(t,t',\varphi) = (\log Z_J(\varphi))_S + (\log Z_h)_S + \mathcal{O}(\varepsilon^0)$.
Therefore, $\bar{V}_S(t,t',\varphi) =V_S(t,t',\varphi) - V_S(t,t',0) = (\log Z_J(\varphi))_S + \mathcal{O}(\varepsilon^0)$,
where $(\log Z_J(\varphi))_S$ does not depend on $t$ and $t'$ and is gauge invariant.
Note that
\begin{equation}
V_S(t,t',0) = w_S(t+t')-w_S(t)-w_S(t') = -(\log Z_h)_S + \mathcal{O}(\varepsilon^0)\,.
\label{abel:V0}
\end{equation}

For calculating the anomalous dimension we need finite $\tau$, $\tau'$ as IR regulators.
We may set $\tau' = \tau$ in order to have a simpler single-scale problem\cite{Grozin:2018vdn}:
\begin{align}
&V_L(1,\varphi) = - 2 e^{\gamma(\varepsilon-2u)} \frac{\Gamma(1-u)}{\Gamma(2+u-\varepsilon)}
\bigl[I_1(u,\varphi) (2+u-2\varepsilon) \cosh\varphi + I_2(u,\varphi) (1-u)\bigr]\,,
\label{abel:VL}\\
&I_1(u,\varphi) = \int_0^1 dt_1 \int_0^1 dt_2 (e^{\varphi/2} t_1 + e^{-\varphi/2} t_2)^{-1+u} (e^{-\varphi/2} t_1 + e^{\varphi/2} t_2)^{-1+u}
\nonumber\\
&{} = \frac{e^{-2 u \varphi}}{4 u^2 \sinh\varphi} \bigl(g_1(u,\varphi) - g_2(u,\varphi)\bigr)\,,
\nonumber\\
&I_2(u,\varphi) = \int_0^1 dt_1 \int_0^1 dt_2 (e^{\varphi/2} t_1 + e^{-\varphi/2} t_2)^u (e^{-\varphi/2} t_1 + e^{\varphi/2} t_2)^{-2+u}
\nonumber\\
&{} = \frac{1}{2 u (1-u)} \left[1 + \frac{e^{-2 u \varphi}}{2 \sinh\varphi}
\bigl(e^{-\varphi} g_1(u,\varphi) - e^\varphi g_2(u,\varphi)\bigr)\right]\,,
\nonumber\\
&I_1(u,0) = I_2(u,0) = \frac{2 - 2^{2u}}{2 u (1-2u)}\,,
\label{abel:I0}\\
&g_1(u,\varphi) = (e^\varphi+1)^{2u} f_1(u,1-e^\varphi) - f_1(u,1-e^{2\varphi})\,,
\nonumber\\
&g_2(u,\varphi) = (e^\varphi+1)^{2u} f_2(u,1-e^\varphi) - f_2(u,1-e^{2\varphi})\,,
\nonumber\\
&f_1(u,x) = \F{2}{1}{-2u,-u\\1-2u}{x} = 1 + 2 \Li2(x) u^2 + \mathcal{O}(u^3)\,,
\nonumber\\
&f_2(u,x) = \F{2}{1}{-2u,1-u\\1-2u}{x} = 1 + 2 \log(1-x) u
\nonumber\\
&\quad{} + \left(\log^2(1-x) - 2 \Li2(x)\right) u^2 + \mathcal{O}(u^3)\,.
\nonumber
\end{align}
Using~(\ref{abel:I0}) we get the equality~(\ref{abel:V0}).
The UV divergence ($u = L\varepsilon$, $\varepsilon \to 0$) of $V_L(1,\varphi)$~(\ref{abel:VL}) is
\begin{equation}
V_L(1,\varphi) = - \frac{2 \varphi \coth\varphi + 1}{u} + \mathcal{O}(1)\,.
\label{abel:UV}
\end{equation}

Alternatively, we can work in momentum space.
The Fourier image of $V_S(t,t',\varphi)$~(\ref{abel:V}) is
\begin{align}
&
\raisebox{-6mm}{\begin{picture}(24,12)
\put(12,6){\makebox(0,0){\includegraphics{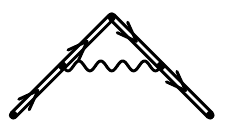}}}
\put(3,4){\makebox(0,0)[r]{$\omega$}}
\put(21,4){\makebox(0,0)[l]{$\omega'$}}
\end{picture}}
=
\raisebox{-6mm}{\begin{picture}(24,12)
\put(12,6){\makebox(0,0){\includegraphics{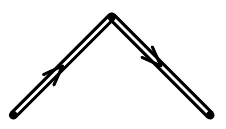}}}
\put(5.5,6.5){\makebox(0,0)[r]{$\omega$}}
\put(18.5,6.5){\makebox(0,0)[l]{$\omega'$}}
\end{picture}}
\times \tilde{V}_S(\omega,\omega',\varphi)\,,
\label{abel:Vw}\\
&\tilde{V}_S(\omega,\omega',\varphi) =
\raisebox{-5.5mm}{\begin{picture}(22,11)
\put(11,5.5){\makebox(0,0){\includegraphics{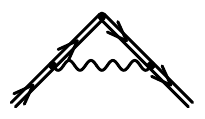}}}
\put(3,3.5){\makebox(0,0)[r]{$\omega$}}
\put(19,3.5){\makebox(0,0)[l]{$\omega'$}}
\end{picture}}
= \sum_{L=1}^\infty d_S^{(L-1)} \tilde{V}_L(\omega'/\omega,\varphi)
\bigl[A_0 (4\omega\omega')^{-\varepsilon}\bigr]^L\,.
\nonumber
\end{align}
We have $\tilde{\bar{V}}_S(\omega,\omega',\varphi) = \tilde{V}_S(\omega,\omega',\varphi) - \tilde{V}_S(\omega,\omega',0) = (\log Z_J(\varphi))_S + \mathcal{O}(\varepsilon^0)$,
where $(\log Z_J(\varphi))_S$ does not depend on $\omega$ and $\omega'$ and is gauge invariant.
Note that
\begin{equation}
\tilde{V}_S(\omega,\omega',0) = - \frac{\omega \tilde{w}_S(\omega) - \omega' \tilde{w}_S(\omega')}{\omega - \omega'}
= -(\log Z_h)_S + \mathcal{O}(\varepsilon^0)
\label{abel:Vw0}
\end{equation}
(this result can be derived by separating partial fractions in the loop integrand
or by Fourier transforming~(\ref{abel:V0})).

The scalar Feynman integrals in $\tilde{V}_L(y,\varphi)$ can be reduced\cite{Grozin:2011rs} to 3 master integrals,
2 ones are trivial~(\ref{abel:I}) and 1 the non-trivial:
\begin{align}
&\tilde{V}_L(y,\varphi) = - \frac{2 e^{\gamma\varepsilon}}{(1+u-\varepsilon) (2\cosh\varphi-y-y^{-1})}
\nonumber\\
&{} \times \bigl\{2 \bigl[(1-\varepsilon) (2\cosh\varphi-y-y^{-1}) \cosh\varphi + u (\cosh\varphi-y) (\cosh\varphi-y^{-1})\bigr] I(y)
\nonumber\\
&\quad{} + \bigl[(\cosh\varphi-y) y^{-u} + (\cosh\varphi-y^{-1}) y^u\bigr] I(2,1+u-\varepsilon)\bigr\}\,,
\nonumber\\
&I(y) = \frac{(4\omega\omega')^u}{i\pi^{d/2}} \int \frac{d^d k}%
{\bigl[-2(k\cdot v+\omega)\bigr] \bigl[-2(k\cdot v'+\omega')\bigr] (-k^2)^{1+u-\varepsilon}}\,.
\label{abel:Iy}
\end{align}
In anomalous-dimension calculations non-zero $\omega$, $\omega'$ are needed only as IR regulators,
and we may set $\omega' = \omega$ in order to have a simpler single-scale problem.
The master integral $I(1)$~(\ref{abel:Iy}) is then expressed\cite{Grozin:2011rs} via a hypergeometric function,
and the result is\cite{Grozin:2016ydd}
\begin{align}
&\tilde{V}_L(1,\varphi) = - 2 e^{\gamma\varepsilon} \frac{I(2,1+u-\varepsilon)}{1+u-\varepsilon}
\bigl\{\bigl[(2+u-2\varepsilon) \cosh\varphi - u\bigr] F(u,\varphi) + 1\bigr\}\,,
\nonumber\\
&F(u,\varphi) = \F{2}{1}{1,1-u\\\frac{3}{2}}{\frac{1-\cosh\varphi}{2}}\,,\quad
F(0,\varphi) = \frac{\varphi}{\sinh\varphi}\,.
\label{abel:Fph}
\end{align}
At $\varphi=0$ ($F(u,\varphi) = 1$) we get~(\ref{abel:Vw0})
(at $\omega'=\omega$ it becomes a derivative).
The UV divergence of $\tilde{V}_L(1,\varphi)$~(\ref{abel:Fph}) is the same as that of $V_L(1,\varphi)$~(\ref{abel:UV}).

The hypergeometric function $F(u,\varphi)$~(\ref{abel:Fph}) has been expanded\cite{Davydychev:2000na}
in $u$ to all orders%
\footnote{There is a typo in this formula.
It has been corrected in\cite{Davydychev:2003mv} and in v4 of the arXiv preprint.},
the coefficients are expressed via Nielsen polylogarithms $S_{nm}(x)$.
The result\cite{Davydychev:2000na} is written for the case of an Euclidean angle,
its analytical continuation to Minkowski angles is\cite{Grozin:2017aty}
\begin{align}
&F(u,\varphi) = \frac{1}{\sinh\varphi (2 \cosh(\varphi/2))^{2 u}}
\biggl[ \frac{\sinh(\varphi u)}{u}
\nonumber\\
&\quad{} - e^{-\varphi u} \sum_{n=1}^\infty u^n \sum_{m=1}^n (-2)^{n-m} S_{m,n-m+1}(-e^{\varphi})
\nonumber\\
&\quad{} + e^{\varphi u} \sum_{n=1}^\infty u^n \sum_{m=1}^n (-2)^{n-m} S_{m,n-m+1}(-e^{-\varphi})
\biggr]\,.
\label{abel:DK}
\end{align}
It is possible to re-express this expansion in terms of Nielsen polylogarithms of just one argument
(see\cite{Kolbig:1986}
or \url{http://functions.wolfram.com/ZetaFunctionsandPolylogarithms/PolyLog3/17/01/}),
but then the symmetry $\varphi\to-\varphi$ will not be explicit.

\subsection{Potential}

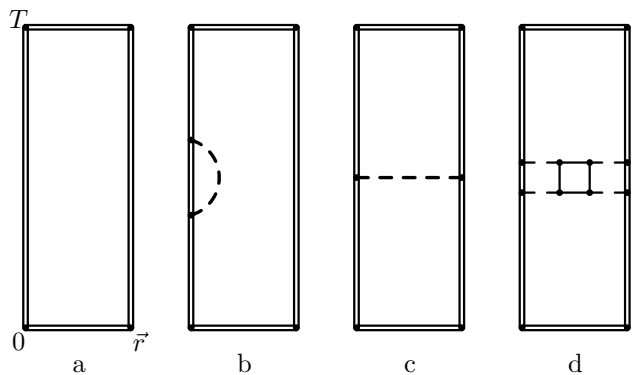
\begin{figure}
\begin{center}
\begin{picture}(83,48)
\put(9,26){\makebox(0,0){\includegraphics{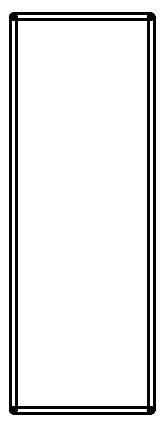}}}
\put(1,4){\makebox(0,0){$0$}}
\put(17,4){\makebox(0,0){$\vec{r}$}}
\put(1,47){\makebox(0,0){$T$}}
\put(9,0){\makebox(0,0)[b]{a}}
\put(31,26){\makebox(0,0){\includegraphics{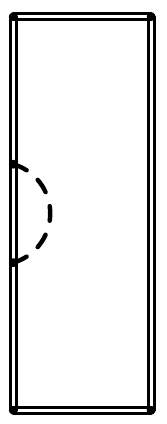}}}
\put(31,0){\makebox(0,0)[b]{b}}
\put(53,26){\makebox(0,0){\includegraphics{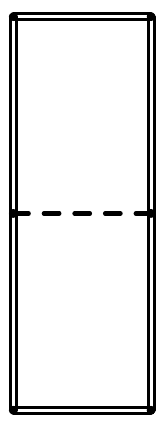}}}
\put(53,0){\makebox(0,0)[b]{c}}
\put(75,26){\makebox(0,0){\includegraphics{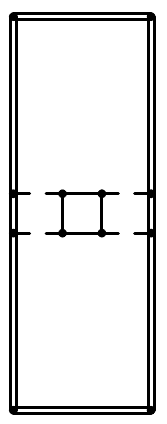}}}
\put(75,0){\makebox(0,0)[b]{d}}
\end{picture}
\end{center}
\caption{The Wilson loop determining the quark-antiquark potential.}
\label{F:V}
\end{figure}

We shall need also some formulas for the static quark-antiquark potential
in order to discuss the conformal anomaly $\Delta(\alpha_s)$.
The Wilson loop in Fig.~\ref{F:V}a described the following sequence of events:
a static (HQET) particle in a color representation $R$ and its antiparticle
are created at a distance $\vec{r}$ at the moment $t=0$;
they stay at these positions for a time $T \gg r$;
and finally they are annihilated at the moment $T$.
This pair has the energy $V(\vec{r}^{\,})$,
and for large $T$ we have
\begin{equation}
\log W = - i V(\vec{r}^{\,}) T\,.
\label{abel:Vdef}
\end{equation}
Due to exponentiation (Sect.~\ref{S:Exp}),
$\log W$ is equal to the sum of webs.
Here we shall consider abelian color structures
for which only 2-leg webs (Fig.~\ref{F:V}b, c) contribute.
We are not interested in contributions where the gluon is attached
to the lower horizontal Wilson line or to the upper one ---
such contributions don't scale as $T$.
The diagram in Fig.~\ref{F:V}b describes not a quark-antiquark potential
but the residual mass term of an HQET particle;
it vanishes in dimensional regularization.
The first contribution of a 4-leg web is shown in Fig.~\ref{F:V}d.

It is convenient to use the Coulomb gauge.
The full Coulomb photon propagator is
\begin{equation}
D(q) = \frac{1}{1 - \Pi(q^2)} \frac{1}{\vec{q}^{\,2}}\,.
\label{abel:Coulomb}
\end{equation}
The photon self-energy $\Pi(q^2)$ is gauge invariant in QED,
so, we may use the same $\Pi(q^2)$~(\ref{abel:Pi}) as in covariant gauges.
Integration in one of the 2 times in Fig.~\ref{F:V}c gives $T$;
integration in the other time gives
\begin{equation}
V(\vec{r}^{\,}) = - e_0^2 \int_{-\infty}^{+\infty} dt\,D(t,\vec{r}^{\,})
= \int \frac{d^{d-1}\vec{q}}{(2\pi)^{d-1}} e^{i\vec{q}\cdot\vec{r}} V(\vec{q}^{\,})\,,
\label{abel:Vr}
\end{equation}
where the momentum-space potential is
\begin{equation}
V(\vec{q}^{\,}) = - e_0^2 D(q)\,,\quad
q = (0,\vec{q}^{\,})\,.
\label{abel:Vq}
\end{equation}
It is finite, because in QED $Z_\alpha = Z_A^{-1}$.
The contribution of a subset $S$ of color structures is
\begin{equation}
V_S(\vec{q}^{\,}) = - (4\pi)^{d/2} e^{\gamma\varepsilon} \sum_{L=1}^\infty
\frac{d_S^{(L-1)}}{(\vec{q}^{\,2})^{1+(L-1)\varepsilon}} A_0^L\,.
\label{abel:VS}
\end{equation}

\section{Leading large-$\beta_0$ order}
\label{S:Lb0}

Let's consider terms with the leading powers of $n_f$ to all orders in $\alpha_s$:
$S = \{(T_F n_f)^L,L\ge0\}$.
It is sufficient to consider QED: $C_F = T_F = 1$, $C_A = 0$, $\beta_0 = - \frac{4}{3} n_f$.
Let's introduce
\begin{equation}
b = \beta_0 \frac{\alpha}{4\pi}\,.
\label{Lb0:b}
\end{equation}
We assume $b\sim1$ and take into account all powers of $b$;
$1/\beta_0\ll1$ is our small parameter,
and we consider only a few terms in expansions in $1/\beta_0$.
This large $\beta_0$ limit is reviewed in chapter~8 of\cite{Grozin:2004yc}.

The photon self energy $\Pi_0(k^2)$ at the leading large $\beta_0$ (L$\beta_0$) order is $\sim1$:
\begin{align}
&\raisebox{-6.5mm}{\includegraphics{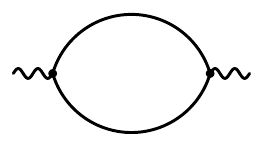}}
\Rightarrow \Pi_0(k^2)
= \Pi_0 A_0 (-k^2)^{-\varepsilon}\,,\quad
\Pi_0 = \beta_0 \frac{D(\varepsilon)}{\varepsilon}\,,
\nonumber\\
&D(\varepsilon) = e^{\gamma\varepsilon}
\frac{(1-\varepsilon) \Gamma(1+\varepsilon) \Gamma^2(1-\varepsilon)}{(1-2\varepsilon) (1-\frac{2}{3}\varepsilon) \Gamma(1-2\varepsilon)}
= 1 + \frac{5}{3} \varepsilon + \cdots
\label{Lb0:Pi}
\end{align}
The charge renormalization in the $\overline{\text{MS}}$ scheme~(\ref{abel:MS}) is
\begin{equation}
\beta_0 A_0 = b Z_\alpha(b) \mu^{2\varepsilon}\,.
\label{Lb0:MS}
\end{equation}
At the L$\beta_0$ order we can solve the RG equation (the $\beta$ function is $b$)
\begin{equation*}
\frac{d\log Z_\alpha(b)}{d\log b} = - \frac{b}{\varepsilon+b}
\end{equation*}
and obtain
\begin{equation}
Z_\alpha(b) = \frac{1}{1+b/\varepsilon}\,.
\label{Lb0:Za}
\end{equation}

\subsection{HQET field anomalous dimension}

\begin{figure}[ht]
\begin{center}
\begin{picture}(100,25)
\put(23,13){\makebox(0,0){\includegraphics{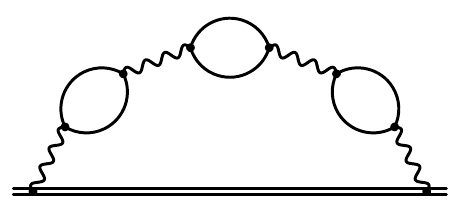}}}
\put(77,12.5){\makebox(0,0){\includegraphics{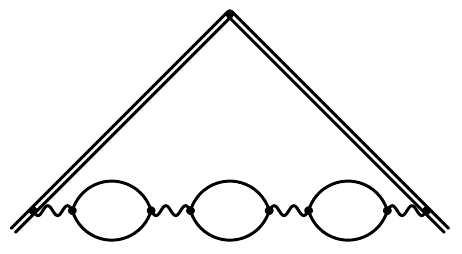}}}
\end{picture}
\end{center}
\caption{Typical diagrams for $\gamma_h$ and $\Gamma(\varphi)$ at L$\beta_0$.}
\label{F:Lb0}
\end{figure}

The diagrams for $w_S$ and $V_S$ include only $\Pi_0$ insertions in the photon propagator
(Fig.~\ref{F:Lb0}): $d_S^{(L)} = \Pi_0^L$.
We can write the 2-leg web $\tilde{w}(\omega)$ in the form
\begin{align}
&\tilde{w}(\omega) = \frac{1}{\beta_0} \sum_{L=1}^\infty \frac{\tilde{f}(\varepsilon,L\varepsilon)}{L}
\bigl[\Pi_0(k^2)\bigr]^L + \mathcal{O}\biggl(\frac{1}{\beta_0^2}\biggr)\quad
(k^2 = (-2\omega)^2)\,,
\label{Lb0:wf}\\
&\tilde{f}(\varepsilon,u) = \frac{u \tilde{w}_L}{D(\varepsilon)}
= \frac{3 \bigl(1 - \frac{2}{3} \varepsilon\bigr)^2 \Gamma(2-2\varepsilon) \Gamma(1-u) \Gamma(1+2u)}%
{(1-\varepsilon) (1-2u) \Gamma^2(1-\varepsilon) \Gamma(1+\varepsilon) \Gamma(2+u-\varepsilon)}\,.
\label{Lb0:fw}
\end{align}
Expressing~(\ref{Lb0:MS}) it via $\alpha(\mu)$ with
$\mu = (-2\omega) D(\varepsilon)^{-1/(2\varepsilon)} \to (-2\omega) e^{-5/6}$
using~(\ref{Lb0:Za}), we have
\begin{equation}
\tilde{w}(\omega) = \frac{1}{\beta_0}
\sum_{L=1}^\infty \frac{\tilde{f}(\varepsilon,L\varepsilon)}{L}
\biggl(\frac{b}{\varepsilon+b}\biggr)^{\!L}
+ \mathcal{O}\biggl(\frac{1}{\beta_0^2}\biggr)\,.
\label{Lb0:wren}
\end{equation}
The function $\tilde{f}(\varepsilon,u)$~(\ref{Lb0:fw}) is regular at $\varepsilon = u = 0$:
\begin{equation}
\tilde{f}(\varepsilon,u) = \sum_{n=0}^\infty \sum_{m=0}^\infty \tilde{f}_{nm} \varepsilon^n u^m\,.
\label{Lb0:feuexp}
\end{equation}
We also expand $\bigl[b/(\varepsilon+b)\bigr]^L$ in $b$ and get a quadruple series for $\tilde{w}(\omega)$.
When selecting $\varepsilon^{-1}$ terms in order to obtain $z_{h1}$,
all coefficients but $\tilde{f}_{n0}$ cancel:
\begin{equation*}
z_{h1}(b) = - \frac{1}{\beta_0} \sum_{n=0}^\infty \frac{\tilde{f}_{n0}}{n+1} (-b)^{n+1}
+ \mathcal{O}\biggl(\frac{1}{\beta_0^2}\biggr)\,,
\end{equation*}
so that
\begin{equation}
\gamma_h(b) = - 2 \frac{b}{\beta_0} \tilde{f}(-b,0)
+ \mathcal{O}\biggl(\frac{1}{\beta_0^2}\biggr)\,.
\label{Lb0:gamma}
\end{equation}
Finally, we obtain\cite{Broadhurst:1994se}
\begin{align}
&\gamma_h(b) = - 6 \frac{b}{\beta_0} \gamma_0(b)
+ \mathcal{O}\biggl(\frac{1}{\beta_0^2}\biggr)\,,
\label{Lb0:gammah}\\
&\gamma_0(b) =
\frac{\bigl(1 + \frac{2}{3} b\bigr)^2 \Gamma(2+2b)}%
{(1+b)^2 \Gamma^3(1+b) \Gamma(1-b)}
= 1
+ \frac{4}{3} b
- \frac{5}{9} b^2
- 2 \biggl(\zeta_3 - \frac{1}{3}\biggr) b^3
\nonumber\\
&{}
+ \biggl(\frac{\pi^4}{10} - 8 \zeta_3 - \frac{7}{3}\biggr) \frac{b^4}{3}
- 2 \biggl(3 \zeta_5 - \frac{\pi^4}{45} - \frac{5}{9} \zeta_3 - \frac{4}{9}\biggr) b^5
\nonumber\\
&{}
+ \biggl(2 \zeta_3^2 + \frac{2}{189} \pi^6 - 8 \zeta_5 - \frac{\pi^4}{54} - \frac{4}{3} \zeta_3 - 1\biggr) b^6
+ \mathcal{O}(b^7)\,.
\label{Lb0:gammah0}
\end{align}
It is absolutely trivial to extend this expansion to any desired order.
This is the Landau-gauge result;
in order to obtain the result in an arbitrary covariant gauge,
one should add the trivial 1-loop term proportional to $a$.
Restoring the color factors
\begin{equation}
\gamma_h = - 6 C_R \frac{\alpha_s}{4\pi} \gamma_0(b) + \cdots\,,\quad
b = - \frac{4}{3} T_F n_f\,,
\label{Lb0:col}
\end{equation}
we reproduce the corresponding terms in~(\ref{Gamma:gamma}).

Alternatively, we can work in coordinate space:
\begin{align*}
&w(\tau) = \frac{1}{\beta_0} \sum_{L=1}^\infty \frac{f(\varepsilon,L\varepsilon)}{L}
\bigl[\Pi_0(k^2) e^{\gamma\varepsilon}\bigr]^L + \mathcal{O}\biggl(\frac{1}{\beta_0^2}\biggr)\quad
(k^2 = (2/\tau)^2)\,,\\
&f(\varepsilon,u) = \frac{u w_L}{D(\varepsilon)}
= \frac{3 \bigl(1 - \frac{2}{3} \varepsilon\bigr)^2 \Gamma(2-2\varepsilon) \Gamma(1+u)}%
{(1-\varepsilon) (1-2u) \Gamma^2(1-\varepsilon) \Gamma(1+\varepsilon) \Gamma(2+u-\varepsilon)}\,,
\end{align*}
see~(\ref{abel:wt}).
Choosing $\mu = (2/\tau) e^{-\gamma} D(\varepsilon)^{1/(2\varepsilon)} \to (2/\tau) e^{-\gamma-5/6}$ we obtain
\begin{equation*}
w(\tau) = \frac{1}{\beta_0} \sum_{L=1}^\infty \frac{f(\varepsilon,L\varepsilon)}{L}
\biggl(\frac{b}{\varepsilon+b}\biggr)^{\!L} + \mathcal{O}\biggl(\frac{1}{\beta_0^2}\biggr)\,;
\end{equation*}
$f(-b,0) = \tilde{f}(-b,0)$,
and we get the result~(\ref{Lb0:gammah0}) again.

\subsection{Cusp anomalous dimension}

We can write $\tilde{\bar{V}}(\omega,\omega,\varphi)$ in the form
($\tilde{\bar{V}}(\varphi) = \tilde{V}(1,\varphi) - \tilde{V}(1,0)$)
\begin{align}
&\tilde{\bar{V}}(\omega,\omega,\varphi)
= \frac{1}{\beta_0} \sum_{L=1}^\infty \frac{\tilde{f}(\varepsilon,L\varepsilon,\varphi)}{L}
\bigl[\Pi_0(k^2)]^L + \mathcal{O}\biggl(\frac{1}{\beta_0^2}\biggr)\quad
(k^2 = (-2\omega)^2)\,,
\nonumber\\
&\tilde{f}(\varepsilon,u,\varphi) = \frac{u \tilde{\bar{V}}(\varphi)}{D(\varepsilon)}
= - \frac{\bigl(1 - \frac{2}{3} \varepsilon\bigr) \Gamma(2-2\varepsilon) \Gamma(1-u) \Gamma(1+2u)}%
{(1-\varepsilon) \Gamma^2(1-\varepsilon) \Gamma(1+\varepsilon) \Gamma(2+u-\varepsilon)}
\nonumber\\
&\quad{} \times
\bigl[\bigl((2+u-2\varepsilon) \cosh\varphi - u\bigr) F(u,\varphi) - 2 (1-\varepsilon)\bigr]\,,
\label{Lb0:fV}
\end{align}
see~(\ref{abel:Fph}).
We re-express the result via the renormalized $b$:
\begin{equation*}
\tilde{\bar{V}}(\omega,\omega,\varphi)
= \frac{1}{\beta_0} \sum_{L=1}^\infty \frac{\tilde{f}(\varepsilon,L\varepsilon,\varphi)}{L}
\biggl(\frac{b}{\varepsilon+b}\biggr)^{\!L}
+ \mathcal{O}\biggl(\frac{1}{\beta_0^2}\biggr)\,,
\end{equation*}
expand $\tilde{f}(\varepsilon,u,\varphi)$ in $\varepsilon$ and $u$
and $\bigl[b/(\varepsilon+b)\bigr]^L$ in $b$.
When selecting $\varepsilon^{-1}$ terms in order to obtain $z_{J1}$,
all coefficients but $\tilde{f}_{n0}(\varphi)$ cancel:
\begin{align*}
&z_{J1}(b,\varphi) = - \frac{1}{\beta_0} \sum_{n=0}^\infty \frac{\tilde{f}_{n0}(\varphi)}{n+1} (-b)^{n+1}
+ \mathcal{O}\biggl(\frac{1}{\beta_0^2}\biggr)\,,\\
&\tilde{f}(\varepsilon,0,\varphi) = \sum_{n=0}^\infty \tilde{f}_{n0}(\varphi) \varepsilon^n
= - 2 (\varphi \coth\varphi - 1) \hat{f}(\varepsilon)\,,\\
&\hat{f}(\varepsilon) = \sum_{n=0}^\infty \hat{f}_n \varepsilon^n\,,\quad
\tilde{f}_{n0}(\varphi) = - 2 (\varphi \coth\varphi - 1) \hat{f}_n\,.
\end{align*}
Therefore at L$\beta_0$ we obtain\cite{Gracey:1994nn,Beneke:1995pq}
\begin{align}
&\Gamma(b,\varphi) = 4 (\varphi \coth\varphi - 1) \frac{b}{\beta_0} \Gamma_0(b)
+ \mathcal{O}\biggl(\frac{1}{\beta_0^2}\biggr)\,,
\nonumber\\
&\Gamma_0(b) = \hat{f}(-b)
= \frac{\bigl(1 + \frac{2}{3} b\bigr) \Gamma(2+2b)}{(1+b) \Gamma^3(1+b) \Gamma(1-b)}
= 1
+ \frac{5}{3} b
- \frac{b^2}{3}
- \biggl(2 \zeta_3 - \frac{1}{3}\biggr) b^3
\nonumber\\
&{} + \biggl(\frac{\pi^4}{10} - 10 \zeta_3 - 1\biggr) \frac{b^4}{3}
- \biggl(6 \zeta_5 - \frac{\pi^4}{18} - \frac{2}{3} \zeta_3 - \frac{1}{3}\biggr) b^5
\nonumber\\
&{} + \biggl(2 \zeta_3^2 + \frac{2}{189} \pi^6 - 10 \zeta_5 - \frac{\pi^4}{90} - \frac{2}{3} \zeta_3 - \frac{1}{3}\biggr) b^6
+ \mathcal{O}(b^7)\,.
\label{Lb0:Gamma0}
\end{align}
It is absolutely trivial to extend this expansion to any desired order.
Restoring the color factors~(\ref{Lb0:col}),
we reproduce the corresponding terms in~(\ref{Cusp:QCD}).

Alternatively, we can work in coordinate space ($\bar{V}(\tau,\tau,\varphi) = V(\tau,\tau,\varphi) - V(\tau,\tau,0)$):
\begin{align*}
&\bar{V}(\tau,\tau,\varphi)
= \frac{1}{\beta_0} \sum_{L=1}^\infty \frac{f(\varepsilon,L\varepsilon,\varphi)}{L}
\bigl[\Pi_0(k^2) e^{\gamma\varepsilon}\bigr]^L + \mathcal{O}\biggl(\frac{1}{\beta_0^2}\biggr)\quad
(k^2 = (2/\tau)^2)\,,\\
&f(\varepsilon,u) = \frac{u w_L}{D(\varepsilon)}
= e^{\gamma(\varepsilon-2u)} \frac{3 \bigl(1 - \frac{2}{3} \varepsilon\bigr)^2 \Gamma(2-2\varepsilon) \Gamma(1+u)}%
{(1-\varepsilon) (1-2u) \Gamma^2(1-\varepsilon) \Gamma(1+\varepsilon) \Gamma(2+u-\varepsilon)}\,,
\end{align*}
see~(\ref{abel:wt}).
Choosing $\mu = (2/\tau) e^{-\gamma} D(\varepsilon)^{1/(2\varepsilon)} \to (2/\tau) e^{-\gamma-5/6}$ we obtain
\begin{equation*}
\bar{V}(\tau,\tau,\varphi) = \frac{1}{\beta_0} \sum_{L=1}^\infty \frac{f(\varepsilon,L\varepsilon)}{L}
\biggl(\frac{b}{\varepsilon+b}\biggr)^{\!L} + \mathcal{O}\biggl(\frac{1}{\beta_0^2}\biggr)\,;
\end{equation*}
$f(-b,0,\varphi) = \tilde{f}(-b,0,\varphi)$,
therefore we get the result~(\ref{Lb0:Gamma0}) again.

\subsection{Potential and conformal anomaly}

Now we consider the potential $V(\vec{q}^{\,})$ at the L$\beta_0$ order.
Choosing $\mu=|\vec{q}^{\,}|$ we have~(\ref{abel:VS})
\begin{equation*}
V(\vec{q}^{\,}) = - \frac{(4\pi)^{d/2} e^{\gamma \varepsilon}}{\beta_0 D(\varepsilon) (\vec{q}^{\,2})^{1-\varepsilon}}
\varepsilon \sum_{L=1}^\infty \biggl(D(\varepsilon) \frac{b}{\varepsilon+b}\biggr)^{\!L}
+ \mathcal{O}\biggl(\frac{1}{\beta_0^2}\biggr)\,.
\end{equation*}
The sum here can be written as
\begin{equation*}
\sum_{L=1}^\infty g(\varepsilon,L\varepsilon) \biggl(\frac{b}{\varepsilon+b}\biggr)^{\!L}\,,\quad
g(\varepsilon,u) = D(\varepsilon)^{u/\varepsilon} = \sum_{n,m=0}^\infty g_{nm} \varepsilon^n u^m\,.
\end{equation*}
This sum is equal to
\begin{equation*}
\frac{b}{\varepsilon} \sum_{n=0}^\infty n!\,g_{0n} b^n + \mathcal{O}(\varepsilon^0)
\end{equation*}
($1/\varepsilon^n$ terms with $n>1$ vanish, so that $V(\vec{q}^{\,})$ is automatically finite),
where
\begin{equation}
g(0,u) = e^{\frac{5}{3}u}\,,\quad
g_{0n} = \frac{1}{n!} \biggl(\frac{5}{3}\biggr)^{\!n}\,.
\label{Lb0:gu}
\end{equation}
Therefore
\begin{equation}
V(\vec{q}^{\,}) = - \frac{(4\pi)^2}{\vec{q}^{\,2}} \frac{b}{\beta_0} V_0(b)
+ \mathcal{O}\biggl(\frac{1}{\beta_0^2}\biggr)\,,\quad
V_0(b) = \frac{1}{1 - \frac{5}{3} b}\,.
\label{Lb0:V}
\end{equation}
The conformal anomaly~(\ref{pi:beta}) at the L$\beta_0$ order is
\begin{align}
&C = \frac{b^2}{\beta_0} C_0(b) + \mathcal{O}\biggl(\frac{1}{\beta_0^2}\biggr)\,,
\nonumber\\
&C_0(b) = \frac{V_0(b) - \Gamma_0(b)}{b^2}
= \frac{28}{9}
+ 2 \biggl( \zeta_3 + \frac{58}{27} \biggr) b
- \biggl( \frac{\pi^4}{10} - 10 \zeta_3 - \frac{652}{27} \biggr) \frac{b^2}{3}
\nonumber\\
&{} + \biggl(6 \zeta_5 - \frac{\pi^4}{18} - \frac{2}{3} \zeta_3 + \frac{3044}{243}\biggr) b^3
- \biggl(2 \zeta_3^2 + \frac{2}{189} \pi^6 - 10 \zeta_5 - \frac{\pi^4}{90} - \frac{2}{3} \zeta_3 - \frac{15868}{729}\biggr) b^4
\nonumber\\
&{} + \cdots
\label{Lb0:C}
\end{align}
It is absolutely trivial to extend this expansion to any desired order.
Restoring the color factors~(\ref{Lb0:col}),
we reproduce the $C_R T_F n_f \alpha_s^2$ and $C_R (T_F n_f)^2 \alpha_s^3$ terms in~(\ref{pi:C}).

\section{Next-to-leading large-$\beta_0$ order}
\label{S:NLb0}

Now let's add to $S$ abelian terms with next-to-leading powers of $n_f$:
$\{C_F (T_F n_f)^{L-1}, L \ge 2\}$.
It is sufficient to consider QED in the large-$\beta_0$ limit,
but now we add the first $1/\beta_0$ correction.

To obtain the photon propagator with the next-to-leading large $\beta_0$ (NL$\beta_0$) accuracy,
we need the photon self-energy up to $1/\beta_0$:
\begin{align}
&\raisebox{-6.5mm}{\includegraphics{pi0.pdf}}
+ 2 \raisebox{-6.5mm}{\includegraphics{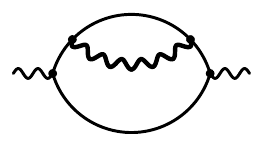}}
+ \raisebox{-6.5mm}{\includegraphics{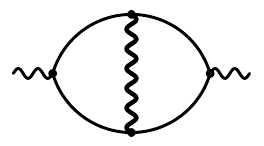}}
\nonumber\\
&{} \Rightarrow \Pi_0(k^2) + \frac{\Pi_1(k^2)}{\beta_0}
+ \mathcal{O}\biggl(\frac{1}{\beta_0^2}\biggr)\,,
\label{NLb0:Pi}
\end{align}
where the photon propagators in $\Pi_1$ are taken at the L$\beta_0$ order.
The NL$\beta_0$ contribution can be written in the form\cite{Palanques-Mestre:1983ogz,Broadhurst:1992si}
\begin{equation}
\Pi_1(k^2) = 3 \varepsilon \sum_{L=2}^\infty \frac{F(\varepsilon,L\varepsilon)}{L} \Pi_0(k^2)^L\,.
\label{NLb0:Pi1}
\end{equation}
Using integration by parts, one can reduce it to
\begin{align}
&F(\varepsilon,u) =
\frac{2 (1-2\varepsilon)^2 (3-2\varepsilon) \Gamma^2(1-2\varepsilon)}%
{9 (1-\varepsilon) (1-u) (2-u) \Gamma^2(1-\varepsilon) \Gamma^2(1+\varepsilon)}
\nonumber\\
&{}\times\biggl[ - u
\frac{2-3\varepsilon-\varepsilon^2 + \varepsilon(2+\varepsilon)u - \varepsilon u^2}{\Gamma^2(1-\varepsilon)}
I(1+u-2\varepsilon)
\nonumber\\
&\hphantom{{}\times\biggl[\biggr.}{} + 2
\frac{2(1+\varepsilon)(3-2\varepsilon) - (4+11\varepsilon-7\varepsilon^2)u + \varepsilon(8-3\varepsilon)u^2 - \varepsilon u^3}%
{(1-u) (2-u) (1-u-\varepsilon) (2-u-\varepsilon)}
\nonumber\\
&\hphantom{{}\times\biggl[\biggr.+2}{} \times
\frac{\Gamma(1+u) \Gamma(1-u+\varepsilon)}{\Gamma(1-u-\varepsilon) \Gamma(1+u-2\varepsilon)}
\biggr]
= \sum_{n,m=0}^\infty F_{nm} \varepsilon^n u^m\,,
\label{NLb0:Feu}
\end{align}
where the integral
\begin{equation*}
I(n) =
\raisebox{-5.5mm}{\begin{picture}(24,12.5)
\put(12,6.25){\makebox(0,0){\includegraphics{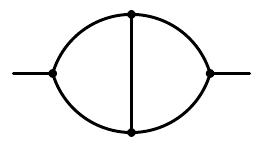}}}
\put(13.5,6.25){\makebox(0,0){$n$}}
\end{picture}}
= \frac{1}{\pi^d} \int \frac{d^d k_1\,d^d k_2}%
{k_1^2 k_2^2 (k_1+p)^2 (k_2+p)^2 \left[(k_1-k_2)^2\right]^n}
\end{equation*}
(euclidean, $p^2=1$) can be expressed via a ${}_3F_2$ function of unit argument\cite{Kotikov:1995cw,Broadhurst:1996ur}
(see the review\cite{Grozin:2012xi} for more references).
The ${}_3F_2$ function can be expanded up to any desired order using known algorithms,
the coefficients are expressed via multiple $\zeta$ values;
therefore, the coefficients $F_{nm}$ can be calculated to any desired order.

The function $F(\varepsilon,u)$ simplifies in some cases.
In particular\cite{Palanques-Mestre:1983ogz},
\begin{equation}
F(\varepsilon,0) =
\frac{(1+\varepsilon) (1-2\varepsilon)^2 (1-\frac{2}{3}\varepsilon)^2 \Gamma(1-2\varepsilon)}%
{(1-\varepsilon)^2 (1-\frac{1}{2}\varepsilon) \Gamma(1+\varepsilon) \Gamma^3(1-\varepsilon)}\,,
\label{NLb0:Fe0}
\end{equation}
so that $F_{n0}$ contain no multiple $\zeta$ values, only $\zeta_n$.
Also\cite{Broadhurst:1992si}
\begin{equation}
F(0,u) = \frac{2}{3}
\frac{\psi'\bigl(2-\frac{u}{2}\bigr) - \psi'\bigl(1+\frac{u}{2}\bigr)
- \psi'\bigl(\frac{3-u}{2}\bigr) + \psi'\bigl(\frac{1+u}{2}\bigr)}%
{(1-u) (2-u)}\,,
\label{NLb0:F0u}
\end{equation}
so that $F_{0m}$ contains\cite{Broadhurst:1992si} only $\zeta_{2n+1}$:
\begin{align}
F_{0m} ={}& - \frac{32}{3} \sum_{s=1}^{[(m+1)/2]} s \left(1 - 2^{-2s}\right) \left(1 - 2^{2s-m-2}\right) \zeta_{2s+1}
\nonumber\\
&{} + \frac{4}{3} (m+1) \left(m + (m+6) 2^{-m-3}\right)\,.
\label{NLb0:F0m}
\end{align}
The two-loop case is, of course, trivial:
\begin{align*}
F(\varepsilon,2\varepsilon) = \frac{2}{9\varepsilon^2} \frac{3-2\varepsilon}{1-\varepsilon}
\biggl[& 2 \frac{(1-2\varepsilon)^2 (2-2\varepsilon+\varepsilon^2)}{(1-3\varepsilon) (2-3\varepsilon)}
\frac{\Gamma(1+2\varepsilon) \Gamma^2(1-2\varepsilon)}{\Gamma^2(1+\varepsilon) \Gamma(1-\varepsilon) \Gamma(1-3\varepsilon)}\\
&{} - 2 + \varepsilon - 2 \varepsilon^2 \biggr]\,.
\end{align*}

Let's write the charge renormalization constant $Z_\alpha$ with the NL$\beta_0$ accuracy as
\begin{align}
&Z_\alpha(b) = \frac{1}{1+b/\varepsilon}
\biggl[1 + \frac{Z_{\alpha1}(b)}{\beta_0} + \mathcal{O}\biggl(\frac{1}{\beta_0^2}\biggr) \biggr]\,,
\nonumber\\
&Z_{\alpha1}(b) = \frac{Z_{\alpha11}(b)}{\varepsilon} + \frac{Z_{\alpha12}(b)}{\varepsilon^2} + \cdots\,,\quad
Z_{\alpha1n} = \mathcal{O}(b^{n+1})\,.
\label{NLb0:Za}
\end{align}
In the abelian theory, $\log(1-\Pi)$ expressed~(\ref{Lb0:MS}) via renormalized $b$
should be equal to $\log Z_\alpha + \mathcal{O}(\varepsilon^0)$.
Equating the coefficients of $\varepsilon^{-1}$ in the $1/\beta_0$ terms in this relation,
we see that $Z_{\alpha11}$~(\ref{NLb0:Za}) is given by the coefficient of $\varepsilon^{-1}$ in
\begin{equation*}
- \biggl(1 + \frac{b}{\varepsilon}\biggr) \Pi_1\,.
\end{equation*}
It is convenient to choose $\mu^2 = (-k^2) D(\varepsilon)^{-1/\varepsilon} \to (-k^2) e^{-5/3}$,
then
\begin{equation*}
\Pi_1 = 3 \varepsilon \sum_{L=2}^\infty \frac{F(\varepsilon,L\varepsilon)}{L}
\biggl(\frac{b}{\varepsilon+b}\biggr)^{\!L}\,.
\end{equation*}
We expand in $b$ and expand $F(\varepsilon,u)$ in $\varepsilon$ and $u$;
selecting $\varepsilon^{-1}$ terms, we find that all coefficients but $F_{n0}$ cancel:
\begin{equation}
Z_{\alpha11} = - 3 \sum_{n=0}^\infty \frac{F_{n0} (-b)^{n+2}}{(n+1) (n+2)}\,.
\label{NLb0:Z1}
\end{equation}
The $\beta$ function with NL$\beta_0$ accuracy is
\begin{equation}
\beta(b) = b + \frac{b^2}{\beta_0} B_1(b) + \mathcal{O}\biggl(\frac{1}{\beta_0^2}\biggr)\,,
\label{NLb0:beta}
\end{equation}
where\cite{Palanques-Mestre:1983ogz,Broadhurst:1992si}
\begin{align}
&b^2 B_1(b) = - \frac{d Z_{\alpha11}(b)}{d\log b}\,,\quad
B_1(b) = 3 \sum_{n=0}^\infty \frac{F_{n0} (-b)^n}{n+1}
\nonumber\\
&{} = 3 + \frac{11}{4} b - \frac{77}{36} b^2
- \biggl(3 \zeta_3 + \frac{107}{48}\biggr) \frac{b^3}{2}
+ \biggl(\frac{\pi^4}{10} - 11 \zeta_3 + \frac{251}{48}\biggr) \frac{b^4}{5}
+ \cdots
\label{NLb0:beta1}
\end{align}
(the coefficients $F_{n0}$ follow from $F(\varepsilon,0)$~(\ref{NLb0:Fe0})).
The corresponding terms in the 5-loop QED $\beta$ function\cite{Baikov:2012zm} are reproduced.
We shall need the full $Z_{\alpha1}$, not just $Z_{\alpha11}$;
integrating the RG equation with the $1/\beta_0$ accuracy we obtain
\begin{equation*}
Z_{\alpha1}(b) = - \varepsilon \int_0^b \frac{b B_1(b)\,d b}{(\varepsilon+b)^2}
= - \frac{3}{2} \frac{b^2}{\varepsilon}
+ \frac{1}{2} \left(4 + F_{10} \varepsilon \right) \frac{b^3}{\varepsilon^2}
- \frac{1}{4} \left(9 + 3 F_{10} \varepsilon + F_{20} \varepsilon^2 \right) \frac{b^4}{\varepsilon^3}
+ \cdots
\end{equation*}

\subsection{HQET field anomalous dimension}

\begin{figure}[ht]
\begin{center}
\begin{picture}(100,26.5)
\put(23,15.25){\makebox(0,0){\includegraphics{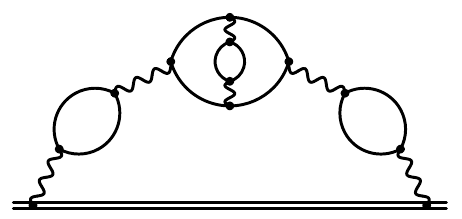}}}
\put(77,13.25){\makebox(0,0){\includegraphics{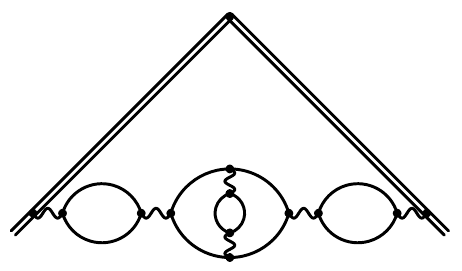}}}
\end{picture}
\end{center}
\caption{Typical diagrams for $\gamma_h$ and $\Gamma(\varphi)$ at NL$\beta_0$.}
\label{F:NLb0}
\end{figure}

At the NL$\beta_0$ order we should expand
the photon propagator $(1-\Pi_0-\Pi_1/\beta_0)^{-1}$ up to $1/\beta_0$ (Fig.~\ref{F:NLb0}).
There is a single $\Pi_1$ insertion and any number of $\Pi_0$ insertions in the photon propagator;
the photon propagator inside $\Pi_1$ contains any number of $\Pi_0$ insertions.
The 2-leg web $\tilde{w}(\omega)$ becomes
\begin{align}
&\tilde{w}(\omega)
= \frac{1}{\beta_0} \sum_{L=1}^\infty \frac{\tilde{f}(\varepsilon,L\varepsilon)}{L}
\biggl(\frac{b}{\varepsilon+b}\biggr)^{\!L}
\nonumber\\
&{}\times\biggl[1 + L \frac{Z_{\alpha1}}{\beta_0}
+ \frac{3\varepsilon}{\beta_0} \sum_{L'=2}^{L-1} \frac{L-L'}{L'} F(\varepsilon,L'\varepsilon) \biggr]
+ \mathcal{O}\biggl(\frac{1}{\beta_0^3}\biggr)\,,
\label{NLb0:ww}
\end{align}
where $L'$ is the number of loops in the $\Pi_1$ insertion,
and the $1/\beta_0$ correction $Z_{\alpha1}$ to the charge renormalization~(\ref{NLb0:Za})
is taken into account.
We expand in $b$ and substitute the expansions~(\ref{NLb0:Feu}) and~(\ref{Lb0:feuexp});
in $z_{h1}$, the coefficient of $\varepsilon^{-1}$, all $\tilde{f}_{nm}$ except $\tilde{f}_{n0}$ cancel.
We obtain\cite{Grozin:2016ydd}
\begin{align}
&\gamma_h(b) = - 6 \biggl[ \frac{b}{\beta_0} \gamma_0(b) - \frac{b^3}{\beta_0^2} \gamma_1(b) \biggr]
+ \mathcal{O}\biggl(\frac{1}{\beta_0^3}\biggr)\,,
\nonumber\\
&\gamma_1(b) = - \frac{3}{2} \bigl[F_{10} + 2 F_{01} - 2 \tilde{f}_{10} \bigr]
+ \bigl[2 F_{20} + 3 (F_{11} + F_{02}) + 3 F_{01} \tilde{f}_{10} - 6 \tilde{f}_{20} \bigr] b
\nonumber\\
&{}
- \biggl[ \frac{3}{4} (3 F_{30} + 4 (F_{21} + F_{12} + F_{03}))
+ (F_{20} + 3 (F_{11} + F_{02})) \tilde{f}_{10}
\nonumber\\
&\quad{}
- \frac{3}{2} \bigl(F_{10} - 2 F_{01}\bigr) \tilde{f}_{20}
- 9 \tilde{f}_{30} \biggr] b^2
+ \cdots
\nonumber\\
&{} = 3 \biggl( 4 \zeta_3 - \frac{17}{4} \biggr)
+ \biggl( - \frac{\pi^4}{5} + 36 \zeta_3 - \frac{103}{9} \biggr) b
\nonumber\\
&{} + \biggl( 24 \zeta_5 - \frac{3}{5} \pi^4 + \frac{59}{2} \zeta_3 + \frac{14579}{864} \biggr) b^2
\nonumber\\
&{} + \biggl( - 48 \zeta_3^3 - \frac{2}{63} \pi^6 + 72 \zeta_5 - \frac{44}{75} \pi^4 + \frac{3229}{45} \zeta_3 - \frac{5191}{540} \biggr) b^3
\nonumber\\
&{} + \biggl( 36 \zeta_7 + \frac{8}{5} \pi^4 \zeta_3 - 144 \zeta_3^2 - \frac{2}{21} \pi^6 + 107 \zeta_5 - \frac{946}{675} \pi^4
+ \frac{9601}{180} \zeta_3 + \frac{22859}{8640} \biggr) b^4
\nonumber\\
&{} + \biggl( - 240 \zeta_3 \zeta_5 - \frac{4}{225} \pi^8 + 108 \zeta_7 + \frac{24}{5} \pi^4 \zeta_3 - \frac{664}{7} \zeta_3^2  - \frac{272}{1323} \pi^6
\nonumber\\
&\hphantom{{}+\biggl(\biggr.}{}
+ \frac{18574}{63} \zeta_5 - \frac{119}{135} \pi^4 - \frac{6263}{63} \zeta_3 + \frac{16103}{1296}
\biggr) b^5 + \cdots
\label{NLb0:gammah}
\end{align}
This expansion can be extended to higher loops,
but the complexity of of expanding $F(\varepsilon,u)$ in $\varepsilon$, $u$ to obtain $F_{nm}$
quickly grows with $n+m$, so, this extension is not quite trivial.
Note that the last (8-loop) term here contains $F_{nm}$ with $n+m=6$, $n>0$, $m>0$,
each of them contains $\zeta_{5,3}$;
but they enter as the combination $F_{51}+F_{42}+F_{33}+F_{24}+F_{15}$ in which this $\zeta_{5,3}$ cancels.
Restoring the color factors,
we reproduce the corresponding terms in ~(\ref{Gamma:gamma}).

\subsection{Cusp anomalous dimension}

The 2-leg web $\tilde{\bar{V}}(\omega,\omega,\varphi)$
is given by a formula similar to~(\ref{NLb0:ww}).
The cusp anomalous dimension at NL$\beta_0$ order is determined by the same coefficients $\hat{f}_n$
as at L$\beta_0$, plus the coefficients $F_{nm}$.
We obtain\cite{Grozin:2016ydd}
\begin{align}
&\Gamma(b,\varphi) = 4 (\varphi \cot\varphi - 1) \biggl[ \frac{b}{\beta_0} \Gamma_0(b) - \frac{b^3}{\beta_0^2} \Gamma_1(b) \biggr]
+ \mathcal{O}\biggl(\frac{1}{\beta_0^3}\biggr)\,,
\nonumber\displaybreak\\
&\Gamma_1(b) = 12 \zeta_3 - \frac{55}{4}
+ \biggl( - \frac{\pi^4}{5} + 40 \zeta_3 - \frac{299}{18} \biggr) b
\nonumber\\
&{}+ \biggl( 24 \zeta_5 - \frac{2}{3} \pi^4 + \frac{233}{6} \zeta_3 + \frac{15211}{864} \biggr) b^2
\nonumber\\
&{}+ \biggl( - 48 \zeta_3^2 - \frac{2}{63} \pi^6 + 80 \zeta_5 - \frac{167}{225} \pi^4 + \frac{1168}{15} \zeta_3 - \frac{971}{240} \biggr) b^3
\nonumber\\
&{}+ \biggl( 36 \zeta_7 + \frac{8}{5} \pi^4 \zeta_3 - 160 \zeta_3^2 - \frac{20}{189} \pi^6 + \frac{377}{3} \zeta_5 - \frac{23}{15} \pi^4
+ \frac{929}{12} \zeta_3 - \frac{8017}{1728} \biggr) b^4
\nonumber\\
&{}+ \biggl( - 240 \zeta_3 \zeta_5 - \frac{4}{225} \pi^8 + 120 \zeta_7 + \frac{16}{3} \pi^4 \zeta_3 - \frac{2776}{21} \zeta_3^2 - \frac{914}{3969} \pi^6
\nonumber\\
&\hphantom{{}+\biggl(\biggr.}{}
+ \frac{6826}{21} \zeta_5 - \frac{1793}{1350} \pi^4 - \frac{31693}{315} \zeta_3 + \frac{79433}{4320} \biggr) b^5
+ \cdots
\label{NLb0:Gamma}
\end{align}
This expansion also can be extended to higher loops,
but the complexity of calculations quickly grows.
At 8 loops the same combination of $F_{nm}$ with $n+m=6$ appears,
so that $\zeta_{5,3}$ cancels here, too.
Restoring the color factors,
we reproduce the corresponding terms in ~(\ref{Cusp:QCD}).

\subsection{Potential and conformal anomaly}

The static potential at the NL$\beta_0$ level is
\begin{align}
V(\vec{q}^{\,}) &{}= - \frac{(4\pi)^2}{\beta_0 \vec{q}^{\,2}} \varepsilon
\sum_{L=1}^\infty g(\varepsilon,L\varepsilon) \biggl(\frac{b}{\varepsilon+b}\biggr)^{\!L}
\biggl[1 + L \frac{Z_{\alpha1}}{\beta_0} + \frac{3 \varepsilon}{\beta_0}
\sum_{L'=2}^{L-1} \frac{L-L'}{L'} F(\varepsilon,L'\varepsilon) \biggr]
\nonumber\\
&{} + \mathcal{O}\biggl(\frac{1}{\beta_0^3}\biggr)
= - \frac{(4\pi)^2}{\vec{q}^{\,2}}
\biggl[ \frac{b}{\beta_0} V_0(b) - \frac{b^3}{\beta_0^2} V_1(b) \biggr]
+ \mathcal{O}\biggl(\frac{1}{\beta_0^3}\biggr)\,,
\label{NLb0:V}
\end{align}
where
\begin{align*}
&V_1(b) = - \frac{3}{2} \left[ F_{10} + 2 F_{01} + 2 g_{01} \right]
+ \frac{1}{2} \left[ F_{20} - 6 F_{02} - 6 \left( F_{10} + 3 F_{01} \right) g_{01} - 30 g_{02} \right] b\\
&{} - \frac{1}{4} \bigl[ F_{30} + 24 F_{03} - 4 \left( F_{20} + 12 F_{02} \right) g_{01}
+ 36 \left( F_{10} + 4 F_{01} \right) g_{02} + 312 g_{03} \bigr] b^2 + \cdots
\end{align*}
contains only the same coefficients $g_{0n}$~(\ref{Lb0:gu}) as the L$\beta_0$ result,
and only $F_{n0}$ and $F_{0m}$ are involved (see~(\ref{NLb0:Fe0}--\ref{NLb0:F0m})).
We obtain\cite{Grozin:2016ydd}
\begin{align}
&V_1(b) = 12 \zeta_3 - \frac{55}{4}
+ \biggl( 78 \zeta_3 - \frac{7001}{72} \biggr) b
+ \biggl( 60 \zeta_5 + \frac{723}{2} \zeta_3 - \frac{147851}{288} \biggr) b^2
\nonumber\\
&{} + \biggl( 770 \zeta_5 + \frac{\pi^4}{200} + \frac{276901}{180} \zeta_3 - \frac{70418923}{25920} \biggr) b^3
\nonumber\\
&{} + \biggl( 1134 \zeta_7 + \frac{32297}{5} \zeta_5 + \frac{41}{1800} \pi^4 + \frac{402479}{60} \zeta_3
- \frac{1249510621}{77760} \biggr) b^4
\nonumber\\
&{} + \biggl( 21735 \zeta_7 + \frac{\zeta_3^2}{7} + \frac{\pi^6}{1323} + \frac{5911849}{126} \zeta_5 + \frac{41}{720} \pi^4
+ \frac{48558187}{1512} \zeta_3 - \frac{10255708489}{93312} \biggr) b^5
\nonumber\\
&{} + \cdots
\label{NLb0:V1}
\end{align}
Thus we have reproduced the $C_F (T_F n_f)^2 \alpha_s^3$ and $C_F^2 T_F n_f \alpha_s^3$ terms in the two-loop potential\cite{Schroder:1998vy},
as well as the $C_F (T_F n_f)^3 \alpha_s^4$ and $C_F^2 (T_F n_f)^2 \alpha_s^4$ terms in the three-loop one\cite{Smirnov:2008pn}.
This expansion can be easily extended to any order;
it contains only $\zeta_n$ because only $F_{n0}$~(\ref{NLb0:Fe0}) and $F_{0m}$~(\ref{NLb0:F0m}) are present.
Note the pattern of the highest weights in~(\ref{NLb0:V1}): 3, 3, 5, 5, 7, 7,
whereas one would expect 3, 4, 5, 6, 7, 8, as in~(\ref{NLb0:Gamma}), (\ref{NLb0:gammah}).

Constructing $\Delta(\alpha_s)$~(\ref{pi:Delta}) from $\Gamma(\varphi)$~(\ref{NLb0:Gamma}) and $V(\vec{q}^{\,})$~(\ref{NLb0:V})
and dividing by $\beta(\alpha_s)$~(\ref{NLb0:beta}), we obtain $C(\alpha_s)$~(\ref{pi:beta}):
\begin{align}
&C = \frac{b^2}{\beta_0} C_0(b) - \frac{b^3}{\beta_0^2} C_1(b)
+ \mathcal{O}\biggl(\frac{1}{\beta_0^3}\biggr)\,,
\nonumber\\
&C_1(b) = \frac{\pi^4}{5} + 38 \zeta_3 - \frac{1711}{24}
+ \biggl( 36 \zeta_5 + \frac{2}{3} \pi^4 + \frac{986}{3} \zeta_3 - \frac{110059}{216} \biggr) b
\nonumber\\
&{} + \biggl( 48 \zeta_3^2 + \frac{2}{63} \pi^6 + 690 \zeta_5 + \frac{233}{360} \pi^4
+ \frac{53135}{36} \zeta_3 - \frac{13910875}{5184} \biggr) b^2
\nonumber\\
&{} + \biggl( 1098 \zeta_7 - \frac{8}{5} \pi^4 \zeta_3 + 160 \zeta_3^2 + \frac{20}{189} \pi^6 + \frac{95276}{15} \zeta_5 + \frac{292}{225} \pi^4
+ \frac{596591}{90} \zeta_3 - \frac{51895439}{3240} \biggr) b^3
\nonumber\\
&{} + \biggl( 240 \zeta_3 \zeta_5 + \frac{4}{225} \pi^8 + 21615 \zeta_7 - \frac{16}{3} \pi^4 \zeta_3
+ \frac{370}{3} \zeta_3^2 + \frac{113}{567} \pi^6
\nonumber\\
&\hphantom{{}+\biggl(\biggr.}{}
+ \frac{419768}{9} \zeta_5 + \frac{1679}{1200} \pi^4 + \frac{23179201}{720} \zeta_3 - \frac{51249331081}{466560} \biggr) b^4
+ \cdots
\label{NLb0:C}
\end{align}
The first term in $C_1$ reproduces the $C_R C_F T_F n_f \alpha_s^3$ term in~(\ref{pi:C}).

\section{Abelian terms with $(T_F n_f)^1$}
\label{S:CF}

Finally, we consider abelian color structures linear in $T_F n_f$:
$S = \{1\} \cup \{C_L^{L-1} T_F n_f,L\ge1\}$.
We can work in QED.
Writing the $L$-loop photon self-energy as
\begin{equation*}
\Pi_{L-1} = \tilde{\Pi}_{L-1} n_f + (n_f^{>1}\text{ terms})\,,
\end{equation*}
we have the photon propagator
\begin{equation}
D^{\mu\nu}(k) = D_0^{\mu\nu}(k) + n_f \sum_{L=1}^\infty \tilde{\Pi}_{L-1} D_L^{\mu\nu}(k) A_0^L
+ (n_f^{>1}\text{ terms})\,.
\label{CF:D}
\end{equation}
In QED $\log\left(1 - \Pi(k^2)\right) = \log Z_\alpha + \mathcal{O}(\varepsilon^0)$;
writing the $L$-loop $\beta$-function coefficient
as $\beta_{L-1} = \bar{\beta}_{L-1} n_f + (n_f^{>1} \text{ terms})$,
we see that $1/\varepsilon$ terms in $\tilde{\Pi}_{L-1}$ are related to $\bar{\beta}_{L-1}$:
\begin{equation}
\tilde{\Pi}_{L-1} = \frac{\bar{\beta}_{L-1}}{L\varepsilon} + \bar{\Pi}_{L-1}
+ \mathcal{O}(\varepsilon)\,.
\label{CF:Pieps}
\end{equation}
Here the $\beta$ function coefficients are\cite{Gorishnii:1990kd}
\begin{equation}
\bar{\beta}_0 = - \frac{4}{3}\,,\quad
\bar{\beta}_1 = - 4\,,\quad
\bar{\beta}_2 = 2\,,\quad
\bar{\beta}_3 = 46\,,
\label{CF:beta}
\end{equation}
and\cite{Ruijl:2017eht}
\begin{align}
&\bar{\Pi}_0 = - \frac{20}{9}\,,\quad
\bar{\Pi}_1 = 16 \zeta_3 - \frac{55}{3}\,,\quad
\bar{\Pi}_2 = - 2 \biggl(80 \zeta_5 - \frac{148}{3} \zeta_3 - \frac{143}{9}\biggr)\,,
\nonumber\\
&\Bar{\Pi}_3 = 2240 \zeta_7 - 1960 \zeta_5 -104 \zeta_3 + \frac{31}{3}\,.
\label{CF:Pi}
\end{align}

\subsection{HQET field anomalous dimension}

The web $\tilde{w}(\omega)$~(\ref{abel:ww}) is
\begin{align}
&\tilde{w}(\omega) = \tilde{w}_1 A_0 (-2\omega)^{-2\varepsilon}
+ n_f \sum_{L=2}^\infty \tilde{\Pi}_{L-2} \tilde{w}_L \bigl[A_0 (-2\omega)^{-2\varepsilon}\bigr]^L
\nonumber\\
&{} + (n_f^{>1} \text{ terms}) + (w_{>2 \text{ legs}} \text{ terms})
\label{CF:ww}\,,
\end{align}
where~(\ref{abel:ww})
\begin{equation}
\tilde{w}_L = \frac{3}{L\varepsilon} + \frac{1}{L} + 3 + \mathcal{O}(\varepsilon)\,.
\label{CF:wL}
\end{equation}
We re-express~(\ref{abel:MS}) $\tilde{w}(\omega)$ via $\alpha(-2\omega)$:
\begin{equation*}
Z_\alpha = 1 - \frac{n_f}{\varepsilon} \sum_{L=1}^\infty \frac{\bar{\beta}_{L-1}}{L}
\biggl(\frac{\alpha}{4\pi}\biggr)^{\!L}
+ (n_f^{>1} \text{ terms})
\end{equation*}
(it is sufficient to include $Z_\alpha$ in the $\tilde{w}_1$ term).
Collecting $\varepsilon^{-1}$ terms we have
\begin{align*}
&z_{h1} = 3 \frac{\alpha}{4\pi}
+ n_f \sum_{L=2}^\infty \frac{3 \bar{\Pi}_{L-2} - \bar{\beta}_{L-2}}{L}
\biggl(\frac{\alpha}{4\pi}\biggr)^{\!L}\\
&{} + (n_f^{>1} \text{ terms}) + (w_{>2 \text{ legs}} \text{ terms})\,.
\end{align*}
Restoring color factors we finally obtain\cite{Grozin:2018vdn}
\begin{align}
\gamma_h ={}& - 2 C_R \frac{\alpha_s}{4\pi} \biggl[3
+ T_F n_f \frac{\alpha_s}{4\pi} 
\sum_{L=0}^\infty (3 \bar{\Pi}_L - \bar{\beta}_L) \biggl(C_F \frac{\alpha_s}{4\pi}\biggr)^{\!L} \biggr]
+ \cdots
\nonumber\\
={}& - 2 C_R \frac{\alpha_s}{4\pi} \biggl\{3
+ T_F n_f \frac{\alpha_s}{4\pi} \biggl[- \frac{16}{3}
+ 3 (16 \zeta_3 - 17) C_F \frac{\alpha_s}{4\pi}
\nonumber\\
&{} - 8 \biggl(60 \zeta_5 - 37 \zeta_3 - \frac{35}{3}\biggr) \biggl(C_F \frac{\alpha_s}{4\pi}\biggr)^{\!2}
\nonumber\\
&{} + 3 (2240 \zeta_7 - 1960 \zeta_5 - 104 \zeta_3 - 5) \biggl(C_F \frac{\alpha_s}{4\pi}\biggr)^{\!3}
+ \mathcal{O}(\alpha_s^4)\biggr]\biggr\} + \cdots\,,
\label{CF:gamma}
\end{align}
where dots mean other color structures.
The corresponding terms from~(\ref{Gamma:gamma}) are reproduced.
This expansion cannot be extended without a highly non-trivial calculation
of higher $\bar{\Pi}_L$ and $\bar{\beta}_L$.

This result can also be obtained in coordinate space:
$w_L$~(\ref{abel:wt}) has the structure identical to~(\ref{CF:wL}),
and if we express $w(\tau)$ via $\alpha((2/\tau) e^{-\gamma})$,
the calculation is exactly the same as above.

\subsection{Cusp anomalous dimension}

The web $\tilde{\bar{V}}(\varphi)$~(\ref{abel:Vw}) is
\begin{equation}
\tilde{\bar{V}}(\varphi) = - 2 \frac{\varphi \coth\varphi - 1}{L\varepsilon} + V(\varphi)
+ \mathcal{O}(\varepsilon)\,,
\label{CF:Vw}
\end{equation}
where $V(\varphi) = V(-\varphi)$ does not depend on $L$
and hence $\bar{\beta}_{L-2}$ cancels in the calculation of $z_{J1}$
(in contrast to $z_{h1}$ where the corresponding term in~(\ref{CF:wL}) contains $1/L$,
and $\bar{\beta}_{L-2}$ does not cancel):
\begin{align*}
&z_{J1} = - 2 (\varphi \coth\varphi - 1) \frac{\alpha}{4\pi} \biggl[1
+ n_f \frac{\alpha}{4\pi} \sum_{L=0}^\infty \frac{\bar{\Pi}_L}{L}
\biggl(\frac{\alpha}{4\pi}\biggr)^{\!L}\biggr]\\
&{} + (n_f^{>1} \text{ terms}) + (w_{>2 \text{ legs}} \text{ terms})\,.
\end{align*}
Restoring color factors we obtain\cite{Grozin:2018vdn}
\begin{align}
&\Gamma(\varphi) = 4 C_R (\varphi \coth\varphi - 1) \frac{\alpha_s}{4\pi} \biggl[1
+ T_F n_f \frac{\alpha_s}{4\pi} \sum_{L=0}^\infty \bar{\Pi}_L \biggl(C_F \frac{\alpha_s}{4\pi}\biggr)^{\!L} \biggr]
+ \cdots
\nonumber\\
&{} = 4 C_R (\varphi \coth\varphi - 1) \frac{\alpha_s}{4\pi} \biggl\{1
+ T_F n_f \frac{\alpha_s}{4\pi} \biggl[
- \frac{20}{9}
+ \biggl(16 \zeta_3 - \frac{55}{3}\biggr) C_F \frac{\alpha_s}{4\pi}
\nonumber\\
&\quad{} - 2 \biggl(80 \zeta_5 - \frac{148}{3} \zeta_3 - \frac{143}{9}\biggr) \biggl(C_F \frac{\alpha_s}{4\pi}\biggr)^{\!2}
\nonumber\\
&\quad{} + \biggl(2240 \zeta_7 - 1960 \zeta_5 - 104 \zeta_3 + \frac{31}{3}\biggr) \biggl(C_F \frac{\alpha_s}{4\pi}\biggr)^{\!3}
+ \mathcal{O}(\alpha_s^4) \biggr] \biggr\} + \cdots\,,
\label{CF:Gamma}
\end{align}
where dots mean other color structures.
The corresponding terms from~(\ref{Cusp:QCD}) are reproduced.
This expansion cannot be extended without a highly non-trivial calculation
of higher $\bar{\Pi}_L$ and $\bar{\beta}_L$.

This result can also be obtained in coordinate space:
$\bar{V}(\varphi)$~(\ref{abel:VL}) has the structure identical to~(\ref{CF:Vw}),
and if we express $\bar{V}(\tau,\tau,\varphi)$ via $\alpha((2/\tau) e^{-\gamma})$,
the calculation is exactly the same as above.

\subsection{Potential and conformal anomaly}

Now we consider the $C_F^{L-1} T_F n_f \alpha_s^L$ terms in the quark--antiquark potential.
In Coulomb gauge they are given by a single Coulomb-gluon propagator\cite{Grozin:2018vdn}:
\begin{align}
&V(\vec{q}^{\,}) = - C_R \frac{4 \pi \alpha_s}{\vec{q}^{\,2}}
\biggl[1 + T_F n_f \frac{\alpha_s}{4\pi} \sum_{L=0}^\infty \bar{\Pi}_L \biggl(C_F \frac{\alpha_s}{4\pi}\biggr)^{\!L} \biggr]
+ \cdots
\nonumber\\
&{} = - \frac{4 \pi \alpha_s}{\vec{q}^{\,2}} C_R \biggl\{ 1 + T_F n_f \frac{\alpha_s}{4\pi} \biggl[ - \frac{20}{9}
+ \biggl(16 \zeta_3 - \frac{55}{3}\biggr) C_F \frac{\alpha_s}{4\pi}
\nonumber\\
&\quad{} - 2 \biggl(80 \zeta_5 - \frac{148}{3} \zeta_3 - \frac{143}{9}\biggr) \biggl(C_F \frac{\alpha_s}{4\pi}\biggr)^{\!2}
\nonumber\\
&\quad{} + \biggl(2240 \zeta_7 - 1960 \zeta_5 - 104 \zeta_3 + \frac{31}{3} \biggr) \biggl(C_F \frac{\alpha_s}{4\pi}\biggr)^{\!3}
+ \mathcal{O}(\alpha_s^4)  \biggr] \biggr\} + \cdots
\label{CF:V}
\end{align}
where $\alpha_s = \alpha_s(|\vec{q}^{\,}|)$ and dots mean other color structures.
The terms up to $\alpha_s^4$ agree with\cite{Smirnov:2008pn}.

Comparing~(\ref{CF:V}) with~(\ref{CF:Gamma}), we see that the $C_F^{L-1} T_F n_f$ color structures
are absent in $\Delta$ to all orders in $\alpha_s$.
In particular, this explains the absence of $C_F$ in the bracket in~(\ref{pi:3}).
It is easy to prove by induction that $C_R C_F^{L-1} \alpha_s^L$ terms
are absent in $C(\alpha_s)$~(\ref{pi:beta}) to all orders\cite{Grozin:2018vdn}.

\section{$C_R C_F^2 (T_F n_f)^2 \alpha_s^5$ and $C_R \bar{d}_{FF} n_f^2 \alpha_s^5$}
\label{S:5l}

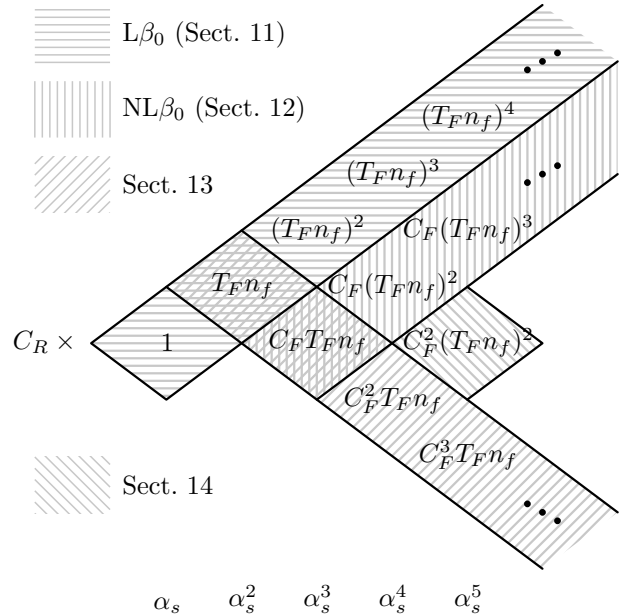
\begin{figure}[ht]
\begin{center}
\begin{picture}(80,82)
\put(44,43.5){\makebox(0,0){\includegraphics{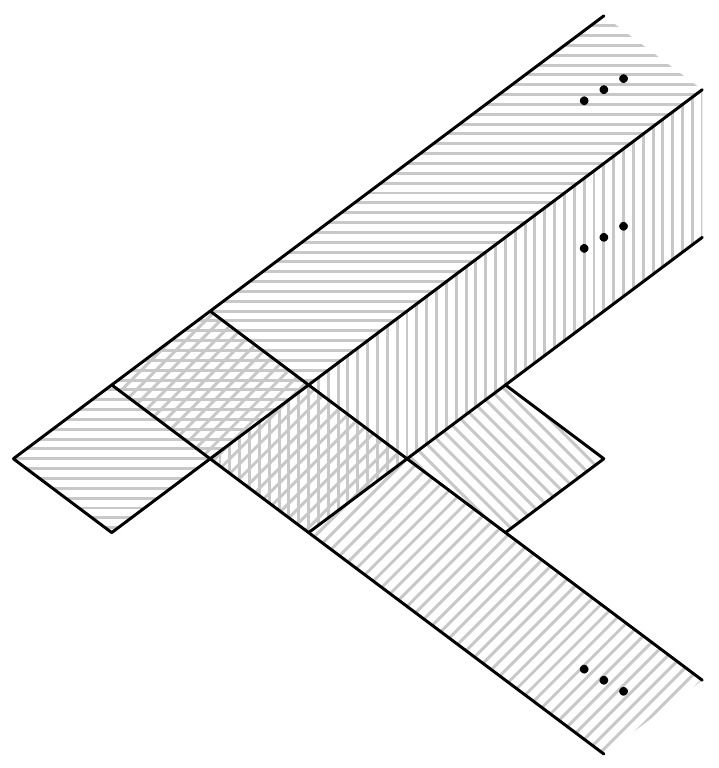}}}
\put(3,36){\makebox(0,0){$C_R\times{}$}}
\put(19,36){\makebox(0,0){$1$}}
\put(29,43.5){\makebox(0,0){$T_F n_f$}}
\put(39,51){\makebox(0,0){$(T_F n_f)^2$}}
\put(49,58.5){\makebox(0,0){$(T_F n_f)^3$}}
\put(59,66){\makebox(0,0){$(T_F n_f)^4$}}
\put(39,36){\makebox(0,0){$C_F T_F n_f$}}
\put(49,43.5){\makebox(0,0){$C_F (T_F n_f)^2$}}
\put(59,51){\makebox(0,0){$C_F (T_F n_f)^3$}}
\put(59,36){\makebox(0,0){$C_F^2 (T_F n_f)^2$}}
\put(49,28.5){\makebox(0,0){$C_F^2 T_F n_f$}}
\put(59,21){\makebox(0,0){$C_F^3 T_F n_f$}}
\put(19,0){\makebox(0,0)[b]{$\alpha_s$}}
\put(29,0){\makebox(0,0)[b]{$\alpha_s^2$}}
\put(39,0){\makebox(0,0)[b]{$\alpha_s^3$}}
\put(49,0){\makebox(0,0)[b]{$\alpha_s^4$}}
\put(59,0){\makebox(0,0)[b]{$\alpha_s^5$}}
\put(0,77){\makebox(0,0)[l]{\includegraphics{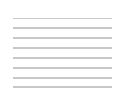}}}
\put(13,77){\makebox(0,0)[l]{L$\beta_0$ (Sect.~\ref{S:Lb0})}}
\put(0,67){\makebox(0,0)[l]{\includegraphics{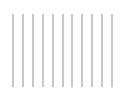}}}
\put(13,67){\makebox(0,0)[l]{NL$\beta_0$ (Sect.~\ref{S:NLb0})}}
\put(0,57){\makebox(0,0)[l]{\includegraphics{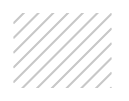}}}
\put(13,57){\makebox(0,0)[l]{Sect.~\ref{S:CF}}}
\put(0,17){\makebox(0,0)[l]{\includegraphics{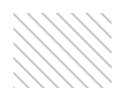}}}
\put(13,17){\makebox(0,0)[l]{Sect.~\ref{S:5l}}}
\end{picture}
\end{center}
\caption{Abelian color structures.}
\label{F:Abel}
\end{figure}

In Sects.~\ref{S:Lb0}--\ref{S:CF} we considered some families
of abelian color structures $C_R C_F^{L-n-1} (T_F n_f)^n \alpha_s^L$
in the HQET field anomalous dimension $\gamma_h$,
the cusp anomalous dimension $\Gamma(\varphi)$,
the quark--antiquark potential $V(\vec{q}^{\,})$
(and hence the conformal anomaly $\Delta$).
These families are shown in Fig.~\ref{F:Abel};
some of them intersect, and we can compare results of the corresponding approaches.
We see that the only 5-loop structure from this class
not considered in the previous sections is $C_R C_F^2 (T_F n_f)^2 \alpha_s^5$.
It can be considered using the general guidelines outlined in Sect.~\ref{S:abel}
if we choose $S = \{1, T_F n_f, C_F T_F n_f, C_F^2 T_F n_f, C_F^2 (T_F n_f)^2\}$.

The gluon self-energy diagram
\begin{equation*}
\includegraphics{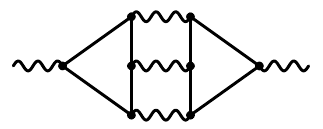}
\end{equation*}
and similar ones with permutations of vertices
contain the color structure $\bar{d}_{FF} n_f^2$, where
\begin{equation}
\bar{d}_{FF} = \frac{d_F^{abcd} d_F^{abcd}}{N_A}\,.
\label{5l:dbar}
\end{equation}
Its contribution ($C_R \bar{d}_{FF} n_f^2 \alpha_s^5$) to $\gamma_h$ and similar quantities
can be obtained by setting $S = \{\bar{d}_{FF} n_f^2\}$.

Alternatively, we can just consider all $C_F^{L-n-1} (T_F n_f)^n$ up to $L=5$
plus $\bar{d}_{FF} n_f^2$,
and re-derive the corresponding results of Sects.~\ref{S:Lb0}--\ref{S:CF}
together with obtaining two new results.
The HQET field anomalous dimension is
\begin{align}
&\gamma_h = C_R \frac{\alpha_s}{4\pi} \biggl\{ - 6
+ \frac{32}{3} T_F n_f \frac{\alpha_s}{4\pi}
- 2 T_F n_f \biggl(\frac{\alpha_s}{4\pi}\biggr)^{\!2}
\biggl[3 C_F (16 \zeta_3 - 17) - \frac{80}{27} T_F n_f\biggr]
\nonumber\\
&{} + 16 T_F n_f \biggl(\frac{\alpha_s}{4\pi}\biggr)^{\!3}
\biggl[C_F^2 \biggl(60 \zeta_5 - 37 \zeta_3 - \frac{35}{3}\biggr)
\nonumber\\
&\quad{} - 2 C_F T_F n_f \biggl(\frac{\pi^4}{15} - 12 \zeta_3 + \frac{103}{27}\biggr)
- \frac{16}{9} (T_F n_f)^2 \biggl(\zeta_3 - \frac{1}{3}\biggr)\biggr]
\nonumber\\
&{} - \biggl(\frac{\alpha_s}{4\pi}\biggr)^{\!4}
\biggl[6 C_F^3 T_F n_f (2240 \zeta_7 - 1960 \zeta_5 - 104 \zeta_3 - 5)
\nonumber\\
&\quad{} + C_F^2 (T_F n_f)^2 \biggl(1792 \zeta_3^2 - \frac{640}{189} \pi^6 + \frac{17920}{3} \zeta_5 + \frac{88}{5} \pi^4
- \frac{68096}{9} \zeta_3 + \frac{59411}{27}\biggr)
\nonumber\\
&\quad{} + 32 \bar{d}_{FF} n_f^2 \biggl(32 \zeta_3^2 - 60 \zeta_5 + \frac{2}{15} \pi^4 + \frac{157}{3} \zeta_3 - 43\biggr)
\nonumber\\
&\quad{} + \frac{4}{3} C_F (T_F n_f)^3 \biggl(256 \zeta_5 - \frac{32}{5} \pi^4 + \frac{944}{3} \zeta_3 + \frac{14579}{81}\biggr)
\nonumber\\
&\quad{} + \frac{256}{81} (T_F n_f)^4 \biggl(\frac{\pi^4}{5} - 16 \zeta_3 - \frac{14}{3}\biggr)\biggr]
\biggr\} + \cdots
\label{5l:gammah}
\end{align}
where dots mean other color structures.
The corresponding parts of~(\ref{Lb0:gammah0}), (\ref{NLb0:gammah}), (\ref{CF:gamma})
are reproduced;
the $C_R C_F^2 (T_F n_f)^2 \alpha_s^5$ and $C_R \bar{d}_{FF} n_f^2 \alpha_s^5$ results are new.

The cusp anomalous dimension is
\begin{align}
&\Gamma(\varphi) = 4 C_R (\varphi \coth\varphi - 1) \frac{\alpha_s}{4\pi} \biggl\{1
- \frac{20}{9} T_F n_f \frac{\alpha_s}{4\pi}
\nonumber\\
&{} + T_F n_f \biggl(\frac{\alpha_s}{4\pi}\biggr)^{\!2}
\biggl[C_F \biggl(16 \zeta_3 - \frac{55}{3}\biggr) - \frac{16}{27} T_F n_f\biggr]
\nonumber\displaybreak\\
&{} + T_F n_f \biggl(\frac{\alpha_s}{4\pi}\biggr)^{\!3}
\biggl[- 2 C_F^2 \biggl(80 \zeta_5 - \frac{148}{3} \zeta_3 - \frac{143}{9}\biggr)
\nonumber\\
&\quad{} + \frac{8}{9} C_F T_F n_f \biggl(\frac{2}{5} \pi^4 - 80 \zeta_3 + \frac{299}{9}\biggr)
+ \frac{64}{27} (T_F n_f)^2 \biggl(2 \zeta_3 - \frac{1}{3}\biggr)\biggr]
\nonumber\\
&{} + \biggl(\frac{\alpha_s}{4\pi}\biggr)^{\!4}
\biggl[C_F^3 T_F n_f \biggl(2240 \zeta_7 - 1960 \zeta_5 - 104 \zeta_3 + \frac{31}{3}\biggr)
\nonumber\\
&\quad{} + \frac{C_F^2 (T_F n_f)^2}{3} \biggl(896 \zeta_3^2 - \frac{320}{189} \pi^6 + 3200 \zeta_5 + \frac{44}{5} \pi^4
- \frac{36512}{9} \zeta_3 + \frac{62971}{54}\biggr)
\nonumber\\
&\quad{} + 16 \bar{d}_{FF} n_f^2 \biggl(\frac{32}{3} \zeta_3^2 - 20 \zeta_5 + \frac{2}{45} \pi^4 + 21 \zeta_3 - \frac{431}{27}\biggr)
\nonumber\\
&\quad{} + \frac{2}{9} C_F (T_F n_f)^3 \biggl(256 \zeta_5 - \frac{64}{9} \pi^4 + \frac{3728}{9} \zeta_3 + \frac{15211}{81}\biggr)
\nonumber\\
&\quad{} + \frac{128}{243} (T_F n_f)^4 \biggl(\frac{\pi^4}{5} - 20 \zeta_3 - 2\biggr)\biggr]
\biggr\} + \cdots
\label{5l:Gamma}
\end{align}
where dots mean other color structures.
The corresponding parts of~(\ref{Lb0:Gamma0}), (\ref{NLb0:Gamma}), (\ref{CF:Gamma})
are reproduced;
the $C_R C_F^2 (T_F n_f)^2 \alpha_s^5$ and $C_R \bar{d}_{FF} n_f^2 \alpha_s^5$ results are new.

The quark-antiquark potential is
\begin{align}
&V(\vec{q}^{\,}) = - \frac{4 \pi \alpha_s}{\vec{q}^{\,2}} \biggl\{1
- \frac{20}{9} T_F n_f \frac{\alpha_s}{4\pi}
+ T_F n_f \biggl(\frac{\alpha_s}{4\pi}\biggr)^{\!2}
\biggl[C_F \biggl(16 \zeta_3 - \frac{55}{3}\biggr) + \frac{400}{81} T_F n_f\biggr]
\nonumber\\
&{} + T_F n_f \biggl(\frac{\alpha_s}{4\pi}\biggr)^{\!3}
\biggl[- 2 C_F^2 \biggl(80 \zeta_5 - \frac{148}{3} \zeta_3 - \frac{143}{9}\biggr)
\nonumber\\
&\quad{} - \frac{2}{3} C_F T_F n_f \biggl(208 \zeta_3 - \frac{7001}{27}\biggr)
- \frac{8000}{729} (T_F n_f)^2\biggr]
\nonumber\\
&{} + \biggl(\frac{\alpha_s}{4\pi}\biggr)^{\!4}
\biggl[C_F^3 T_F n_f \biggl(2240 \zeta_7 - 1960 \zeta_5 - 104 \zeta_3 + \frac{31}{3}\biggr)
\nonumber\\
&\quad{} + C_F^2 (T_F n_f)^2 \biggl(512 \zeta_3^2 + \frac{22400}{9} \zeta_5 - \frac{44}{135} \pi^4
- \frac{25792}{9} \zeta_3 + \frac{13025}{54}\biggr)
\nonumber\\
&\quad{} + 16 \bar{d}_{FF} n_f^2 \biggl(\frac{32}{3} \zeta_3^2 - 20 \zeta_5 + \frac{2}{45} \pi^4 + 21 \zeta_3 - \frac{431}{27}\biggr)
\nonumber\\
&\quad{} + \frac{2}{9} C_F (T_F n_f)^3 \biggl(640 \zeta_5 + 3856 \zeta_3 - \frac{147851}{27}\biggr)
+ \frac{160000}{6561} (T_F n_f)^4 \biggr]
\biggr\} + \cdots
\label{5l:V}
\end{align}
where $\alpha_s = \alpha_s(|\vec{q}^{\,}|)$ and dots mean other color structures.
The corresponding parts of~(\ref{Lb0:V}), (\ref{NLb0:V1}), (\ref{CF:V})
are reproduced;
the $C_R C_F^2 (T_F n_f)^2 \alpha_s^5$ and $C_R \bar{d}_{FF} n_f^2 \alpha_s^5$ results are new.

We can obtain $\Delta(\alpha_s)$~(\ref{pi:Delta}) from $\Gamma(\varphi)$~(\ref{5l:Gamma}) and $V(\vec{q}^{\,})$~(\ref{5l:V}).
The $C_R \bar{d}_{FF} n_f^2 \alpha_s^5$ contribution has canceled,
just like all $C_R C_F^{L-2} T_F n_f \alpha_s^L$ ones (Sect.~\ref{S:CF}),
and for the same reason: there is just a single $\Pi(k^2)$ insertion
both in the diagram for $\Gamma(\varphi)$ and in the diagram for $V(\vec{q}^{\,})$.
If it had not canceled, it would be impossible to introduce $C(\alpha_s)$~(\ref{pi:beta}).
We obtain
\begin{align}
&C(\alpha_s) = C_R T_F n_f \biggl(\frac{\alpha_s}{4\pi}\biggr)^{\!2}
\biggl\{- \frac{112}{27}
\nonumber\\
&{} + \frac{\alpha_s}{4\pi} \biggl[
\frac{C_F}{3} \biggl(\frac{4}{5} \pi^4 + 152 \zeta_3 - \frac{1711}{6}\biggr)
+ \frac{32}{9} T_F n_f \biggl(\zeta_3 + \frac{58}{27}\biggr)
\biggr]
\nonumber\\
&{} + \biggl(\frac{\alpha_s}{4\pi}\biggr)^{\!2} \biggl[
- C_F^2 \biggl(160 \zeta_3^2 + \frac{80}{189} \pi^6 + \frac{3200}{3} \zeta_5 - \frac{74}{45} \pi^4 - \frac{8848}{9} \zeta_3
- \frac{21037}{54}\biggr)
\nonumber\\
&\quad{} - 2 C_F T_F n_f \biggl(32 \zeta_5 + \frac{16}{27} \pi^4 + \frac{7888}{27} \zeta_3 - \frac{110059}{243}\biggr)
\nonumber\\
&\quad{} + \frac{32}{81} (T_F n_f)^2 \biggl(\frac{\pi^4}{5} - 20 \zeta_3 - \frac{1304}{27}\biggr)
\biggr]\biggr\} + \cdots
\label{5l:C}
\end{align}
where dots mean other color structures.
The corresponding parts of~(\ref{pi:C}), (\ref{Lb0:C}), (\ref{NLb0:C})
are reproduced;
all $C_R C_F^{L-1} \alpha_s^L$ terms vanish (Sect.~\ref{S:CF});
the $C_R C_F^2 T_F n_f \alpha_s^5$ result is new.

\begin{figure}[ht]
\begin{center}
\begin{picture}(80,82)
\put(54,43.5){\makebox(0,0){\includegraphics{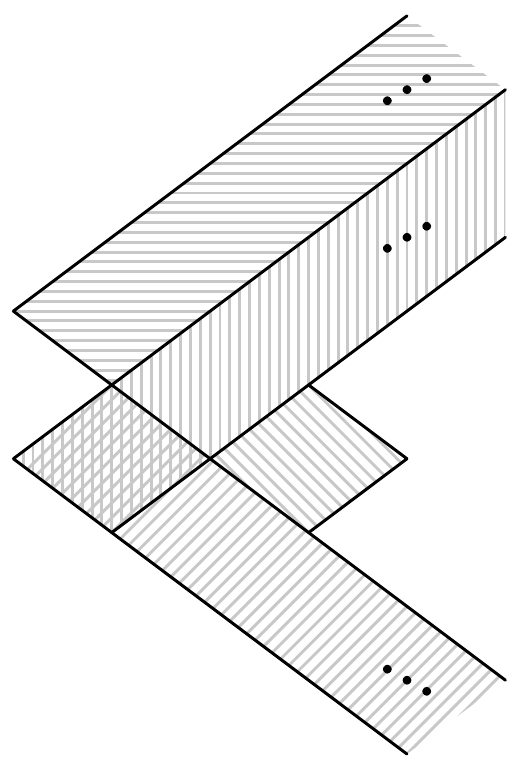}}}
\put(23,43.5){\makebox(0,0){$C_R\times{}$}}
\put(39,51){\makebox(0,0){$T_F n_f$}}
\put(49,58.5){\makebox(0,0){$(T_F n_f)^2$}}
\put(59,66){\makebox(0,0){$(T_F n_f)^3$}}
\put(39,36){\makebox(0,0){$C_F$}}
\put(49,43.5){\makebox(0,0){$C_F T_F n_f$}}
\put(59,51){\makebox(0,0){$C_F (T_F n_f)^2$}}
\put(49,28.5){\makebox(0,0){$C_F^2$}}
\put(59,36){\makebox(0,0){$C_F^2 T_F n_f$}}
\put(59,21){\makebox(0,0){$C_F^3$}}
\put(39,0){\makebox(0,0)[b]{$\alpha_s^2$}}
\put(49,0){\makebox(0,0)[b]{$\alpha_s^3$}}
\put(59,0){\makebox(0,0)[b]{$\alpha_s^4$}}
\put(0,67){\makebox(0,0)[l]{\includegraphics{acap1.pdf}}}
\put(13,67){\makebox(0,0)[l]{L$\beta_0$ (\ref{Lb0:C})}}
\put(0,20){\makebox(0,0)[l]{\includegraphics{acap2.pdf}}}
\put(13,20){\makebox(0,0)[l]{NL$\beta_0$ (\ref{NLb0:C})}}
\put(0,77){\makebox(0,0)[l]{\includegraphics{acap3.pdf}}}
\put(13,77){\makebox(0,0)[l]{0}}
\put(0,10){\makebox(0,0)[l]{\includegraphics{acap4.pdf}}}
\put(13,10){\makebox(0,0)[l]{(\ref{5l:C})}}
\end{picture}
\end{center}
\caption{Abelian color structures for $C(\alpha_s)$.}
\label{F:AbelC}
\end{figure}

Let's summarize what's known about the conformal anomaly $C(\alpha_s)$.
The abelian structures $C_R C_F^{L-n-1} (T_F n_f)^n \alpha_s^L$ ($L\ge2$) are shown in Fig.~\ref{F:AbelC}.
All contributions with $n=0$ vanish
(Sect.~\ref{S:CF}; they are shown in the corresponding shading in the figure).
The contributions $C_R (T_F n_f)^{L-1} \alpha_s^L$ ($L\ge2$)
are the leading large-$\beta_0$ ones (Sect.~\ref{S:Lb0});
several of them are presented in~(\ref{Lb0:C}),
but a practically infinite number of them are easily available.
The contributions $C_R C_F (T_F n_f)^{L-2} \alpha_s^L$ ($L\ge2$)
are next-to-leading large-$\beta_0$ ones (Sect.~\ref{S:NLb0}).
The first of them ($L=2$) also belongs to the first family, and hence vanishes.
The contributions with $L\le7$ are presented in~(\ref{NLb0:C});
several more can be obtained using known algorithms,
but the calculational complexity grows fast with $L$.
The $C_R C_F^2 T_F n_f \alpha_s^4$ result is obtained here.
No further $C_R C_F^{L-n-1} (T_F n_f)^n \alpha_s^L$ terms can be obtained
without calculating $\Pi(k^2)$ beyond the 4-loop results\cite{Ruijl:2017eht},
and this is a highly non-trivial task.
The non-abelian terms $C_R C_A \alpha_s^2$ and $C_R C_A T_F n_l \alpha_s^3$ are also known~(\ref{pi:C}).

\section{Conclusion}
\label{S:Conc}

At 4 loops, the HQET field anomalous dimension,
as well as the small-angle expansion
and the large-angle asymptotics of the cusp anomalous dimension,
are completely known.
The full angle dependence is known for all color structures
except $C_R C_A^2 T_F n_f$, $C_R C_A^3$, and $d_{RA}$.
The highest weight terms in the QCD results for the Bremsstrahlung function
(the $\varphi^2$ term of the small-angle expansion)
and for the light-like cusp anomalous dimension
(the large $\varphi$ asymptotics)
coincide with the corresponding results in $\mathcal{N}=4$ SYM ---
the principle of maximum transcendentality.

All $\alpha_s^4/\varepsilon$ terms in the on-shell renormalization constant
of the massive quark field in QCD are known analytically.
The results for the color structures $C_F C_A^3$ and $d_{FA}$ are new.

At 5 loops, the abelian color structures
$C_R (T_F n_f)^4$, $C_R C_F (T_F n_f)^3$, $C_R C_F^2 (T_F n_f)^2$, $C_R C_F^3 T_F n_f$, and $C_R \bar{d}_{FF} n_f^2$
in the HQET field anomalous dimension and the cusp anomalous dimension are known.
The results for $C_R C_F^2 (T_F n_f)^2$ and $C_R \bar{d}_{FF} n_f^2$ are new.
The structures $C_R (T_F n_f)^{L-1} \alpha_s^L$ are known to all loop orders $L$;
$C_R C_F (T_F n_f)^{L-2} \alpha_s^L$ are presented explicitly up to $L=8$ loops,
and more terms can be obtained using known algorithms.
All abelian results for $\Gamma(\varphi)$ at 5 and more loops,
obtained in Sects.~\ref{S:Lb0}--\ref{S:5l},
have the 1-loop $\varphi$ dependence $\varphi \coth\varphi - 1$,
and hence they trivially give contributions to the light-like cusp anomalous dimension $K(\alpha_s)$.

The conjecture about the cusp anomalous dimension proposed in\cite{Grozin:2015kna}
works up to 3 loops and for some color structures at 4 loops,
but breaks for some other 4-loop structures
(its formulation at $L \ge 4$ is not quite unambiguous).
The reason why it works in some highly non-trivial cases is still unknown
(all known results for $L \ge 5$ have 1-loop angle dependence,
and the conjecture holds for them by construction,
so, they add no new information).

The cusp anomalous dimension with euclidean angle close to $\pi$ is related to the static potential.
This relation follows from conformal symmetry, and is strictly valid for $\mathcal{N}=4$ SYM.
In QCD its breaking is given by the conformal anomaly $\Delta(\alpha_s)$.
It is conjectured (though not proven) that $\Delta(\alpha_s) = \beta(\alpha_s) C(\alpha_s)$.
The coefficient function $C(\alpha_s)$ is known at 3 loops, most 4-loop color structures are also known.
The structures $C_R (T_F n_f)^{L-1} \alpha_s^L$ is it are known to all loop orders $L$,
and $C_R C_F (T_F n_f)^{L-2} \alpha_s^L$ --- up to $L=7$ (this expansion can be extended).
The structures $C_R C_F^{L-2} T_F n_f \alpha_s^L$ vanish for all $L$.

New results for the 4-loop contributions $C_R C_F^2 (T_F n_f)^2 \alpha_s^5$ and $C_R \bar{d}_{FF} n_f^2 \alpha_s^5$
(see~(\ref{5l:dbar})) to the quark-antiquark static potential $V(\vec{q}^{\,})$ are presented.
The last one cancels in $\Delta(\alpha_s)$
(thus giving one more confirmation of its factorization into $\beta(\alpha_s) C(\alpha_s)$);
the first one produced the new $C_R C_F^2 T_F n_l \alpha_s^4$ term in $C(\alpha_s)$.

There is an additional problem here:
the coefficient of $C_R C_A^3 \alpha_s^4$ has a logarithmic singularity at $\delta\to0$,
and the definition of $\Delta(\alpha_s)$ breaks down.
It is supposed that resummation of leading powers of this logarithm will lead to a finite result,
probably, containing $\log\alpha_s$, but the details are not clear.
If this is so, the cusp anomalous dimension will contain a logarithmic dependence on $\alpha_s$,
which is very unusual for anomalous dimensions in quantum field theory.
This question needs further clarification.

I am grateful to
D.\,J.~Broadhurst,
R.~Br\"user,
K.\,G.~Chetyrkin,
J.\,M.~Henn,
G.\,P.~Korchemsky,
A.\,V.~Kotikov,
R.\,N.~Lee,
P.~Marquard,
A.\,F.~Pikelner,
A.\,V.~Smirnov,
V.\,A.~Smirnov,
M.~Stahlhofen,
M.~Steinhauser
for collaboration on various projects related to the present review,
and to A.\,L.~Kataev and V.\,S.~Molokoedov for discussing\cite{Kataev:2022iqf}.
All Feynman diagrams in this article have been produced using \texttt{feyn.gle}\cite{Grozin:2022fde}.
The work has been supported by Russian Science Foundation under grant 20-12-00205.

\bibliographystyle{ws-ijmpa}
\bibliography{cusp}

\end{document}